\documentclass[a4paper]{article}

\usepackage{epsfig}
\usepackage{hyperref}
\usepackage{a4}
\usepackage{latexsym, amssymb} 
\usepackage{amsmath}

\newcommand{\lapprox}{%
\mathrel{%
\setbox0=\hbox{$<$}
\raise0.6ex\copy0\kern-\wd0
\lower0.65ex\hbox{$\sim$}
}}
\newcommand{\gapprox}{%
\mathrel{%
\setbox0=\hbox{$>$}
\raise0.6ex\copy0\kern-\wd0
\lower0.65ex\hbox{$\sim$}
}}

\newcommand{\ba}{\begin{array}}
\newcommand{\ea}{\end{array}}
\newcommand{\bd}{\begin{displaymath}}
\newcommand{\ed}{\end{displaymath}}
\newcommand{\be}{\begin{equation}}
\newcommand{\ee}{\end{equation}}
\newcommand{\bea}{\begin{eqnarray}}
\newcommand{\eea}{\end{eqnarray}}

\catcode`@=11 % This allows us to modify PLAIN macros.
\def \gsim{\mathrel{\mathpalette\@versim>}}
\def \lsim{\mathrel{\mathpalette\@versim<}}
\def \@versim#1#2{\lower0.4ex\vbox{\baselineskip\z@skip\lineskip\z@skip
     \lineskiplimit\z@\ialign{$\m@th#1\hfil##\hfil$%
     \crcr#2\crcr\sim\crcr}}}
\catcode`@=12 % at signs are no longer letters

\makeatletter
 
 \@addtoreset{equation}{section}
\makeatother

\begin{document}

\begin{flushright}
{\small KEK-TH-1785}
\end{flushright}

\begin{center}

{\large\bf Study on the azimuthal angle correlation between two jets in the top quark pair production}\\[15mm]
Kaoru Hagiwara\footnote{E-mail: kaoru.hagiwara@kek.jp}
and Junya Nakamura\footnote{E-mail: junya.nakamura@uni-tuebingen.de}\\ \bigskip
{\em $^{1,2}$KEK Theory Center and Sokendai, \\
Tsukuba, Ibaraki 305-0801, Japan\\ \bigskip
$^2$  Institute for Theoretical Physics,
 University of T\"ubingen, \\
 Auf der Morgenstelle 14,
 72076 T\"ubingen, Germany.}
\\[20mm] 
\end{center}

\begin{abstract} 
An azimuthal angle correlation between the two hardest jets is studied in the $t\bar{t}$ production process at the 14 TeV LHC. The event samples are generated by merging the tree level matrix elements for the $t\bar{t}$~plus up to 2 or 3 partons with parton showers. The generated event samples show a strong correlation in the azimuthal angle difference between the two hardest jets, as predicted in the analysis based on the tree level matrix elements for the $t\bar{t}+2$ partons. The effects of merging the matrix elements for the $t\bar{t}+3$ partons on the correlation are studied in detail. It is found that they play important roles in improving the prediction of the correlation.

\end{abstract}

\vskip 1 true cm

\newpage
\setcounter{footnote}{0}

\def\baselinestretch{1.5}
%==============================================================================
%==============================================================================
\tableofcontents

\newpage

\section{Introduction}\label{intro}
Since the discovery of the Higgs boson was announced in the summer of 2012, the LHC measurements of its properties have so far been supporting the standard model (SM) predictions~\cite{Aad:2013wqa, Aad:2014lwa, Chatrchyan:2013mxa, Khachatryan:2014ira}. 
The Higgs sector of the SM respects the charge-conjugation and parity (CP) symmetry and the Higgs boson should be CP even. Therefore if an admixture of the CP odd component is observed, it will be a direct evidence of CP violation in the Higgs sector and thus physics beyond the SM.\\

From the analyses on the tree level matrix elements, it has been shown that the azimuthal angle difference between the two partons (gluon, quarks or antiquarks) produced in association with the Higgs boson produced by gluon fusion is very sensitive to the CP property of the Higgs boson~\cite{Plehn:2001nj, DelDuca:2001fn, Klamke:2007cu, Hagiwara:2009wt}. 
Several analyses including effects of higher order corrections show that the correlation between the two partons found at the tree level matrix elements can be observed as a correlation between the two hardest {\it jets} despite smearing, see e.g. refs.~\cite{DelDuca:2006hk, Ruwiedel:2007zz, Andersen:2008gc, Andersen:2010zx}. 
However the attempts to observe the CP odd admixture precisely in this approach are expected to be difficult due to large theoretical uncertainties in Monte Carlo event simulation, particularly in our use of a parton shower generator which can simulate the QCD radiation only in the soft and/or collinear limit. \\

It has been pointed out in ref.~\cite{Hagiwara:2013jp} that the two partons produced in association with a top quark pair has a large azimuthal angle correlation near the threshold $m_{t\bar{t}}^{}\sim 2m_t$ and the correlation is similar to that of the two partons produced together with the CP odd Higgs boson via gluon fusion. 
The claim of ref.~\cite{Hagiwara:2013jp} is that experimental techniques to measure such an angular correlation between jets can be established first by using these SM processes which have large cross sections. More precisely, we measure the azimuthal angle difference between two jets produced in association with a top quark pair and tune a Monte Carlo event generator to reproduce the data quantitatively. If an event generator tuned in this way is used, the theoretical uncertainty on the prediction of the azimuthal angle correlation between two jets produced in association with the Higgs boson can be reduced significantly. This will help achieve accurate measurements of the CP property of the Higgs boson. \\

In the present paper we attempt to create a bridge between the proposal of ref.~\cite{Hagiwara:2013jp} and actual experimental measurements, by studying our present capability and limitation of simulating the top quark pair plus multi-jet production process,
so that experimentalists can use the real data to improve our simulation tools to be used to probe more fundamental physics such as the CP property of the Higgs boson. 
The simplest method to include leading higher order corrections to the top quark pair production is to apply a parton shower generator to the exclusive top quark pair events, where the parton shower scale evolution of the top quark pair events produces the top quark pair plus multi-jet events at a hadronization scale.
The event samples generated in this way are expected to reproduce qualitatively the multi-jet event rates and the jet $p_T^{}$ and rapidity distribution, since the successive emission of parton showers follows the QCD prediction in the soft and/or collinear region and the overall jet rates have been fitted to the data in $e^+e^-$ and hadronic collisions. 
Those events, however, do not have correct correlations among jets since a parton shower generator emits azimuthally symmetric radiation about a parent momentum direction. 
To reproduce azimuthal angle correlations between two jets, at least the $t\bar{t}+2$ partons matrix elements have to be embedded. 
In order to consistently combine the event samples for different parton multiplicity generated by tree level matrix elements with parton showers,
a tree level merging algorithm is required. There exists several tree level merging algorithms proposed in literatures, including the CKKW~\cite{Catani:2001cc, Krauss:2002up, Hoeche:2009rj, Hamilton:2009ne}, the CKKW-L~\cite{Lonnblad:2001iq, Lavesson:2005xu, Lonnblad:2011xx}, the GKS~\cite{Giele:2011cb}, the MLM~\cite{Mangano:2001xp, Mangano:2006rw}, the pseudo shower~\cite{Mrenna:2003if} and the shower $k_{\perp}^{}$~\cite{Alwall:2008qv} algorithms. Comparisons between these different algorithms have also been studied~\cite{Mrenna:2003if, Alwall:2008qv, Alwall:2007fs, Lavesson:2007uu}.
Our objective in the present paper is to implement tree level merging algorithms and study theoretical issues on predicting the azimuthal angle correlation between the two hardest jets. \\

We generate the event samples for the top quark pair production at the 14 TeV LHC, by merging the tree level matrix elements for the $t\bar{t}$ plus up to 2 or 3 partons with the parton shower model in PYTHIA8. By using the generated event samples, the azimuthal angle difference between the two hardest jets (i.e. the two highest transverse momentum jets), $\Delta \phi=\phi_1^{}-\phi_2^{}$, is studied. \\

As tree level merging algorithms, the CKKW-L merging algorithm~\cite{Lonnblad:2001iq, Lavesson:2005xu, Lonnblad:2011xx} and a new tree level merging algorithm are implemented. Our new algorithm differs from the CKKW-L algorithm in the strategy for phase space separation. It is designed so that the contribution from the $t\bar{t}+0, 1$ parton matrix elements to the event samples with two or more jets, which we call the contamination, can be more suppressed above the merging scale. Therefore, more accurate predictions on correlations between two jets are expected. We confirm this by numerically comparing the two algorithms.
The contamination is studied by varying a relation between the merging scale and the scale of a jet definition. We find that the contamination is not negligible when the merging scale is set equal to or slightly smaller than the scale of the anti-$k_T^{}$ jet definition. \\

We produce the $\Delta \phi$ distribution by using the generated event samples. The distribution shows a strong correlation in $\Delta \phi$, as predicted in the previous analysis~\cite{Hagiwara:2013jp} based on the $t\bar{t}+2$ partons tree level matrix elements. This observation confirms that the correlation found in the $t\bar{t}+2$ partons tree level matrix elements~\cite{Hagiwara:2013jp} is still visible after including the dominant QCD higher order corrections and thus can be observed in the experiments. \\

We observe a clear difference in the $\Delta \phi$ distribution between the event samples generated by merging the matrix elements for the $t\bar{t}$ plus up to 2 partons and those generated by merging the matrix elements for the $t\bar{t}$ plus up to 3 partons. Furthermore, the difference is found slightly larger, when the rapidity range for jets is more restricted. 
We study the effects of the $t\bar{t}+3$ partons matrix elements on the correlation in detail and find out the origins of the difference. 
We show that the $t\bar{t}+3$ partons matrix elements play important roles in predicting $\Delta \phi$ accurately. \\

We present a method for merging the matrix element event samples which include a $t\bar{t}$ decay as a part of the hard process with the parton shower. 
In this method, correlations between the decay products of the $t\bar{t}$ are predicted correctly, while a merging algorithm is performed as if the $t\bar{t}$ was not decayed~\footnote{This is necessary, since QCD radiation off a top quark takes place faster than its decay.}. 
The effect of the $t\bar{t}$ dilepton decay on the azimuthal angle correlation is studied. We find that the effect is small, when the two hardest jets are picked up from all jets not including the two hardest b jets. \\

We note in passing that this is not the first attempt to estimate higher order corrections to the azimuthal angle difference between the two partons produced in association with a top quark pair.
The azimuthal angle difference between two jets in the $t\bar{t}$ production process has been studied with the aim of a scalar top quark search in ref.~\cite{Buckley:2014fqa}, of a gluino search in ref.~\cite{Mukhopadhyay:2014dsa} and of investigating top quark mass effects in the effective Higgs-gluon coupling in refs.~\cite{Buschmann:2014twa} and \cite{Buschmann:2014sia}. \\

In Section~\ref{algorithm}, our implementation of the tree level merging algorithms is described in detail. In Section~\ref{result1}, the azimuthal angle correlation is studied.
In Section~\ref{conclusion}, we summarize our findings.

%%%%%%%%%%%%%%%%%%%%%%%%%%%%%%%%%%%%%%%%%%%%%%%%%%%%%%%%%%%%%%%%%%%%%%%%%%%%%%%%%%%%%%%%%%%%%%%%%%%%%%%%%%%%%%%%%%%%%%%%%%%%%%%%%%%%%%%%%%%%%%%%%%%%%%%%%%%%%%%%%%%%%%%%%%%%%%%%%%%%%%%%%%%%%%%%%%%%%%%%%%%%%%%%%%%%%%%%%%%%%%%%%%%%%%%%%%%%%%%%%%%%%%%%%%%%%%%%%%%%%%%%%%%%%%%%%%%%%%%%%%%%%%%%%%%%%%%%%%%%%%%%%%%%%%%%%%%%%%%%%%%%%%%%%%%%%%%%%%%%%%%%%%%%%%%%%%%%%%%%%%%%%%%%%%%%%%%%%%%%%%%%%%%%%%%%%%%%%%%%%%%%%%%%%%%%%%%%%%%%%%%%%%%%%%%%%%%%%%%%%%%%%%%%%%%%%%%%%%%%%%%%%%%%%%%%%%%%%%%%%%%%%%%%%%%%%%%%%%%%%%%%%%%%%%%%%%%%%%%%%%%%%%%%%%%%%%%%%%%%%%%%%%%%%%%%%%%%%%%%%%%%%%%%%%%%%%%%%%%%%%%%%%%%%%%%%%%%%%%%%%%%%%%%%%%%%%%%%%%%%%%%%%%%%%%%%%%%%%%%%%%%
\section{Implementation of merging algorithms}\label{algorithm}

Our implementation of tree level merging algorithms is described in this section. In Section~\ref{algorithmsub}, the basic idea of tree level merging algorithms~\cite{Catani:2001cc} is reviewed. Our notations used throughout the paper are also introduced. 
In Section~\ref{sec:CKKW-L}, the CKKW-L merging algorithm~\cite{Lonnblad:2001iq, Lavesson:2005xu, Lonnblad:2011xx} is reviewed at first. After that, we introduce a new tree level merging algorithm, which differs from the CKKW-L merging algorithm in the strategy for phase space separation.
The procedure of constructing the PYTHIA8 parton shower history is described in Section~\ref{history} and that of calculating the weight function is explained in Section~\ref{sec:weight-functions}. These are ingredients of the merging algorithms. 
A method for consistently merging the matrix elements event samples which include decays of the top and antitop quarks as a part of the hard process is presented in Section~\ref{sec:algorithm-top-decays}.
In Section~\ref{sec:event-generation}, a procedure for the event generation of the top quark pair production and setups for it such as scale choices are explained. Our implementation is carefully tested in Section~\ref{check}.

%%%%%%%%%%%%%%%%%%%%%%%%%%%%%%%%%%%%%%%%%%%%%%%%%%%%%%%%%%%%%%%%%%%%%%%%%%%%%%%%%%%%%%%%%%%%%%%%%%%%%%%%%%%%%%%%%%%%%%%%%%%%%%%%%%%%%%%%%%%%%%%%%%%%%%%%%%%%%%%%%%%%%%%%%%%%%%%%%%%%%%%%%%%%%%%%%%%%%%%%%%%%%%%%%%%%%%%%%%%%%%%%%%%%%%%%%%%%%%%%%%%%%%%%%%%%%%%%%%%%%%%%%%%%%%%%%%%%%%%%%%%%%%%%%%%%%%%%%%%%%%%%%%%%%%%%%%%%%%%%%%%%%%%%%%%%%%%%%%%%%%%%%%%%%%%%%%%%%%%%%%%%%%%%%%%%%%%%%%%%%%%%%%%%%%%%%%%%%%%%%%%%%%%%%%%%%%%%%%%%%%%%%%%%%%%%%%%%%%%%%%%%%%%%%%%%%%%%%%%%%%%%%%%%%%%%%%%%%%%%%%%%%%%%%%%%%%%%%%%%%%%%%%%%%%%%%%%%%%%%%%%%%%%%%%%%%%%%%%%%%%%%%%%%%%%%%%%%%%%%%%%%%%%%%%%%%%%%%%%%%%%%%%%%%%%%%%%%%%%%%%%%%%%%%%%%%%%%%%%%%%%%%%%%%%%%%%%%%%%%%%%%
\subsection{Tree level merging algorithms}\label{algorithmsub}
In this section, the basic idea of tree level merging algorithms~\cite{Catani:2001cc} is reviewed. Our notations used throughout the paper are also presented. \\

Let us start with the DGLAP evolution equation~\cite{Gribov:1972, Altarelli:1977, Dokshitzer:1977} with the Sudakov form factor~\cite{Marchesini:1987cf}
\begin{align}
t \frac{d}{dt} \frac{q(x,t)}{\Delta(t)} 
&= \int_0^{\epsilon(t)} \frac{dz}{z} \frac{\alpha_s^{}}{2\pi} \hat{P}_{qq}^{}(z)  \frac{q(x/z, t)}{\Delta(t)}, \label{dglap-with-Sudakov}
\end{align}
where the Sudakov form factor is given by
\begin{align}
\Delta(t)=\exp{\biggl(-\int^t_{\mu^2_{}}\frac{dt^{\prime}_{}}{t^{\prime}_{}}\int^{\epsilon(t^{\prime}_{})}_0 dz \frac{\alpha_s^{}}{2\pi}\hat{P}_{qq}^{}(z)\biggr)}. \label{Sudakov-original}
\end{align}
Here only the quark parton distribution function $q(x,t)$ (PDF) and the splitting function without the virtual correction $\hat{P}_{qq}^{}(z)$ for $q\to qg$ are introduced in order to simplify our writing. The generalization is, however, simple. Integrating eq.~(\ref{dglap-with-Sudakov}) over $t_{\Lambda}^{}<t <t_X^{}$ gives
\begin{align}
\frac{q(x,t_X^{})}{\Delta(t_X^{})}=\frac{q(x,t_{\Lambda}^{})}{\Delta(t_{\Lambda}^{})}
+ \int^{t_X^{}}_{t_{\Lambda}^{}} \frac{dt}{t}\int^{\epsilon(t)}_0 d\hat{p}_{qq}^{}(z) \frac{q(x/z,t )}{\Delta(t)},\label{dglap-with-sudakov-integrate}
\end{align}
where a short hand notation is used
\begin{align}
d\hat{p}_{qq}^{}(z)= \frac{\alpha_s^{}}{2\pi} \frac{dz}{z} \hat{P}_{qq}^{}(z).\label{short-notation-1}
\end{align}
After the infinite iterations of eq.~(\ref{dglap-with-sudakov-integrate}), it is found that
\begin{align}
\frac{q(x,t_X^{})}{\Delta(t_X^{})}=& \frac{q(x,t_{\Lambda}^{})}{\Delta(t_{\Lambda}^{})}
+  \int^{t_X^{}}_{t_{\Lambda}^{}} \frac{dt_1^{}}{t_1^{}}\int^{\epsilon(t_1^{})}_0 d\hat{p}_{qq}^{}(z_1^{}) \frac{q(x/z_1^{},t_{\Lambda}^{} )}{\Delta(t_{\Lambda}^{})} \nonumber \\
&+  \int^{t_X^{}}_{t_{\Lambda}^{}} \frac{dt_1^{}}{t_1^{}}\int^{\epsilon(t_1^{})}_0 d\hat{p}_{qq}^{}(z_1^{}) 
\int^{t_1^{}}_{t_{\Lambda}^{}} \frac{dt_2^{}}{t_2^{}}\int^{\epsilon(t_2^{})}_0 d\hat{p}_{qq}^{}(z_2^{}) 
 \frac{ q\bigl(x/(z_1^{}z_2^{} \bigr),t_{\Lambda}^{} ) }{\Delta(t_{\Lambda}^{})} + \cdots
.
\end{align}
By dividing this equation by $q(x,t_X^{})/\Delta(t_X^{})$, we find~\cite{Marchesini:1987cf}
\begin{align}
1=& \frac{q(x,t_{\Lambda}^{})}{q(x,t_X^{})} \frac{\Delta(t_X^{})}{\Delta(t_{\Lambda}^{})}
+  \int^{t_X^{}}_{t_{\Lambda}^{}} \frac{dt_1^{}}{t_1^{}}\int^{\epsilon(t_1^{})}_0 d\hat{p}_{qq}^{}(z_1^{}) \frac{ q(x/z_1^{},t_{\Lambda}^{} ) }{ q(x,t_X^{}) }
\frac{\Delta(t_X^{})}{\Delta(t_{\Lambda}^{})} \nonumber \\
&+  \int^{t_X^{}}_{t_{\Lambda}^{}} \frac{dt_1^{}}{t_1^{}}\int^{\epsilon(t_1^{})}_0 d\hat{p}_{qq}^{}(z_1^{}) 
\int^{t_1^{}}_{t_{\Lambda}^{}} \frac{dt_2^{}}{t_2^{}}\int^{\epsilon(t_2^{})}_0 d\hat{p}_{qq}^{}(z_2^{}) 
\frac{ q\bigl(x/(z_1^{}z_2^{}),t_{\Lambda}^{} \bigr) }{ q(x,t_X^{}) }\frac{\Delta(t_X^{})}{\Delta(t_{\Lambda}^{})} + \cdots \label{infinite-iteration-DGLAP-with-Sudakov-previous}
\end{align}
The first term of the right hand side (RHS) in this equation can be regarded as the probability of generating no radiation from a quark $q(x)$ in a proton during the scale evolution of the proton between $t_X^{}$ and $t_{\Lambda}^{}$ ($t_X^{}>t_{\Lambda}^{}$). The second term represents the integrated probability of generating exclusively one radiation from the quark during the scale evolution, and so on. 
The left hand side (LHS) ensures the probability conservation. \\

We write the no radiation probability from a quark $q(x)$ in a proton during the scale evolution of the proton between $t_1^{}$ and $t_2^{}$ ($t_1^{}>t_2^{}$) in the following form~\cite{Marchesini:1987cf}
\begin{align}
\Pi_q^{}(t_1^{}, t_2^{}; x) = \frac{q(x,t_{2}^{})}{q(x,t_1^{})} \frac{\Delta(t_1^{})}{\Delta(t_{2}^{})}.
\end{align}
Then, we can express eq.~(\ref{infinite-iteration-DGLAP-with-Sudakov-previous}) as
\begin{align}
1=&
\Pi_{q}^{}( t_X^{}, t_{\Lambda}^{}; x )+
\int^{t_X^{}}_{t_{\Lambda}^{}} \frac{dt_1^{}}{t_1^{}}\int^{\epsilon(t_1^{})}_0 d\hat{p}_{qq}^{}(z_1^{})\ 
\Pi_{q}^{}( t_X^{}, t_1^{}; x)\ 
\frac{q(x/z_1^{}, t_1^{})}{q(x, t_1^{})}\ 
\Pi_q^{}( t_1^{}, t_{\Lambda}^{}; x/z_1^{} ) \nonumber \\
&
+
\int^{t_X^{}}_{t_{\Lambda}^{}} \frac{dt_1^{}}{t_1^{}}\int^{\epsilon(t_1^{})}_0 d\hat{p}_{qq}^{}(z_1^{}) 
\int^{t_1^{}}_{t_{\Lambda}^{}} \frac{dt_2^{}}{t_2^{}}\int^{\epsilon(t_2^{})}_0 d\hat{p}_{qq}^{}(z_2^{}) 
\Pi_{q}^{}( t_X^{}, t_1^{}; x)
\frac{q(x/z_1^{}, t_1^{})}{q(x, t_1^{})}
\Pi_{q}^{}( t_1^{}, t_{2}^{}; x/z_1^{} ) 
\nonumber \\
&
\hspace{0.3cm}
\times
\frac{q\bigl(x/(z_1^{}z_2^{}), t_2^{}\bigr)}{q(x/z_1^{}, t_2^{})}
\Pi_{q}^{}\bigl( t_2^{}, t_{\Lambda}^{}; x/(z_1^{}z_2^{}) \bigr) \nonumber \\
&+\cdots.\label{infinite-iteration-DGLAP-with-Sudakov}
\end{align}
It is an easy task to guess the explicit form of $\Pi_q^{}(t_1^{}, t_2^{}; x)$ from the above equation~\cite{Sjostrand:1985xi},
\begin{align}
\Pi_q^{}(t_1^{}, t_2^{}; x) = \exp{\biggl( -\int^{t_1^{}}_{t_{2}^{}} \frac{dt}{t}\int^{\epsilon(t)}_0 d\hat{p}_{qq}^{}(z)\frac{q(x/z, t)}{q(x, t)} \biggr)}. \label{no-rad-prob-ISR}
\end{align}
Given the scale $t_X^{}$ and the energy fraction $x$ of a quark in a proton, eq.~(\ref{infinite-iteration-DGLAP-with-Sudakov}) allows us to generate radiations from the incoming quark by evolving the proton from the scale $t_X^{}$.
This is known as backward evolution~\cite{Sjostrand:1985xi, Gottschalk:1986bk, Marchesini:1987cf}.
\\

The above equation in eq.~(\ref{infinite-iteration-DGLAP-with-Sudakov}) derived from the DGLAP equation concerns only radiation from an incoming quark in a proton i.e. initial state radiation. Here we generalize the equation to the one which can predict radiation from outgoing partons i.e. final state radiation as well as initial state radiation.
What we should notice for this purpose is that the PDFs play a role in constraining scale evolution of the proton, or in other words constraining radiation from the quark in the proton during scale evolution of the proton. 
Hence, when radiation from outgoing partons is considered, we replace the PDFs with a function which is obtained from the kinematic information of the outgoing partons. The function constrains radiation from the outgoing partons, through the energy and momentum conservation for instance.\\

We let $\{p\}_{X+n}^{}$ denotes a complete specification of an event sample consisting of $X+n$ partons~\footnote{This expression is inspired by refs.~\cite{Giele:2011cb, Giele:2007di}.}. The information of the two incoming partons is also included. 
Then, we introduce a function for the evolution of a $\{p\}_{X+n}^{}$ as
\begin{align}
f\bigl(z, t;  \{p\}_{X+n}^{} \bigr),\label{intro-const-function}
\end{align}
which constrains the evolution of the $\{p\}_{X+n}^{}$ at the evolution scale $t$ and the energy fraction $z$. 
With the constraint function, eq.~(\ref{infinite-iteration-DGLAP-with-Sudakov}) can be generalized to 
\begin{align}
1=&
\Pi\bigl( t_X^{}, t_{\Lambda}^{}; \{p\}_X^{} \bigr)+
\int^{t_X^{}}_{t_{\Lambda}^{}} \frac{dt_1^{}}{t_1^{}}\int^1_0 d\hat{p}(z_1^{})\ 
\Pi\bigl( t_X^{}, t_1^{};  \{p\}_X^{} \bigr)\ 
f\bigl(z_1^{}, t_1^{};  \{p\}_X^{} \bigr)\ 
\Pi\bigl( t_1^{}, t_{\Lambda}^{};  \{p\}_{X+1}^{} \bigr) \nonumber \\
&
+
\int^{t_X^{}}_{t_{\Lambda}^{}} \frac{dt_1^{}}{t_1^{}}\int^1_0 d\hat{p}(z_1^{}) 
\Pi\bigl( t_X^{}, t_1^{}; \{p\}_X^{} \bigr)\ 
f\bigl(z_1^{}, t_1^{};  \{p\}_X^{} \bigr)  \nonumber \\
& \hspace{0.5cm}\times
\int^{t_1^{}}_{t_{\Lambda}^{}} \frac{dt_2^{}}{t_2^{}}\int^1_0 d\hat{p}(z_2^{}) 
\Pi\bigl( t_1^{}, t_{2}^{}; \{p\}_{X+1}^{} \bigr)
f\bigl(z_2^{}, t_2^{};  \{p\}_{X+1}^{})\ 
\Pi\bigl( t_2^{}, t_{\Lambda}^{}; \{p\}_{X+2}^{} \bigr) \nonumber \\
& + \cdots,
\label{infinite-iteration-DGLAP-with-Sudakov-general}
\end{align}
and accordingly
\begin{align}
\Pi\bigl(t_1^{}, t_2^{}; \{p\}_{X+n}^{}) = \exp{\biggl( -\int^{t_1^{}}_{t_{2}^{}} \frac{dt}{t}\int^1_0 d\hat{p}(z)f\bigl(z, t;  \{p\}_{X+n}^{}\bigr) \biggr)}, \label{no-rad-prob-general}
\end{align}
which is defined as the no radiation probability for a $\{p\}_{X+n}^{}$ as a whole, during the scale evolution of it between $t_1^{}$ and $t_2^{}$ ($t_1^{}>t_2^{}$).
In other words, this is the probability that a $\{p\}_{X+n}^{}$ remains the same during the scale evolution. 
Note that 
\begin{align}
d\hat{p}(z)=\frac{dz}{z}\frac{\alpha_s^{}}{2\pi}\hat{P}(z) &\mathrm{\ \ \ \ for\ initial\ state\ radiation},\nonumber \\
d\hat{p}(z)=dz\frac{\alpha_s^{}}{2\pi}\hat{P}(z) &\mathrm{\ \ \ \   for\ final\ state\ radiation}, \label{short-hand-splitting-function}
\end{align}
where the appropriate splitting function(s) should be used for $\hat{P}(z)$ according to a branching process $\{p\}_{X+n}^{} \to \{p\}_{X+n+1}^{}$. For initial state radiation, the constraint function $f(z, t;  \{p\}_{X+n}^{})$ always includes the PDFs. The soft gluon singularity at $z=1$ in the splitting functions will be avoided by introducing $\theta$ functions in the constraint function. \\

Now that we have the integrated form of the DGLAP equation with the Sudakov form factor in our notations as eqs.~(\ref{infinite-iteration-DGLAP-with-Sudakov-general}) (\ref{no-rad-prob-general}), we discuss the basic idea of tree level merging algorithms~\cite{Catani:2001cc}.
Let us first define the cross section of a hard process producing $X$ by $\sigma(X)$. For example, when $X=q\bar{q}$ in $e^+_{}e^-_{}$ annihilation is considered, $\sigma(X)$ is given by
\begin{align}
\sigma(e^+_{}e^-_{} \to q\bar{q}) = \frac{1}{2s}\int d\Phi_{q\bar{q}}^{} \overline{\sum_{\lambda}}\bigl|{\cal M}_{e^+_{}e^-_{} \to q\bar{q}}^{}\bigr|^2_{},
\end{align}
or, when $X=Z$ in proton proton collisions is considered, $\sigma(X)$ is given by
\begin{align}
\sigma(pp \to Z) = \int^1_0 dx_1^{} \int^1_0 dx_2^{}\ q(x_1^{}, \mu_F^{2}) \bar{q}(x_2^{}, \mu_F^{2}) \frac{1}{2sx_1^{}x_2^{}}\int d\Phi_{Z}^{} \overline{\sum_{\lambda}}\bigl|{\cal M}_{q\bar{q}\to Z}^{}\bigr|^2_{}. 
\end{align}
The DGLAP evolution of the hard process can be expressed by multiplying $\sigma(X)$ by eq.~(\ref{infinite-iteration-DGLAP-with-Sudakov-general}), namely
\begin{align}
\sigma(X)
&=\sigma(X)\ \Pi\bigl( t_X^{}, t_{\Lambda}^{}; \{p\}_X^{} \bigr) \nonumber \\
&+\sigma(X) \int^{t_X^{}}_{t_{\Lambda}^{}} \frac{dt_1^{}}{t_1^{}}\int^1_0 d\hat{p}(z_1^{})\ 
\Pi\bigl( t_X^{}, t_1^{};  \{p\}_X^{} \bigr)\ 
f\bigl(z_1^{}, t_1^{};  \{p\}_X^{} \bigr)\ 
\Pi\bigl( t_1^{}, t_{\Lambda}^{};  \{p\}_{X+1}^{} \bigr) \nonumber \\
&+\sigma(X) \int^{t_X^{}}_{t_{\Lambda}^{}} \frac{dt_1^{}}{t_1^{}}\int^1_0 d\hat{p}(z_1^{}) 
\Pi\bigl( t_X^{}, t_1^{}; \{p\}_X^{} \bigr)\ 
f\bigl(z_1^{}, t_1^{};  \{p\}_X^{} \bigr)  \nonumber \\
& \hspace{0.5cm}\times
\int^{t_1^{}}_{t_{\Lambda}^{}} \frac{dt_2^{}}{t_2^{}}\int^1_0 d\hat{p}(z_2^{}) 
\Pi\bigl( t_1^{}, t_{2}^{}; \{p\}_{X+1}^{} \bigr)
f\bigl(z_2^{}, t_2^{};  \{p\}_{X+1}^{})\ 
\Pi\bigl( t_2^{}, t_{\Lambda}^{}; \{p\}_{X+2}^{} \bigr) \nonumber \\
& + \cdots.\label{X-DGLAP-evolution}
\end{align}
The core idea of merging algorithms is to replace the terms constructed by the leading order cross section times the universal radiation probability with the exact tree level cross sections. For the second term of the RHS in the above equation, for instance, it proceeds as
\begin{align}
\sigma(X) \int^{t_X^{}}_{t_{\Lambda}^{}} \frac{dt_1^{}}{t_1^{}}\int^1_0 d\hat{p}(z_1^{})\ 
\to \sigma(X+1).
\end{align}
The equation in eq.~(\ref{X-DGLAP-evolution}) has the following expression
\begin{align}
\sigma(X)
&=\sigma(X)\ \Pi\bigl( t_X^{}, t_{\mathrm{cut}}^{}; \{p\}_X^{} \bigr) \nonumber \\
&+\sigma\bigl(X+1; \{p\}_{X+1}^{} > t_{\mathrm{cut}}^{} \bigr) \ 
\Pi\bigl( t_X^{}, t_1^{}; \{p\}_X^{} \bigr)\ 
f^{\prime}_{}\bigl(z_1^{}, t_1^{};  \{p\}_X^{} \bigr)\ 
\Pi\bigl( t_1^{}, t_{\mathrm{cut}}^{}; \{p\}_{X+1}^{} \bigr) \nonumber \\
&+\sigma\bigl(X+2; \{p\}_{X+2}^{} > t_{\mathrm{cut}}^{} \bigr) \
\Pi\bigl( t_X^{}, t_1^{}; \{p\}_X^{} \bigr)\ 
f^{\prime}_{}\bigl(z_1^{}, t_1^{};  \{p\}_X^{} \bigr)\
\Pi\bigl( t_1^{}, t_{2}^{}; \{p\}_{X+1}^{} \bigr) \nonumber \\
&
\hspace{0.5cm}
\times f^{\prime}_{}\bigl(z_2^{}, t_2^{};  \{p\}_{X+1}^{})\ 
\Pi\bigl( t_2^{}, t_{\mathrm{cut}}^{}; \{p\}_{X+2}^{} \bigr)\ \nonumber \\
& + \cdots,\label{improved-DGLAP-1}
\end{align}
where $t_{\Lambda}^{}$ is replaced with $t_{\mathrm{cut}}^{}$. 
The soft and collinear divergences in tree level matrix elements are regularized in the following way. By using the definition of the evolution variable $t$, calculate the minimum value $t_{\mathrm{min}}^{}$ from $\{p\}_{X+n}^{}$ and require $t_{\mathrm{min}}^{}>t_{\mathrm{cut}}^{}$. This is expressed as $\{p\}_{X+n}^{}>t_{\mathrm{cut}}^{}$ in the above equation~\footnote{Here it is assumed that the hard process cross section $\sigma(X)$ is finite everywhere in its phase space.}. The cut off scale $t_{\mathrm{cut}}^{}$ is called the merging scale. Notice that the constraint functions in eq.~(\ref{improved-DGLAP-1}) are different from those in eq.~(\ref{X-DGLAP-evolution}), since some part of the constraints are already included in the exact tree level cross sections. \\

Eq.~(\ref{improved-DGLAP-1}) summarizes the basic idea of tree level merging algorithms with our notations. Tree level merging algorithms are designed as to improve the DGLAP equation with a help from exact tree level cross sections, or in other words exact tree level matrix elements.

%%%%%%%%%%%%%%%%%%%%%%%%%%%%%%%%%%%%%%%%%%%%%%%%%%%%%%%%%%%%%%%%%%%%%%%%%%%%%%%%%%%%%%%%%%%%%%%%%%%%%%%%%%%%%%%%%%%%%%%%%%%%%%%%%%%%%%%%%%%%%%%%%%%%%%%%%%%%%%%%%%%%%%%%%%%%%%%%%%%%%%%%%%%%%%%%%%%%%%%%%%%%%%%%%%%%%%%%%%%%%%%%%%%%%%%%%%%%%%%%%%%%%%%%%%%%%%%%%%%%%%%%%%%%%%%%%%%%%%%%%%%%%%%%%%%%%%%%%%%%%%%%%%%%%%%%%%%%%%%%%%%%%%%%%%%%%%%%%%%%%%%%%%%%%%%%%%%%%%%%%%%%%%%%%%%%%%%%%%%%%%%%%%%%%%%%%%%%%%%%%%%%%%%%%%%%%%%%%%%%%%%%%%%%%%%%%%%%%%%%%%%%%%%%%%%%%%%%%%%%%%%%%%%%%%%%%%%%%%%%%%%%%%%%%%%%%%%%%%%%%%%%%%%%%%%%%%%%%%%%%%%%%%%%%%%%%%%%%%%%%%%%%%%%%%%%%%%%%%%%%%%%%%%%%%%%%%%%%%%%%%%%%%%%%%%%%%%%%%%%%%%%%%%%%%%%%%%%%%%%%%%%%%%%%%%%%%%%%%%%%%%%
\subsection{The CKKW-L algorithm and its extension}\label{sec:CKKW-L}

In this section, at first the event generation procedure following the improved DGLAP evolution equation in eq.~(\ref{improved-DGLAP-1}) by using the CKKW-L merging algorithm~\cite{Lonnblad:2001iq, Lavesson:2005xu, Lonnblad:2011xx} is reviewed. 
Then, we present a new merging algorithm, which differs from the CKKW-L algorithm in the strategy for phase space separation. 
Our strategy is introduced with the aim of predicting jet angular correlations more accurately. 
The independence on parton shower starting scale is discussed at the end.\\

We will examine each term, one by one, in the right hand side (RHS) of eq.~(\ref{improved-DGLAP-1}) in the following. In order to generate an event sample according to the probability of the first term
\begin{align}
\sigma(X)\ \Pi\bigl( t_X^{}, t_{\mathrm{cut}}^{}; \{p\}_X^{} \bigr),
\end{align}
at first an event sample $\{p\}_X^{}$ is generated with the cross section $\sigma(X)$. The next step is to calculate the Sudakov form factor $\Pi( t_X^{}, t_{\mathrm{cut}}^{}; \{p\}_X^{} )$~\footnote{Strictly speaking, this factor defined in eq.~(\ref{no-rad-prob-general}) is not identical to the Sudakov form factor defined in eq.~(\ref{Sudakov-original}) in our notations. However, we call it the Sudakov form factor, too.}. The CKKW-L algorithm uses a parton shower generator to calculate Sudakov form factors. We execute a shower generator on the $\{p\}_X^{}$, setting the shower starting scale to the $t_X^{}$. 
If the first evolution scale randomly chosen is higher than the $t_{\mathrm{cut}}^{}$, then we throw away the event sample as a whole and go to the next new event sample. This procedure is equivalent to accepting the event sample with the probability equal to $\Pi( t_X^{}, t_{\mathrm{cut}}^{}; \{p\}_X^{} )$, since it is the no radiation probability during the scale evolution of the $\{p\}_X^{}$ between the scales $t_X^{}$ and $t_{\mathrm{cut}}^{}$. If the first evolution scale is lower than the $t_{\mathrm{cut}}^{}$, this evolution is continued until the cut off scale $t_{\mathrm{had}}^{}$ of the shower model and the event sample contributes to the inclusive event samples.\\

Next, let us look at the second term
\begin{align}
\sigma\bigl(X+1; \{p\}_{X+1}^{} > t_{\mathrm{cut}}^{} \bigr) \ 
\Pi\bigl( t_X^{}, t_1^{};  \{p\}_X^{} \bigr)\ 
f^{\prime}_{}\bigl(z_1^{}, t_1^{};  \{p\}_X^{} \bigr)\ 
\Pi\bigl( t_1^{}, t_{\mathrm{cut}}^{};  \{p\}_{X+1}^{} \bigr).\label{CKKW-L-2ndterm}
\end{align}
At first an event samples $\{p\}_{X+1}^{}$ is generated with the tree level cross section $\sigma(X+1; \{p\}_{X+1}^{} > t_{\mathrm{cut}}^{} )$. In order to calculate the first Sudakov form factor $\Pi( t_X^{}, t_1^{};  \{p\}_X^{} )$, we need the intermediate event $\{p\}_X^{}$ and the scale $t_1^{}$. 
In the CKKW-L algorithm, these are obtained from the $\{p\}_{X+1}^{}$ by executing a program which does the exact inverse of the shower generation of a parton shower generator which is used.
The program should produce a $\{p\}_X^{}$ and a scale $t_1^{}$ from the $\{p\}_{X+1}^{}$ as if the parton shower generator had evolved the $\{p\}_X^{}$ and then had generated the $\{p\}_{X+1}^{}$ with the evolution scale $t_1^{}$. This backward flow is often called a parton shower history. The construction of the parton shower history completely depends on the shower generator which is used. In Section~\ref{history}, the construction of the PYTHIA8 parton shower history is described. For now let us assume that we successfully construct a shower history of the $\{p\}_{X+1}^{}$ and obtain a $\{p\}_X^{}$ and a scale $t_1^{}$. In order to calculate the first Sudakov form factor, the shower generator is executed on the $\{p\}_{X}^{}$, starting from the scale $t_X^{}$, as before. If the first evolution scale randomly chosen is higher than the $t_1^{}$, we throw away the event sample $\{p\}_{X+1}^{}$ as whole and go to the next new event sample. 
If not, the constraint function $f^{\prime}_{}(z_1^{}, t_1^{};  \{p\}_X^{} )$ is calculated and the event sample $\{p\}_{X+1}^{}$ is re-weighted according to it. The calculation of the constraint function is discussed in Section~\ref{sec:weight-functions}.
If the event sample is still survived after the re-weighting, 
the second Sudakov form factor $\Pi( t_1^{}, t_{\mathrm{cut}}^{};  \{p\}_{X+1}^{} )$ is calculated by executing the shower generator on the $\{p\}_{X+1}^{}$ from the scale $t_1^{}$ and throwing away the event sample as a whole if the first evolution scale is higher than the $t_{\mathrm{cut}}^{}$. If the first evolution scale is lower than the $t_{\mathrm{cut}}^{}$, this evolution is continued until the cut off scale $t_{\mathrm{had}}^{}$ and the event sample contributes to the inclusive event samples. \\

The event generation following the third and higher terms in the RHS of eq.~(\ref{improved-DGLAP-1}) can be performed in the same way. Although merging algorithms can treat tree level cross sections of any number of partons, there are limitations in their calculations. If we decide not to calculate $\sigma(X+3, \{p\}_{X+2}^{} > t_{\mathrm{cut}}^{} )$ for instance, what we should do is to remove the last Sudakov form factor $\Pi( t_2^{}, t_{\mathrm{cut}}^{}; \{p\}_{X+2}^{} )$ from the third term in the RHS of eq.~(\ref{improved-DGLAP-1})~\cite{Lonnblad:2001iq}. 
The third term, which used to be the probability of generating a $\{p\}_{X+2}^{}$ exclusively above the $t_{\mathrm{cut}}^{}$, is now the probability of generating a $\{p\}_{X+2}^{}$ exclusively above a $t_{2}^{}$ ($>t_{\mathrm{cut}}^{}$). The evolution equation is closed at the third term, i.e. the dots in eq.~(\ref{improved-DGLAP-1}) disappears. \\

Up to now it has been assumed that the definition of the merging scale $t_{\mathrm{cut}}^{}$ is equivalent to that of the shower evolution variable $t$. It is discussed in refs.~\cite{Lonnblad:2001iq, Lonnblad:2011xx} that the definition of the merging scale $t_{\mathrm{cut}}^{}$ can be arbitrary, as long as it regulates the singularity in tree level matrix elements. 
We let $Q_{\mathrm{cut}}^2$ denotes a definition of the merging scale. 
This can be the $k_{\perp}^{}$ definition of the jet clustering algorithm~\cite{Catani:1993hr} for instance. With an arbitrary definition of the merging scale, the Sudakov form factor in the first term of the RHS of eq.~(\ref{improved-DGLAP-1}) is calculated as follows. We execute a shower generator on the $\{p\}_X^{}$, starting from the $t_X^{}$, as before. After the first evolution, we obtain a $\{p\}_{X+1}^{}$ and calculate the minimum value $Q_{\mathrm{min}}^2$ on the $\{p\}_{X+1}^{}$ by using the definition of the merging scale $Q_{\mathrm{cut}}^2$. If $Q_{\mathrm{min}}^2>Q_{\mathrm{cut}}^2$, we throw away the event sample as a whole and go to the next new event sample. We express this as $\{p\}_{X+1}^{} > Q_{\mathrm{cut}}^2$. If not i.e. $Q_{\mathrm{min}}^2<Q_{\mathrm{cut}}^2$, this evolution is continued until the cut off scale $t_{\mathrm{had}}^{}$ and the event sample contributes to the inclusive event samples. We express this as $\{p\}_{X+1}^{} < Q_{\mathrm{cut}}^2$. This procedure is applied to the calculation of the last Sudakov from factor at each term in the RHS of eq.~(\ref{improved-DGLAP-1}). \\

\begin{figure}[t]
\centering
\includegraphics[scale=0.45]{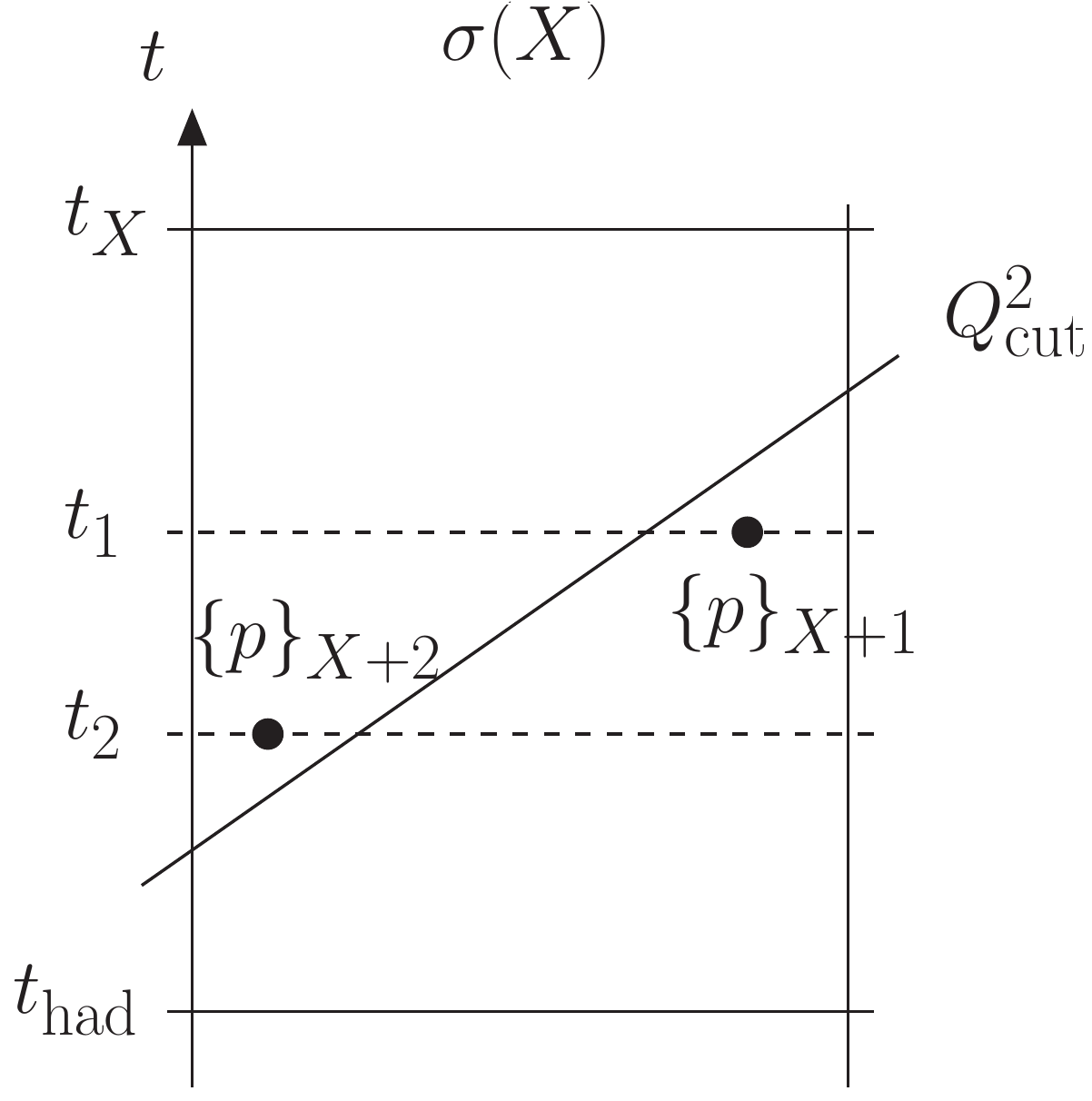}
\hspace{0.9cm}
\includegraphics[scale=0.45]{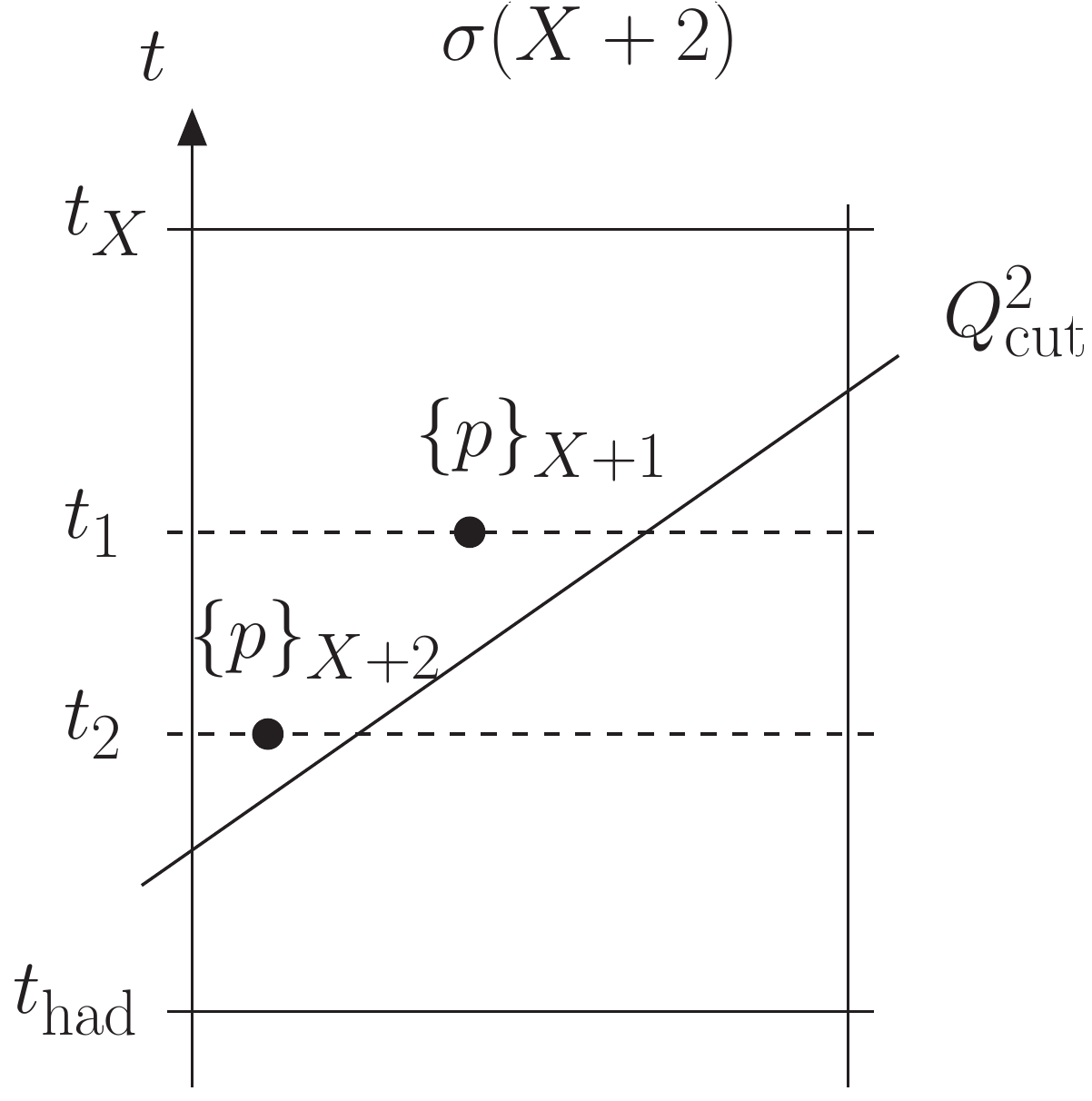}
\caption{\small The vertical axis represents the evolution variable $t$, the horizontal axis represents some variables in the shower model which determine kinematics. The diagonal line indicates the merging scale $Q_{\mathrm{cut}}^2$. 
The left and right panels represent the same event $\{p\}_{X+2}^{}$, however their origin can be different. In the CKKW-L algorithm, the contribution in the left panel originates from $\sigma(X)$ and one in the right panel originates from $\sigma(X+2)$. In our merging algorithm, the both originate from $\sigma(X+2)$. See text in Section~\ref{sec:CKKW-L} for a detailed explanation.}
\label{fig:ckkwl}
\end{figure}

However, it can be easily imagined that, even though $\{p\}_{X+1}^{} < Q_{\mathrm{cut}}^2$ is satisfied after the first evolution of the $\{p\}_{X}^{}$, the second evolution may generate a $\{p\}_{X+2}^{}$ which is $\{p\}_{X+2}^{} > Q_{\mathrm{cut}}^2$. This is considered as double counting with the contribution from the third term in the RHS of eq.~(\ref{improved-DGLAP-1}). 
The solution to the double counting issue in the CKKW-L algorithm can be explained as follows, by using illustrations in Figure~\ref{fig:ckkwl}~\footnote{This illustration is inspired by refs.~\cite{Lonnblad:2001iq, Lonnblad:2011xx}}. 
The vertical axis represents the evolution variable $t$, the horizontal axis represents some kinematic variables in a shower model, such as the energy fraction variable $z$. One point in the figure corresponds to one phase space point of a $\{p\}_{X+n}^{}$, and it is uniquely determined once a value for the vertical axis and a value for the horizontal axis are chosen.
The diagonal line indicates the merging scale $Q_{\mathrm{cut}}^2$. Note that this line will be perpendicular to the vertical axis, when the definition of the merging scale is equivalent to that of the evolution variable i.e. $Q_{\mathrm{cut}}^2=t_{\mathrm{cut}}^{}$.
The left panel shows a contribution of the first term in the RHS of eq.~(\ref{improved-DGLAP-1}) after the first and second evolution. 
This event sample satisfies $\{p\}_{X+1}^{} < Q_{\mathrm{cut}}^2$ after the first evolution, thus it is already accepted. However, after the second evolution, it becomes $\{p\}_{X+2}^{} > Q_{\mathrm{cut}}^2$. 
The right panel shows a contribution of the third term in the RHS of eq.~(\ref{improved-DGLAP-1}) after no evolution.
As it is illustrated in the right panel, the CKKW-L algorithm requires intermediate events to satisfy the merging scale cut, i.e. $\{p\}_{X+1}^{} > Q_{\mathrm{cut}}^2$ as well as $\{p\}_{X+2}^{} > Q_{\mathrm{cut}}^2$.
The sum of the two contributions in the left and right panels
does not lead to double counting, since the $\{p\}_{X+1}^{}$ in the left panel an d the $\{p\}_{X+1}^{}$ in the right panel live in the different phase space regions. For the CKKW-L algorithm, eq.~(\ref{improved-DGLAP-1}) can be written as follows,
\begin{align}
&\sigma(X)_{\mathrm{CKKW\mathchar`-L}}^{} \nonumber\\
&=\sigma(X)\ \Pi\Bigl( t_X^{}, \{p\}_{X+1}^{} < Q_{\mathrm{cut}}^2; \{p\}_X^{} \Bigr) \nonumber \\
&+\sigma\Bigl(X+1; \{p\}_{X+1}^{} > Q_{\mathrm{cut}}^2 \Bigr) \ 
\Pi\bigl( t_X^{}, t_1^{};  \{p\}_X^{} \bigr)\ 
f^{\prime}_{}\bigl(z_1^{}, t_1^{};  \{p\}_X^{} \bigr)\ 
\Pi\Bigl( t_1^{}, \{p\}_{X+2}^{} < Q_{\mathrm{cut}}^2;  \{p\}_{X+1}^{} \Bigr) \nonumber \\
&+\sigma\Bigl(X+2; \{p\}_{X+2}^{} > Q_{\mathrm{cut}}^2, \{p\}_{X+1}^{} > Q_{\mathrm{cut}}^2 \Bigr) \
\Pi\bigl( t_X^{}, t_1^{}; \{p\}_X^{} \bigr)\ 
f^{\prime}_{}\bigl(z_1^{}, t_1^{};  \{p\}_X^{} \bigr)\
\Pi\bigl( t_1^{}, t_{2}^{}; \{p\}_{X+1}^{} \bigr) \nonumber \\
&
\hspace{0.5cm}
\times f^{\prime}_{}\bigl(z_2^{}, t_2^{};  \{p\}_{X+1}^{})\ 
\Pi\Bigl( t_2^{}, \{p\}_{X+3}^{} < Q_{\mathrm{cut}}^2; \{p\}_{X+2}^{} \Bigr)\ \nonumber \\
& + \cdots.\label{improved-DGLAP-2}
\end{align}

In this study, we use a different method to avoid the double counting issue. When a purpose of merging is to predict angular correlations between two hard jets, it can be one disadvantage of the CKKW-L algorithm that the contribution shown in the left panel of Figure~\ref{fig:ckkwl} is described by $\sigma(X)$. The $\{p\}_{X+2}^{}$ in the left panel has potential to produce a $X+$ 2 hard jets event. However, angular correlations between the two jets are not correct, since this event originates from $\sigma(X)$, not $\sigma(X+2)$. Therefore, our algorithm is designed so that the contribution shown in the left panel originates from $\sigma(X+2)$, not $\sigma(X)$. The improved evolution equation in eq.~(\ref{improved-DGLAP-1}) for our merging algorithm can be written as follows,
\begin{align}
&\sigma(X)_{\mathrm{CKKW\mathchar`-L+}}^{} \nonumber\\
&=\sigma(X)\ \Pi\Bigl( t_X^{}, \{p\}_{X+1}^{} < Q_{\mathrm{cut}}^2, \{p\}_{X+2}^{} < Q_{\mathrm{cut}}^2, \cdots; \{p\}_X^{} \Bigr) \nonumber \\
\vspace{0.5cm}
&+\sigma\Bigl(X+1; \{p\}_{X+1}^{} > Q_{\mathrm{cut}}^2 \Bigr) \ 
\Pi\bigl( t_X^{}, t_1^{};  \{p\}_X^{} \bigr)\ 
f^{\prime}_{}\bigl(z_1^{}, t_1^{};  \{p\}_X^{} \bigr)\nonumber \\
&\hspace{0.5cm}\times 
\Pi\Bigl( t_1^{}, \{p\}_{X+2}^{} < Q_{\mathrm{cut}}^2, \{p\}_{X+3}^{} < Q_{\mathrm{cut}}^2, \cdots;  \{p\}_{X+1}^{} \Bigr) \nonumber \\
&+\sigma\Bigl(X+2; \{p\}_{X+2}^{} > Q_{\mathrm{cut}}^2 \Bigr) \
\Pi\bigl( t_X^{}, t_1^{}; \{p\}_X^{} \bigr)\ 
f^{\prime}_{}\bigl(z_1^{}, t_1^{};  \{p\}_X^{} \bigr)\
\Pi\bigl( t_1^{}, t_{2}^{}; \{p\}_{X+1}^{} \bigr) \nonumber \\
&\hspace{0.5cm}\times 
f^{\prime}_{}\bigl(z_2^{}, t_2^{};  \{p\}_{X+1}^{}) \Pi\Bigl( t_2^{}, \{p\}_{X+3}^{} < Q_{\mathrm{cut}}^2, \{p\}_{X+4}^{} < Q_{\mathrm{cut}}^2, \cdots;  \{p\}_{X+2}^{} \Bigr) \nonumber \\
&+ \cdots.\label{improved-DGLAP-5}
\end{align}
Notice that the cut $\{p\}_{X+1}^{} > Q_{\mathrm{cut}}^2$ in the third term of the RHS of eq.~(\ref{improved-DGLAP-2}) now disappears and a cut $\{p\}_{X+2}^{} < Q_{\mathrm{cut}}^2$ is newly added in the first term. Merging scale cuts expressed by the dots in the last Sudakov form factor at each term depends on the maximal number of partons predicted by the tree level cross section. In order to make the expression clearer, let us consider the case that we do not calculate $\sigma\bigl(X+3)$. The evolution equation is given by
\begin{align}
&\sigma(X)_{\mathrm{CKKW\mathchar`-L+}}^{N=2} \nonumber\\
&=\sigma(X)\ \Pi\Bigl( t_X^{}, \{p\}_{X+1}^{} < Q_{\mathrm{cut}}^2, \{p\}_{X+2}^{} < Q_{\mathrm{cut}}^2; \{p\}_X^{} \Bigr) \nonumber \\
&+\sigma\Bigl(X+1; \{p\}_{X+1}^{} > Q_{\mathrm{cut}}^2 \Bigr) \ 
\Pi\bigl( t_X^{}, t_1^{};  \{p\}_X^{} \bigr)\ 
f^{\prime}_{}\bigl(z_1^{}, t_1^{};  \{p\}_X^{} \bigr)\ 
\Pi\Bigl( t_1^{}, \{p\}_{X+2}^{} < Q_{\mathrm{cut}}^2;  \{p\}_{X+1}^{} \Bigr) \nonumber \\
&+\sigma\Bigl(X+2; \{p\}_{X+2}^{} > Q_{\mathrm{cut}}^2 \Bigr) \
\Pi\bigl( t_X^{}, t_1^{}; \{p\}_X^{} \bigr)\ 
f^{\prime}_{}\bigl(z_1^{}, t_1^{};  \{p\}_X^{} \bigr)\
\Pi\bigl( t_1^{}, t_{2}^{}; \{p\}_{X+1}^{} \bigr)\ f^{\prime}_{}\bigl(z_2^{}, t_2^{};  \{p\}_{X+1}^{} \bigr),\label{improved-DGLAP-3}
\end{align}
where $N$ denotes the maximal number of partons predicted by the tree level cross sections. 
This equation implies that both of the contributions in Figure~\ref{fig:ckkwl} originate from $\sigma(X+2)$. 
Since they never originate from $\sigma(X)$ or $\sigma(X+1)$, there is no double counting. For the sake of completeness, we give the evolution equation for $N=3$,
\begin{align}
&\sigma(X)_{\mathrm{CKKW\mathchar`-L+}}^{N=3} \nonumber\\
&=\sigma(X)\ \Pi\Bigl( t_X^{}, \{p\}_{X+1}^{} < Q_{\mathrm{cut}}^2, \{p\}_{X+2}^{} < Q_{\mathrm{cut}}^2, \{p\}_{X+3}^{} < Q_{\mathrm{cut}}^2; \{p\}_X^{} \Bigr) \nonumber \\
\vspace{0.5cm}
&+\sigma\Bigl(X+1; \{p\}_{X+1}^{} > Q_{\mathrm{cut}}^2 \Bigr) \ 
\Pi\bigl( t_X^{}, t_1^{};  \{p\}_X^{} \bigr)\ 
f^{\prime}_{}\bigl(z_1^{}, t_1^{};  \{p\}_X^{} \bigr)\nonumber \\
&\hspace{0.5cm}\times 
\Pi\Bigl( t_1^{}, \{p\}_{X+2}^{} < Q_{\mathrm{cut}}^2, \{p\}_{X+3}^{} < Q_{\mathrm{cut}}^2;  \{p\}_{X+1}^{} \Bigr) \nonumber \\
&+\sigma\Bigl(X+2; \{p\}_{X+2}^{} > Q_{\mathrm{cut}}^2 \Bigr) \
\Pi\bigl( t_X^{}, t_1^{}; \{p\}_X^{} \bigr)\ 
f^{\prime}_{}\bigl(z_1^{}, t_1^{};  \{p\}_X^{} \bigr)\
\Pi\bigl( t_1^{}, t_{2}^{}; \{p\}_{X+1}^{} \bigr) \nonumber \\
&\hspace{0.5cm}\times 
f^{\prime}_{}\bigl(z_2^{}, t_2^{};  \{p\}_{X+1}^{}) \Pi\Bigl( t_2^{}, \{p\}_{X+3}^{} < Q_{\mathrm{cut}}^2;  \{p\}_{X+2}^{} \Bigr) \nonumber \\
&+\sigma\Bigl(X+3; \{p\}_{X+3}^{} > Q_{\mathrm{cut}}^2 \Bigr) \
\Pi\bigl( t_X^{}, t_1^{}; \{p\}_X^{} \bigr)\ 
f^{\prime}_{}\bigl(z_1^{}, t_1^{};  \{p\}_X^{} \bigr)\
\Pi\bigl( t_1^{}, t_{2}^{}; \{p\}_{X+1}^{} \bigr) \nonumber \\
&\hspace{0.5cm}\times 
f^{\prime}_{}\bigl(z_2^{}, t_2^{};  \{p\}_{X+1}^{}) \Pi\bigl( t_2^{}, t_3^{};  \{p\}_{X+2}^{} \bigr) f^{\prime}_{}\bigl(z_3^{}, t_3^{};  \{p\}_{X+2}^{} \bigr).\label{improved-DGLAP-4}
\end{align}
Notice that a cut $\{p\}_{X+3}^{} < Q_{\mathrm{cut}}^2$ is newly added in the last Sudakov form factor in the first and second terms of the RHS. \\

In this paper we call the above algorithm given in eq.~(\ref{improved-DGLAP-5}) the CKKW-L+ merging algorithm and use it for the event generation. In Section~\ref{mergingscale}, the CKKW-L+ algorithm will be numerically compared with the CKKW-L algorithm. \\

Finally, let us discuss the scale $t_X^{}$. The DGLAP evolution of a hard process $\sigma(X)$ in eq.~(\ref{X-DGLAP-evolution}) indicates that the scale $t_X^{}$ is the factorization scale of the hard process. Thus in the merging algorithms the $t_X^{}$ must be determined from the $\{p\}_{X}^{}$ which is constructed as a parton shower history from a $\{p\}_{X+n}^{}$. 
Since we obtain the scales $t_1^{}, t_2^{}, \cdots$ as shower histories from the event samples generated with tree level matrix elements, it happens that $t_X^{}< t_1^{}, t_2^{}, \cdots$. This should be considered as an important prediction of the tree level matrix elements, since this never happens in the DGLAP shower evolution. 
Even in such a case, the maximal shower evolution scale for the event sample must be set to the $t_X^{}$. 
Let us suppose the case that $ t_1^{} > t_X^{} > t_2^{} > \cdots$.
The first Sudakov form factor $\Pi ( t_X^{}, t_1^{};  \{p\}_X^{} )$ is obviously unity i.e. no veto. A care is needed for the calculation of the second Sudakov form factor. 
The shower starting scale for the calculation of the second Sudakov form factor $\Pi( t_1^{}, t_2^{};  \{p\}_{X+1}^{} )$ or $\Pi( t_1^{}, \{p\}_{X+2}^{} < Q_{\mathrm{cut}}^2, \cdots;  \{p\}_{X+1}^{} )$ 
must be set to the $t_X^{}$, not the $t_1^{}$.
The maximal shower evolution scale $t_X^{}$ is often called parton shower starting scale. \\

In the shower evolution without merging algorithms as in eq.~(\ref{X-DGLAP-evolution}), predictions such as a jet $p_T^{}$ distribution can depend strongly on the shower starting scale $t_X^{}$, since it determines hardness of radiation. It has been confirmed that this dependence is reduced significantly once we use merging algorithms~\cite{Alwall:2008qv}. The reason of the independence can be understood from the improved DGLAP equation in eq.~(\ref{improved-DGLAP-5}). Let us rewrite eq.~(\ref{improved-DGLAP-4}), which is obtained from eq.~(\ref{improved-DGLAP-5}) by setting $N=3$, in the following form
\begin{align}
\sigma_{\mathrm{inc}}^{}(X)= \sigma_{\mathrm{exc}}^{}(X+0)+\sigma_{\mathrm{exc}}^{}(X+1)+\sigma_{\mathrm{exc}}^{}(X+2)+\sigma_{\mathrm{exc}}^{}(X+3) \label{inc-cross-sec-1}
\end{align}
or equivalently
\begin{align}
1= \frac{\sigma_{\mathrm{exc}}^{}(X+0)}{\sigma_{\mathrm{inc}}^{}(X)}+\frac{\sigma_{\mathrm{exc}}^{}(X+1)}{\sigma_{\mathrm{inc}}^{}(X)}+\frac{\sigma_{\mathrm{exc}}^{}(X+2)}{\sigma_{\mathrm{inc}}^{}(X)}+\frac{\sigma_{\mathrm{exc}}^{}(X+3)}{\sigma_{\mathrm{inc}}^{}(X)}.\label{inc-cross-sec-2}
\end{align}
First of all, the hardness of radiation parametrized by the evolution variable $t$ as $t_1^{}, t_2^{}, \cdots$ is determined from the tree level matrix elements, thus is nothing to do with the $t_X^{}$. 
Since the first Sudakov form factor at each term such as $\Pi( t_X^{}, t_1^{};  \{p\}_X^{} )$ and $\Pi( t_X^{}, \{p\}_{X+1}^{} < Q_{\mathrm{cut}}^2, \cdots; \{p\}_X^{})$ depends on the $t_X^{}$, the exclusive cross sections $\sigma_{\mathrm{exc}}^{}(X+i)$ and accordingly the inclusive cross section $\sigma_{\mathrm{inc}}^{}(X)$ are affected by the $t_X^{}$. 
However, because the first Sudakov form factor can be considered as an overall factor, the ratio of the exclusive cross section to the inclusive cross section, i.e. $\sigma_{\mathrm{exc}}^{}(X+i)/\sigma_{\mathrm{inc}}^{}(X)$, will be little affected. 
Stable distribution can be expected from the above two reasons.
The dependence on the shower starting scale is numerically evaluated in Section~\ref{check}.

%%%%%%%%%%%%%%%%%%%%%%%%%%%%%%%%%%%%%%%%%%%%%%%%%%%%%%%%%%%%%%%%%%%%%%%%%%%%%%%%%%%%%%%%%%%%%%%%%%%%%%%%%%%%%%%%%%%%%%%%%%%%%%%%%%%%%%%%%%%%%%%%%%%%%%%%%%%%%%%%%%%%%%%%%%%%%%%%%%%%%%%%%%%%%%%%%%%%%%%%%%%%%%%%%%%%%%%%%%%%%%%%%%%%%%%%%%%%%%%%%%%%%%%%%%%%%%%%%%%%%%%%%%%%%%%%%%%%%%%%%%%%%%%%%%%%%%%%%%%%%%%%%%%%%%%%%%%%%%%%%%%%%%%%%%%%%%%%%%%%%%%%%%%%%%%%%%%%%%%%%%%%%%%%%%%%%%%%%%%%%%%%%%%%%%%%%%%%%%%%%%%%%%%%%%%%%%%%%%%%%%%%%%%%%%%%%%%%%%%%%%%%%%%%%%%%%%%%%%%%%%%%%%%%%%%%%%%%%%%%%%%%%%%%%%%%%%%%%%%%%%%%%%%%%%%%%%%%%%%%%%%%%%%%%%%%%%%%%%%%%%%%%%%%%%%%%%%%%%%%%%%%%%%%%%%%%%%%%%%%%%%%%%%%%%%%%%%%%%%%%%%%%%%%%%%%%%%%%%%%%%%%%%%%%%%%%%%%%%%%%%%%
\subsection{Construction of the PYTHIA8 parton shower history}\label{history}

In Section~\ref{sec:CKKW-L}, it is explained that the CKKW-L merging algorithms require the construction of the parton shower history and a construction program must do the exact inverse of the shower generation of a parton shower generator which we use.
In this section, the construction of the PYTHIA8 parton shower history is described. 
A parton shower history is constructed by successively clustering two partons into one parton, and one history consists of a set of intermediate events with the corresponding clustering scales which are ordered. For instance, a history of an event sample $\{p\}_{X+n}^{}$ consists of $\{p\}_{X+(n-1)}^{}, \{p\}_{X+(n-2)}^{}, \cdots, \{p\}_{X+i}^{}, \cdots, \{p\}_{X+1}^{}, \{p\}_{X}^{}$ with $t_n^{} < t_{n-1}^{} < \cdots < t_{i+1}^{} < \cdots < t_2^{}< t_1^{}$.
Because the detailed definition of the evolution variable and the kinematics construction are different for initial state radiation (ISR) and final state radiation (FSR) in PYTHIA8, the clustering procedure is also different for a clustering of incoming and outgoing partons and that of two outgoing partons. The former procedure is described in Section~\ref{isr} and the latter in Section~\ref{fsr}. 
In these sections, we use the knowledges and the notations given in the original publications~\cite{Sjostrand:2004ef, Corke:2010yf, Norrbin:2000uu} for the shower model in PYTHIA8. Some technical details in our implementation are summarized in Section~\ref{sec:remarks}.

%%%%%%%%%%%%%%%%%%%%%%%%%%%%%%%%%%%%%%%%%%%%%%%%%%%%%%%%%%%%%%%%%%%%%%%%%%%%%%%%%%%%%%%%%%%%%%%%%%%%%%%%%%%%%%%%%%%%%%%%%%%%%%%%%%%%%%%%%%%%%%%%%%%%%%%%%%%%%%%%%%%%%%%%%%%%%%%%%%%%%%%%%%%%%%%%%%%%%%%%%%%%%%%%%%%%%%%%%%%%%%%%%%%%%%%%%%%%%%%%%%%%%%%%%%%%%%%%%%%%%%%%%%%%%%%%%%%%%%%%%%%%%%%%%%%%%%%%%%%%%%%%%%%%%%%%%%%%%%%%%%%%%%%%%%%%%%%%%%%%%%%%%%%%%%%%%%%%%%%%%%%%%%%%%%%%%%%%%%%%%%%%%%%%%%%%%%%%%%%%%%%%%%%%%%%%%%%%%%%%%%%%%%%%%%%%%%%%%%%%%%%%%%%%%%%%%%%%%%%%%%%%%%%%%%%%%%%%%%%%%%%%%%%%%%%%%%%%%%%%%%%%%%%%%%%%%%%%%%%%%%%%%%%%%%%%%%%%%%%%%%%%%%%%%%%%
\subsubsection{Construction of the ISR history}\label{isr}

\begin{figure}[t]
\centering
\includegraphics[scale=0.55]{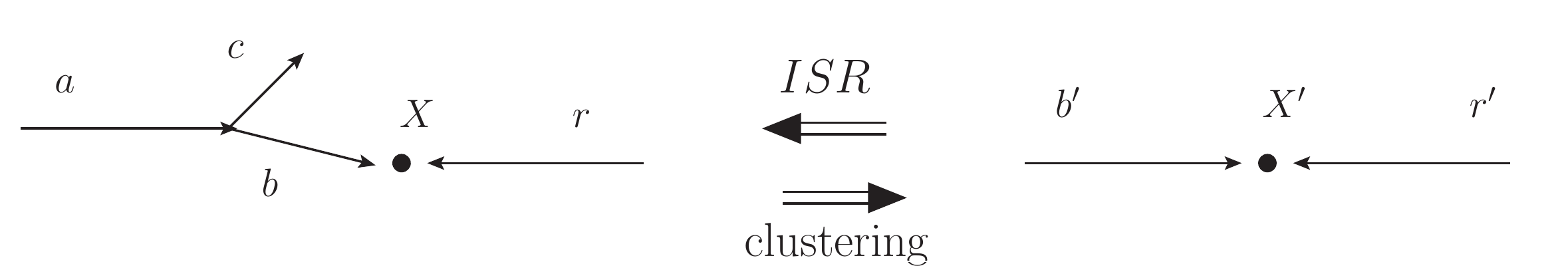}
\caption{\small Illustrating that an initial state shower model evolves a process $b^{\prime}_{}r^{\prime}_{}\to X^{\prime}_{}$ and then generates a process $ar\to Xc$ (from right to left), and its inverse (from left to right)}
\label{ISR}
\end{figure}

Let us suppose that an incoming parton $a$ and an outgoing parton $c$ in a process $ar\to Xc$ are clustered into a new parton $b$ and hence an intermediate process $b^{\prime}_{}r^{\prime}_{}\to X^{\prime}_{}$ together with a clustering scale $p_{\perp \mathrm{clus}}^{}$ is produced. This is illustrated in Figure~\ref{ISR} (from left to right). 
The clustering has to proceed as if the initial state shower model in PYTHIA8 had evolved the hard process $b^{\prime}_{}r^{\prime}_{}\to X^{\prime}_{}$ and then had generated the process $ar\to Xc$ with the evolution scale $p_{\perp \mathrm{evol}}^{}=p_{\perp \mathrm{clus}}^{}$.
The clustering scale $p_{\perp \mathrm{clus}}^{}$ is derived from
\begin{subequations}\label{pythiaISR}
\begin{align}
&p_b^{}=p_a^{}-p_c^{},\\
&z=\frac{m_{br}^2}{m_{ar}^2}=\frac{(p_b^{}+p_r^{})^2}{(p_a^{}+p_r^{})^2},\\
&p_{\perp \mathrm{clus}}^2=-(1-z)(p_b^{})^2.
\end{align}
\end{subequations}
Here the $z$ can be interpreted as the energy fraction $E_b^{}/E_a^{}$ in the center of mass frame of proton proton collisions. \\

The new incoming parton $b$ after the clustering is not moving along the z-axis and it is a spacelike particle i.e. $(p_b^{})^2<0$. Thus we need to make the $b$ on-shell (massless) and moving along the z-axis.
The $X$ and $X^{\prime}_{}$ denote all the other particles in the final state, hence $X=X^{\prime}_{}=t\bar{t}+g$ for instance. The four-momenta of the $b^{\prime}_{}$, $r^{\prime}_{}$ and $X^{\prime}_{}$ are derived as follows.
\begin{enumerate}
\item Read the azimuthal angle $\phi_c^{}$ of the c. 
\item Rotate the $c$ and $X$ in azimuth by $-\phi_c^{}$.
\item Calculate the four-momentum of the $b$ as $p_b^{}=p_a^{}-p_c^{}$.
\item Boost the $b$, $r$ and $X$ to the $b+r$ rest frame, and then rotate them in polar angle so that the $b$ and $r$ move along the z-axis.
\item Rotate the $X$ in azimuth by $+\phi_c$.
\item Newly obtain the four-momenta of massless incoming partons $b$ and $r$ in the $b+r$ rest frame, i.e. $p_b=(m_{br}/2,0,0,m_{br}/2)$ and $p_r=(m_{br}/2,0,0,-m_{br}/2)$.
\item Boost the $b$, $r$ and $X$ along the z-axis so that the $r$ has its original momentum. 
\end{enumerate}
With the above algorithm, the non-zero transverse momentum of the parton $b$ is translated into the kinematics of the $X$. 
As required, the $b^{\prime}_{}$ is on mass shell (massless) and it moves along the z-axis. 
The kinematics of the $r$ does not change i.e. $p_r^{}=p_{r^{\prime}_{}}^{}$. The above algorithm is carefully tested as follows. 
We apply the algorithm to an event $ar\to Xc$ which has been generated by the PYTHIA8 initial state evolution $a\to bc$ of a hard process event $b^{\prime}_{}r^{\prime}_{}\to X^{\prime}_{}$, and then we confirm that the algorithm correctly reproduces the event $b^{\prime}_{}r^{\prime}_{}\to X^{\prime}_{}$.

%%%%%%%%%%%%%%%%%%%%%%%%%%%%%%%%%%%%%%%%%%%%%%%%%%%%%%%%%%%%%%%%%%%%%%%%%%%%%%%%%%%%%%%%%%%%%%%%%%%%%%%%%%%%%%%%%%%%%%%%%%%%%%%%%%%%%%%%%%%%%%%%%%%%%%%%%%%%%%%%%%%%%%%%%%%%%%%%%%%%%%%%%%%%%%%%%%%%%%%%%%%%%%%%%%%%%%%%%%%%%%%%%%%%%%%%%%%%%%%%%%%%%%%%%%%%%%%%%%%%%%%%%%%%%%%%%%%%%%%%%%%%%%%%%%%%%%%%%%%%%%%%%%%%%%%%%%%%%%%%%%%%%%%%%%%%%%%%%%%%%%%%%%%%%%%%%%%%%%%%%%%%%%%%%%%%%%%%%%%%%%%%%%%%%%%%%%%%%%%%%%%%%%%%%%%%%%%%%%%%%%%%%%%%%%%%%%%%%%%%%%%%%%%%%%%%%%%%%%%%%%%%%%%%%%%%%%%%%%%%%%%%%%%%%%%%%%%%%%%%%%%%%%%%%%%%%%%%%%%%%%%%%%%%%%%%%%%%%%%%%%%%%%%%%%%%%%%%%%%%%%%%%%%%%%%%%%%%%%%%%%%%%%%%%%%%%%%%%%%%%%%%%%%%%%%%%%%%%%%%%%%%%%%%%%%%%%%%%%%%%%%%
\subsubsection{Construction of the FSR history}\label{fsr}

\begin{figure}[t]
\centering
\includegraphics[scale=0.55]{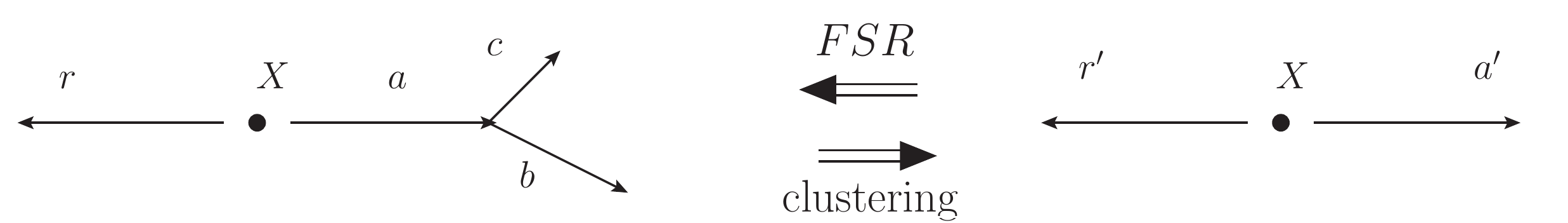}
\caption{\small 
Illustrating that a final state shower model evolves the set of partons $a^{\prime}_{}$ and $r^{\prime}_{}$ and then generates partons $b$, $c$ and $r$ (from right to left), and its inverse (from left to right).
}
\label{FSR}
\end{figure}

Let us consider the case that there are two outgoing partons $b$ and $c$ and one parton $r$ which is either incoming or outgoing. Then let us suppose that the $b$ and $c$ are clustered into a new outgoing parton $a$ and hence the set of partons $a^{\prime}_{}$ and $r^{\prime}_{}$ together with a clustering scale $p_{\perp \mathrm{clus}}^{}$ is produced. 
This is illustrated in Figure~\ref{FSR} (from left to right). The clustering has to proceed as if the final state shower model in PYTHIA8 had evolved the set of the partons $a^{\prime}_{}$ and $r^{\prime}_{}$ and then had generated the partons $b$, $c$ and $r$ with the evolution scale $p_{\perp \mathrm{evol}}^{}=p_{\perp \mathrm{clus}}^{}$.
The parton $r$ may not be uniquely determined. In our algorithm, the $r$ is randomly chosen. The new outgoing parton $a$ after the clustering is off mass shell. Thus we need to put the $a$ on mass shell. There are two approaches depending on whether the parton $r$ is outgoing or incoming. \\

When the $r$ is an outgoing parton, the parton $a$ is put on mass shell by giving the four-momentum of the parton $a$ to the parton $r$. The four-momentum of the $a+r$ is kept unchanged i.e. $p_a^{}+p_r^{}=p_{a^{\prime}_{}}^{}+p_{r^{\prime}_{}}^{}$. The kinematics of all the other partons indicated by $X$ in Figure~\ref{FSR} including the incoming partons will not be affected. The four-momenta of the partons $a^{\prime}_{}$ and $r^{\prime}_{}$ are derived as follows. 
\begin{enumerate}
\item Boost the $a$ and $r$ to the $a+r$ rest frame, $p_{a, 0}^{}$ and $p_{r, 0}^{}$.
\item In the $a+r$ rest frame, calculate the energies and the absolute values of the momenta of the $a$ and $r$ which are put on mass shell with the on-shell masses $m_a^{}$ and $m_r^{}$,
\begin{subequations}
\begin{align}
m_{ar}^2=(p_{a, 0}^{}+p_{r, 0}^{})^2,\\
E_{a, \mathrm{new}}^{}=\frac{m_{ar}^2+m_a^2-m_r^2}{2m_{ar}^{}},\\
E_{r, \mathrm{new}}^{}=\frac{m_{ar}^2-m_a^2+m_r^2}{2m_{ar}^{}},\\
|\vec{p}_{a, \mathrm{new}}^{}|=|\vec{p}_{r, \mathrm{new}}^{}|=\sqrt{E_{a, \mathrm{new}}^2-m_a^2}. 
\end{align}
\end{subequations}
\item Modify the magnitudes of the momenta of the $a$ and $r$
to the $|\vec{p}_{a, \mathrm{new}}^{}|$ and $|\vec{p}_{r, \mathrm{new}}^{}|$, respectively, while the directions of the momenta are kept unchanged, 
\begin{align}
p_a^i=\frac{p_{a, 0}^i}{|\vec{p}_{a, 0}^{}|} |\vec{p}_{a, \mathrm{new}}^{}|,\ \ \ \ \ \ 
p_r^i=\frac{p_{r, 0}^i}{|\vec{p}_{r, 0}^{}|} |\vec{p}_{r, \mathrm{new}}^{}|. \ \ \ \ \ (i=1,2,3)
\end{align}
\item Boost these back to the original $a+r$ frame. \\
\end{enumerate}

When the $r$ is an incoming parton, the parton $a$ is put on mass shell by reducing the four-momenta of the $a$ and $r$, while the four-momentum $p_r^{}-p_a^{}$ is kept unchanged. Thus, the four-momentum $p_a^{}+p_r^{}$ will not be conserved in this case. The four-momenta of the partons $a^{\prime}_{}$ and $r^{\prime}_{}$ are derived from
\begin{subequations}
\begin{align}
\alpha&=\frac{p_r^3}{E_r^{}}=1\ \mathrm{or}-1,\\
p_{r^{\prime}_{}}^{\mu}&=\Biggl( E_r^{}-\frac{(p_a^{})^2-m_a^2}{2(E_a^{}-\alpha p_a^3)},\ 0,\ 0,\ \alpha \biggl(E_r^{}-\frac{(p_a^{})^2-m_a^2}{2(E_a^{}-\alpha p_a^3)}\biggr) \Biggr),\\
p_{a^{\prime}_{}}^{\mu}&=\Bigl( E_{a}^{}+E_{r^{\prime}_{}}^{}-E_{r}^{},\ p_a^1,\ p_a^2,\ p_{a}^{3}+p_{r^{\prime}_{}}^3-p_{r}^3 \Bigr).
\end{align}
\end{subequations}
\\

For either case, the clustering scale $p_{\perp \mathrm{clus}}^{}$ is derived from,
\begin{subequations}\label{const-fsr-scale}
\begin{align}
&p_a^{}=p_b^{}+p_c^{},\\
&p_0^{}=p_{a^{\prime}_{}}^{}+p_{r^{\prime}_{}}^{}, \\
&z=\frac{p_0^{}\cdot p_b^{}}{p_0^{}\cdot p_a^{}}\biggl(1-\frac{p_0^{}\cdot p_c^{}}{p_0^{}\cdot p_b^{}}\frac{m_a^2}{2p_0^{}\cdot p_a^{} + m_r^2-m_a^2-p_0^2}\biggr),\\
&p_{\perp \mathrm{clus}}^2=z(1-z)\bigl((p_a^{})^2-m_a^2\bigr),
\end{align}
\end{subequations}
where $m_a^{}$ and $m_r^{}$ are the on-shell masses of the partons $a$ and $r$, respectively. Notice that the $p_0^{}$ is constructed from the four-momenta of the partons $a^{\prime}_{}$ and $r^{\prime}_{}$, which are obtained from the above algorithm. 
The mass effect is taken into account. This is particularly relevant in the clustering which includes a top quark. 
Our algorithm is carefully tested as follows. We apply the algorithm to a process which has been generated by the PYTHIA8 final state evolution of a hard process, and then we confirm that the algorithm correctly reproduces the hard process.

%%%%%%%%%%%%%%%%%%%%%%%%%%%%%%%%%%%%%%%%%%%%%%%%%%%%%%%%%%%%%%%%%%%%%%%%%%%%%%%%%%%%%%%%%%%%%%%%%%%%%%%%%%%%%%%%%%%%%%%%%%%%%%%%%%%%%%%%%%%%%%%%%%%%%%%%%%%%%%%%%%%%%%%%%%%%%%%%%%%%%%%%%%%%%%%%%%%%%%%%%%%%%%%%%%%%%%%%%%%%%%%%%%%%%%%%%%%%%%%%%%%%%%%%%%%%%%%%%%%%%%%%%%%%%%%%%%%%%%%%%%%%%%%%%%%%%%%%%%%%%%%%%%%%%%%%%%%%%%%%%%%%%%%%%%%%%%%%%%%%%%%%%%%%%%%%%%%%%%%%%%%%%%%%%%%%%%%%%%%%%%%%%%%%%%%%%%%%%%%%%%%%%%%%%%%%%%%%%%%%%%%%%%%%%%%%%%%%%%%%%%%%%%%%%%%%%%%%%%%%%%%%%%%%%%%%%%%%%%%%%%%%%%%%%%%%%%%%%%%%%%%%%%%%%%%%%%%%%%%%%%%%%%%%%%%%%%%%%%%%%%%%%%%%%%%%%%%%%%%%%%%%%%%%%%%%%%%%%%%%%%%%%%%%%%%%%%%%%%%%%%%%%%%%%%%%%%%%%%%%%%%%%%%%%%%%%%%%%%%%%%%%
\subsubsection{Some technical details}\label{sec:remarks}
In this section, we write some technical details in our implementation of the shower history construction:
\begin{itemize}
\item The clustering $2\to1$ must respect the QCD $1\to2$ vertices.
\item When there are more than one candidates for a clustering pair at a clustering step, the one which has the lowest clustering scale is always chosen~\footnote{A more sophisticated approach is proposed in ref.~\cite{Lonnblad:2011xx}.}.
\item Sequential clustering scales are required to be ordered, that is, a clustering scale at a clustering step is required to be higher than the scale at the previous clustering step. However, it is not required that $t_X^{} > t_1^{}, t_2^{}, \cdots$.
\item If the hard process $\{p\}_X^{}$ cannot be obtained at the end of sequential clusterings, we take the following approach. 
Let us assume the case that the $\{p\}_X^{}$ is not obtained from a $\{p\}_{X+n}^{}$ after sequential $n$ times clusterings. The program for the shower history construction is executed again on the $\{p\}_{X+n}^{}$, but this time a clustering pair whose clustering scale is not the lowest but the second lowest is chosen at the first clustering step. If the $\{p\}_X^{}$ is not obtained yet, the program is executed again on the $\{p\}_{X+n}^{}$ and a clustering pair whose clustering scale is the third lowest is chosen at the first clustering step, and so on. If this approach still does not help, the shower history construction for the event sample is abandoned. 
\end{itemize}

%%%%%%%%%%%%%%%%%%%%%%%%%%%%%%%%%%%%%%%%%%%%%%%%%%%%%%%%%%%%%%%%%%%%%%%%%%%%%%%%%%%%%%%%%%%%%%%%%%%%%%%%%%%%%%%%%%%%%%%%%%%%%%%%%%%%%%%%%%%%%%%%%%%%%%%%%%%%%%%%%%%%%%%%%%%%%%%%%%%%%%%%%%%%%%%%%%%%%%%%%%%%%%%%%%%%%%%%%%%%%%%%%%%%%%%%%%%%%%%%%%%%%%%%%%%%%%%%%%%%%%%%%%%%%%%%%%%%%%%%%%%%%%%%%%%%%%%%%%%%%%%%%%%%%%%%%%%%%%%%%%%%%%%%%%%%%%%%%%%%%%%%%%%%%%%%%%%%%%%%%%%%%%%%%%%%%%%%%%%%%%%%%%%%%%%%%%%%%%%%%%%%%%%%%%%%%%%%%%%%%%%%%%%%%%%%%%%%%%%%%%%%%%%%%%%%%%%%%%%%%%%%%%%%%%%%%%%%%%%%%%%%%%%%%%%%%%%%%%%%%%%%%%%%%%%%%%%%%%%%%%%%%%%%%%%%%%%%%%%%%%%%%%%%%%%%%%%%%%%%%%%%%%%%%%%%%%%%%%%%%%%%%%%%%%%%%%%%%%%%%%%%%%%%%%%%%%%%%%%%%%%%%%%%%%%%%%%%%%%%%%%%
\subsection{Weight functions}\label{sec:weight-functions}

In eq.~(\ref{intro-const-function}) of Section~\ref{algorithmsub}, we have introduced a function $f(z,t;\{p\}_{X+n}^{})$ which constraints the DGLAP shower evolution of the $\{p\}_{X+n}^{}$ at the evolution scale $t$ and the energy fraction $z$.
In the improved DGLAP equation, some part of the constraint is already included in the tree level cross sections.
This has been implied by using $f^{\prime}_{}(z,t;\{p\}_{X+n}^{})$ instead of $f(z,t;\{p\}_{X+n}^{})$ in eqs.~(\ref{improved-DGLAP-1}) or (\ref{improved-DGLAP-5}). 
To make it clear, let us write down the second term in the right hand side (RHS) of the DGLAP evolution of the hard process $\sigma(X)$ in eq.~(\ref{X-DGLAP-evolution}), which gives the integrated probability of exclusively generating one radiation during the evolution between $t_X^{}$ and $t_{\Lambda}^{}$,
\begin{align}
\int_0^1 dx_1^{} \int_0^1 dx_2^{} g(x_1^{}, t_X^{}) g(x_2^{}, t_X^{})
\hat{\sigma}\bigl(gg \to X; \hat{s}=x_1^{}x_2^{}s \bigr)
\int^{t_X^{}}_{t_{\Lambda}^{}} \frac{dt_1^{}}{t_1^{}}\int^1_0 d\hat{p}(z_1^{})
f\bigl(z_1^{}, t_1^{};  \{p\}_{X}^{} \bigr),\label{const-function-one-rad}
\end{align}
where now the hard process is explicitly written and the Sudakov form factors are omitted. 
This term is improved with a help from the tree level cross section 
\begin{align}
\int_0^1 dx_1^{} \int_0^1 dx_2^{} g(x_1^{}, t_1^{}) g(x_2^{}, t_1^{})
\hat{\sigma}\bigl(gg \to X+g; \hat{s}=x_1^{}x_2^{}s \bigr)
f^{\prime}_{}\bigl(z_1^{}, t_1^{};  \{p\}_{X}^{} \bigr),\label{const-function-one-improved-before}
\end{align}
where the merging scale cut $\{p\}_{X+g}^{}>Q^2_{\mathrm{cut}}$ is implicit. Furthermore, we write this as
\begin{align}
\int_0^1 dx_1^{} \int_0^1 dx_2^{} g(x_1^{}, t_{\Lambda}^{}) g(x_2^{}, t_{\Lambda}^{})
\hat{\sigma}\bigl(gg \to X+g; \hat{s}=x_1^{}x_2^{}s \bigr)
f^{\prime}_{}\bigl(z_1^{}, t_1^{};  \{p\}_{X}^{} \bigr)
f^{\prime}_{}\bigl( t_{\Lambda}^{};  \{p\}_{X+g}^{} \bigr).\label{const-function-one-improved}
\end{align}
A fixed scale $t_{\Lambda}^{}$ is now used in the PDFs and this change is included in the function $f^{\prime}_{}( t_{\Lambda}^{};  \{p\}_{X+g}^{} )$. 
We call $f^{\prime}_{}(z, t;  \{p\}_{X+n}^{} )$ a weight function hereafter, since it is used to re-weight an event sample in the merging algorithms, see the discussion below eq.~(\ref{CKKW-L-2ndterm}) for the detail.
The first task in this section is to derive the weight function $f^{\prime}_{}(z_1^{}, t_1^{};  \{p\}_{X}^{} )$ explicitly.\\

There are three radiation patterns in the PYTHIA8 parton shower, namely final state radiation with an outgoing recoiling parton, initial state radiation with an incoming recoiling parton and final state radiation with an incoming recoiling parton. This is also discussed in Section~\ref{history}. For the first case i.e. final state radiation with an outgoing recoiling parton, the kinematics of the incoming partons will not be changed. Hence eq.~(\ref{const-function-one-rad}) is evaluated as
\begin{align}
&\int_0^1 dx_1^{} \int_0^1 dx_2^{} g(x_1^{}, t_X^{}) g(x_2^{}, t_X^{})
\hat{\sigma}\bigl(gg \to X; \hat{s}=x_1^{}x_2^{}s \bigr)
\int^{t_X^{}}_{t_{\Lambda}^{}} \frac{dt_1^{}}{t_1^{}}\int^1_0 d\hat{p}(z_1^{}) \nonumber \\
&=
\int_0^1 dx_1^{} \int_0^1 dx_2^{} g(x_1^{}, t_X^{}) g(x_2^{}, t_X^{})
\hat{\sigma}\bigl(gg \to X+g; \hat{s}=x_1^{}x_2^{}s \bigr) \nonumber \\
&=
\int_0^1 dx_1^{} \int_0^1 dx_2^{} g(x_1^{}, t_{\Lambda}^{}) g(x_2^{}, t_{\Lambda}^{})
\hat{\sigma}\bigl(gg \to X+g; \hat{s}=x_1^{}x_2^{}s \bigr)
\frac{g(x_1^{}, t_X^{})}{g(x_1^{}, t_{\Lambda}^{})}
\frac{g(x_2^{}, t_X^{})}{g(x_2^{}, t_{\Lambda}^{})}, \label{const-function-one-rad-FF}
\end{align}
thus we obtain
\begin{align}
f^{\prime}_{}\bigl(z_1^{}, t_1^{};  \{p\}_{X}^{} \bigr)
f^{\prime}_{}\bigl( t_{\Lambda}^{};  \{p\}_{X+g}^{} \bigr)
=
\frac{g(x_1^{}, t_X^{})}{g(x_1^{}, t_{\Lambda}^{})}
\frac{g(x_2^{}, t_X^{})}{g(x_2^{}, t_{\Lambda}^{})},
\end{align}
or equivalently
\begin{align}
f^{\prime}_{}\bigl(z_1^{}, t_1^{};  \{p\}_{X}^{} \bigr)
f^{\prime}_{}\bigl( t_{\Lambda}^{};  \{p\}_{X+g}^{} \bigr)
=
\frac{g(x_1, t_1^{}) }{g(x_1, t_{\Lambda}^{}) }
\frac{g(x_1, t_X^{}) }{g(x_1, t_1^{})}
\frac{g(x_2, t_X^{}) }{g(x_2, t_1^{})}
\frac{g(x_2, t_1^{}) }{g(x_2, t_{\Lambda}^{}) }. \label{con-func-FF}
\end{align}
\\

For the second case i.e. initial state radiation with an incoming recoiling parton, the kinematics of a radiating incoming parton will be changed. Thus the weight function should include the parton distribution functions (PDFs). The PDF factor is explicitly given in eq.~(\ref{infinite-iteration-DGLAP-with-Sudakov}). When we assume that the incoming parton which has the energy fraction $x_1^{}$ is the radiating parton, eq.~(\ref{const-function-one-rad}) is evaluated as
\begin{align}
&\int_0^1 dx_1^{} \int_0^1 dx_2^{} g(x_1^{}, t_X^{}) g(x_2^{}, t_X^{})
\hat{\sigma}\bigl(gg \to X; \hat{s}=x_1^{}x_2^{}s \bigr)
\int^{t_X^{}}_{t_{\Lambda}^{}} \frac{dt_1^{}}{t_1^{}}\int^1_0 d\hat{p}(z_1^{}) 
\frac{g(x_1^{}/z_1^{}, t_1^{})}{g(x_1^{}, t_1^{})}\nonumber \\
&=
\int_0^1 dx_1^{} \int_0^1 dx_2^{} g(x_1^{}, t_X^{}) g(x_2^{}, t_X^{})
\hat{\sigma}\bigl(gg \to X+g; \hat{s}=(x_1^{}/z_1^{})x_2^{}s \bigr)
\frac{1}{z_1^{}}
\frac{g(x_1^{}/z_1^{}, t_1^{})}{g(x_1^{}, t_1^{})}
\nonumber \\
&=
\int_0^1 dw_1^{} \int_0^1 dx_2^{} g(z_1^{}w_1^{}, t_X^{}) g(x_2, t_X^{})
\hat{\sigma}\bigl(gg \to X+g; \hat{s}=w_1^{}x_2^{}s \bigr)
\frac{g(w_1^{}, t_1^{})}{g(z_1^{}w_1^{}, t_1^{})} \nonumber \\
&=
\int_0^1 dw_1^{} \int_0^1 dx_2^{} g(w_1^{}, t_1^{}) g(x_2, t_1^{})
\hat{\sigma}\bigl(gg \to X+g; \hat{s}=w_1^{}x_2^{}s \bigr)
\frac{ g(z_1^{}w_1^{}, t_X^{})}{g(z_1^{}w_1^{}, t_1^{})} 
\frac{g(x_2, t_X^{}) }{g(x_2, t_1^{})}
\nonumber \\
&=
\int_0^1 dw_1^{} \int_0^1 dx_2^{} g(w_1^{}, t_{\Lambda}^{}) g(x_2, t_{\Lambda}^{})
\hat{\sigma}\bigl(gg \to X+g; \hat{s}=w_1^{}x_2^{}s \bigr)
\frac{g(w_1^{}, t_1^{})}{ g(w_1^{}, t_{\Lambda}^{}) }
\frac{ g(z_1^{}w_1^{}, t_X^{})}{g(z_1^{}w_1^{}, t_1^{})} 
\frac{g(x_2, t_X^{}) }{g(x_2, t_1^{})}
\frac{g(x_2, t_1^{}) }{ g(x_2, t_{\Lambda}^{}) }, \label{const-function-one-rad-II}
\end{align}
thus we obtain
\begin{align}
f^{\prime}_{}\bigl(z_1^{}, t_1^{};  \{p\}_{X}^{} \bigr)
f^{\prime}_{}\bigl( t_{\Lambda}^{};  \{p\}_{X+g}^{} \bigr)
=
\frac{g(w_1^{}, t_1^{})}{ g(w_1^{}, t_{\Lambda}^{}) }
\frac{ g(z_1^{}w_1^{}, t_X^{})}{g(z_1^{}w_1^{}, t_1^{})} 
\frac{g(x_2, t_X^{}) }{g(x_2, t_1^{})}
\frac{g(x_2, t_1^{}) }{ g(x_2, t_{\Lambda}^{}) }.\label{con-func-II}
\end{align}
Note that $1/z_1^{}$ at the second line of eq.~(\ref{const-function-one-rad-II}) comes from $d\hat{p}(z_1^{})$, see eq.~(\ref{short-hand-splitting-function}). \\

For the third case i.e. final state radiation with an incoming recoiling parton, the kinematics of the incoming recoiling parton will be changed. Thus the weight function includes the PDFs~\cite{Corke:2010yf}. 
When we assume that the incoming parton which has the energy fraction $x_1^{}$ is the recoiling parton, eq.~(\ref{const-function-one-rad}) is evaluated as, by letting $w_1^{}$ denotes the energy fraction of the recoiling parton after the radiation, 
\begin{align}
&\int_0^1 dx_1^{} \int_0^1 dx_2^{} g(x_1^{}, t_X^{}) g(x_2^{}, t_X^{})
\hat{\sigma}\bigl(gg \to X; \hat{s}=x_1^{}x_2^{}s \bigr)
\int^{t_X^{}}_{t_{\Lambda}^{}} \frac{dt_1^{}}{t_1^{}}\int^1_0 d\hat{p}(z_1^{}) 
\frac{w_1^{}}{x_1^{}}
\frac{g(w_1^{}, t_1^{})}{g(x_1^{}, t_1^{})}\nonumber \\
&=
\int_0^1 dw_1^{} \int_0^1 dx_2^{} g(x_1^{}, t_X^{}) g(x_2^{}, t_X^{})
\hat{\sigma}\bigl(gg \to X; \hat{s}=x_1^{}x_2^{}s \bigr)
\int^{t_X^{}}_{t_{\Lambda}^{}} \frac{dt_1^{}}{t_1^{}}\int^1_0 d\hat{p}(z_1^{}) 
\frac{g(w_1^{}, t_1^{})}{g(x_1^{}, t_1^{})}\nonumber \\
&=
\int_0^1 dw_1^{} \int_0^1 dx_2^{} g(w_1^{}, t_1^{}) g(x_2^{}, t_X^{})
\hat{\sigma}\bigl(gg \to X; \hat{s}=x_1^{}x_2^{}s \bigr)
\int^{t_X^{}}_{t_{\Lambda}^{}} \frac{dt_1^{}}{t_1^{}}\int^1_0 d\hat{p}(z_1^{}) 
\frac{g(x_1^{}, t_X^{})}{g(x_1^{}, t_1^{})}\nonumber \\
&=
\int_0^1 dw_1^{} \int_0^1 dx_2^{} g(w_1^{}, t_1^{}) g(x_2^{}, t_X^{})
\hat{\sigma}\bigl(gg \to X+g; \hat{s}=w_1^{}x_2^{}s \bigr)
\frac{g(x_1^{}, t_X^{})}{g(x_1^{}, t_1^{})}\nonumber \\
&=
\int_0^1 dw_1^{} \int_0^1 dx_2^{} g(w_1^{}, t_{\Lambda}^{}) g(x_2, t_{\Lambda}^{})
\hat{\sigma}\bigl(gg \to X+g; \hat{s}=w_1^{}x_2^{}s \bigr)
\frac{g(w_1^{}, t_1^{})}{ g(w_1^{}, t_{\Lambda}^{}) }
\frac{g(x_1^{}, t_X^{})}{g(x_1^{}, t_1^{})}
\frac{g(x_2, t_X^{}) }{g(x_2, t_1^{})}
\frac{g(x_2, t_1^{}) }{ g(x_2, t_{\Lambda}^{}) },
\end{align}
thus we obtain
\begin{align}
f^{\prime}_{}\bigl(z_1^{}, t_1^{};  \{p\}_{X}^{} \bigr)
f^{\prime}_{}\bigl( t_{\Lambda}^{};  \{p\}_{X+g}^{} \bigr)
=
\frac{g(w_1^{}, t_1^{})}{ g(w_1^{}, t_{\Lambda}^{}) }
\frac{g(x_1^{}, t_X^{})}{g(x_1^{}, t_1^{})}
\frac{g(x_2, t_X^{}) }{g(x_2, t_1^{})}
\frac{g(x_2, t_1^{}) }{ g(x_2, t_{\Lambda}^{}) }.\label{con-func-FI}
\end{align}
\\

By looking at the weight functions of the three radiation patters in eqs.~(\ref{con-func-FF}), (\ref{con-func-II}) and (\ref{con-func-FI}), the general expression can be found as follows.  Let us suppose an event sample $\{p\}_{X+n}^{}$, which gives a shower history $\{p\}_{X+(n-1)}^{}, \{p\}_{X+(n-2)}^{}, \cdots, \{p\}_{X+i}^{}, \cdots, \{p\}_{X+1}^{}, \{p\}_{X}^{}$ with $t_n^{} < t_{n-1}^{} < \cdots < t_{i+1}^{} < \cdots < t_2^{}< t_1^{}$. Let us also define the energy fractions and the parton types of the incoming partons in the $\{p\}_{X+i}^{}$ by $x_1^{(i)}$, $x_2^{(i)}$ and $f_1^{(i)}$, $f_2^{(i)}$, respectively. The weight function for the $\{p\}_{X+i}^{}$ is given by~\cite{Lonnblad:2011xx}
\begin{align}
f^{\prime}_{}\bigl(z_{i+1}^{}, t_{i+1}^{};  \{p\}_{X+i}^{} \bigr)
=
\frac{ \alpha_s^{}( t_{i+1}^{} ) }{ \alpha_s^{}( t_{\Lambda}^{} ) } 
\frac{ f_1^{(i)}( x_1^{(i)}, t_i^{} ) }{ f_1^{(i)}( x_1^{(i)}, t_{i+1}^{} ) }
\frac{ f_2^{(i)}( x_2^{(i)}, t_i^{} ) }{ f_2^{(i)}( x_2^{(i)}, t_{i+1}^{} ) }, \label{general-const-function}
\end{align}
from which the total weight function for the $\{p\}_{X+n}^{}$ is 
\begin{align}
\prod_{i=0}^{n}f^{\prime}_{}\bigl(z_{i+1}^{}, t_{i+1}^{};  \{p\}_{X+i}^{} \bigr),\label{total-const-function}
\end{align}
where $t_0^{}=t_X^{}$ and $t_{n+1}^{}=t_{\Lambda}^{}$. It is assumed that the scale $t_{\Lambda}^{}$ is used as the scales of the PDFs and those of the strong couplings for generating the $\{p\}_{X+n}^{}$. A factor consisting of the strong couplings is present in the function, because the evolution scale $t$ is used as the scale in the strong coupling in the PYTHIA8 parton shower evolution and thus this strategy should be also used in the improved evolution equations~\cite{Catani:2001cc}. We take into account the fact that The PYTHIA8 shower model uses the different strong couplings for initial state radiation and final state radiation.
When a radiation is classified as the initial state radiation (the final state radiation) with a scale $t_{i+1}^{}$ by the shower history construction,  $\alpha_s^{}(m_z^{})=1.37$ $(1.383)$ is used for the factor in eq.~(\ref{general-const-function}). 
It can be easily confirmed that the results in eqs.~(\ref{con-func-FF}), (\ref{con-func-II}) and (\ref{con-func-FI}) are derived from eq.~(\ref{total-const-function}) by setting $n=1$.

%%%%%%%%%%%%%%%%%%%%%%%%%%%%%%%%%%%%%%%%%%%%%%%%%%%%%%%%%%%%%%%%%%%%%%%%%%%%%%%%%%%%%%%%%%%%%%%%%%%%%%%%%%%%%%%%%%%%%%%%%%%%%%%%%%%%%%%%%%%%%%%%%%%%%%%%%%%%%%%%%%%%%%%%%%%%%%%%%%%%%%%%%%%%%%%%%%%%%%%%%%%%%%%%%%%%%%%%%%%%%%%%%%%%%%%%%%%%%%%%%%%%%%%%%%%%%%%%%%%%%%%%%%%%%%%%%%%%%%%%%%%%%%%%%%%%%%%%%%%%%%%%%%%%%%%%%%%%%%%%%%%%%%%%%%%%%%%%%%%%%%%%%%%%%%%%%%%%%%%%%%%%%%%%%%%%%%%%%%%%%%%%%%%%%%%%%%%%%%%%%%%%%%%%%%%%%%%%%%%%%%%%%%%%%%%%%%%%%%%%%%%%%%%%%%%%%%%%%%%%%%%%%%%%%%%%%%%%%%%%%%%%%%%%%%%%%%%%%%%%%%%%%%%%%%%%%%%%%%%%%%%%%%%%%%%%%%%%%%%%%%%%%%%%%%%%%%%%%%%%%%%%%%%%%%%%%%%%%%%%%%%%%%%%%%%%%%%%%%%%%%%%%%%%%%%%%%%%%%%%%%%%%%%%%%%%%%%%%%%%%%%
\subsection{The merging algorithm with top decays}\label{sec:algorithm-top-decays}

The decay of the top quark is characterized by its decay width $\Gamma(t\to bW^+)\sim 1.5$ GeV. The large scale discrepancy between the decay width and the production scale of a top quark ($\sim m_t^{}$) indicates that QCD radiation off a top quark takes place faster than the decay of the top quark. 
Although gluon radiation off a top quark is suppressed by its large mass, it can be important for more accurate predictions. The parton shower generator in PYTHIA8 models gluon radiation off heavy particles including the top quark~\cite{Norrbin:2000uu}. Therefore, the construction of the PYTHIA8 parton shower history also has to take into account the clustering which includes a top quark i.e. $t+g\to t$.\\

\begin{figure}[t]
\centering
\includegraphics[scale=0.4]{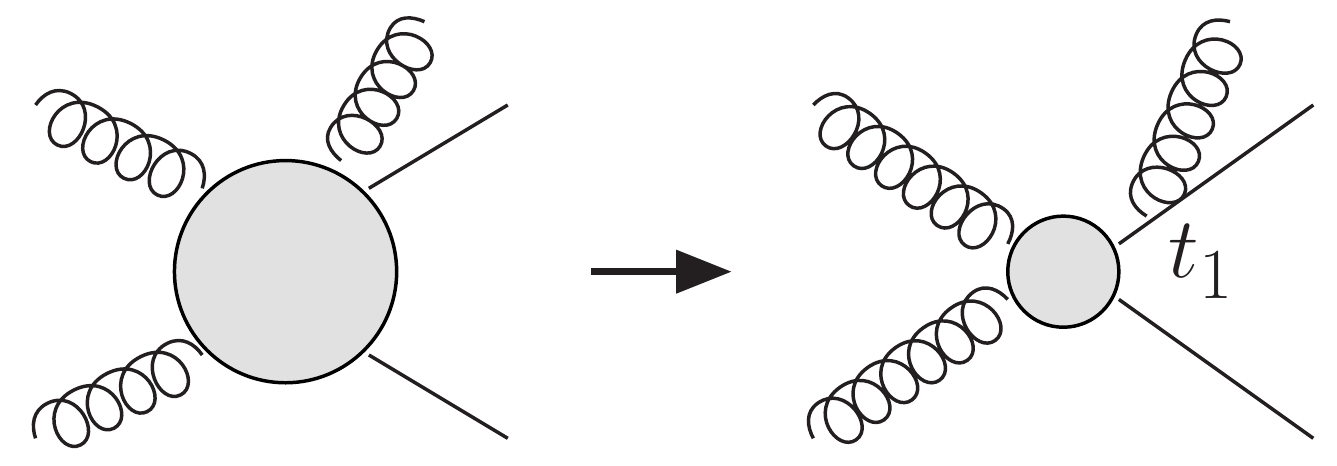}
\hspace{1.6cm}
\includegraphics[scale=0.4]{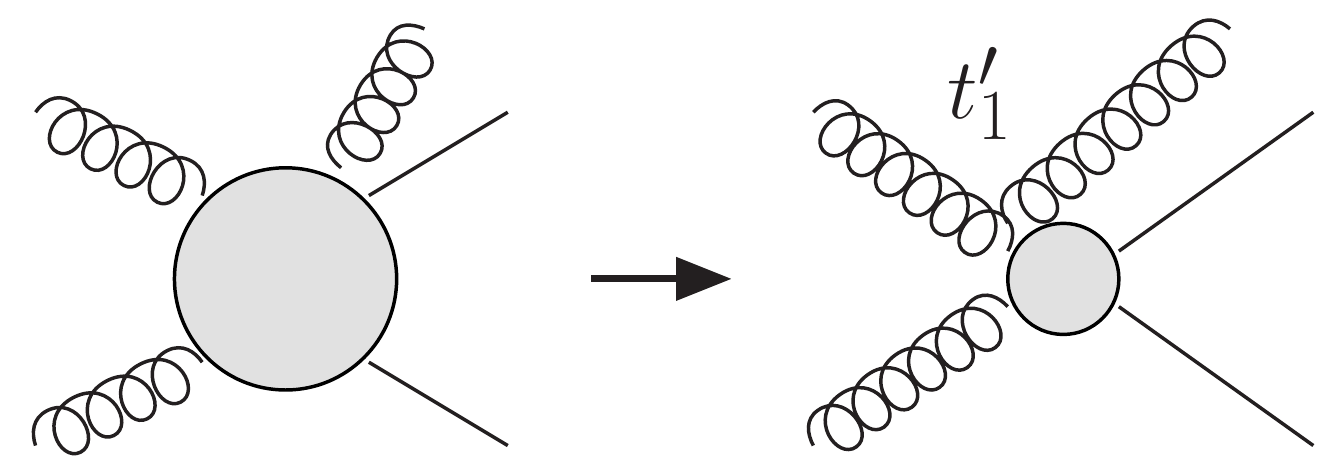}
\caption{\small 
Schematic pictures showing that two different shower histories are constructed for an event $gg\to t\bar{t}g$.
In the left picture, the clustering $t+g\to t$, which gives the lowest clustering scale $t_1^{}$ of all the candidates for a clustering pair in the event, is performed. 
The right picture shows the case that the clustering $t+g\to t$ is not allowed in the program. A different clustering which gives a higher scale $t_1^{\prime}$ than the $t_1^{}$ will be performed instead for the same event.}
\label{fig:top-gluon}
\end{figure}

Let us first examine the case that the clustering $t+g\to t$ is not implemented in the program for the shower history construction. 
Figure~\ref{fig:top-gluon} presents schematic pictures showing that two different shower histories are constructed for an event $gg\to t\bar{t}g$. 
In the left picture, the clustering $t+g\to t$ which gives the lowest clustering scale $t_1^{}$ of all the candidates for a clustering pair in the event is performed. If the clustering $t+g\to t$ is not implemented in the program, a different clustering which gives a higher scale $t_1^{\prime}$ than the $t_1^{}$ will be performed for the same event, as illustrated in the right picture. 
This difference in the parton shower history affects the calculation of the Sudakov form factors in the evolution equation in eq.~(\ref{improved-DGLAP-5}) and thus can induce some problems such as the unnecessary dependence on the merging scale.\\

The clustering $t+g\to t$ in the construction of the PYTHIA8 parton shower history is necessary, no matter whether top quarks are decayed in simulation, since the decays occur after gluon radiation off the top quarks as discussed above. The simplest approach to simulate the $t\bar{t}$ production including a $t\bar{t}$ decay may be as follows. At first, the event samples of the $t\bar{t}$ production process are generated with the merging algorithms i.e. $X=t\bar{t}$ in eq.~(\ref{improved-DGLAP-5}), by assuming that the $t\bar{t}$ is stable. Then, the top and antitop quarks are decayed independently following the differential partial decay width, $d\Gamma(t\to l^+\bar{\nu}b)$ or $d\Gamma(t\to q\bar{q}^{\prime}b)$. QCD radiation off the decay products will be generated at the end. In this simple approach, correlations between the decay products of the top quark and those of the antitop quark are not produced correctly, and the off shell effects of the top and antitop quarks are also absent. One solution to these issues is to generate the event samples which include a $t\bar{t}$ decay as a part of the hard process, according to the exact tree level matrix elements. \\

\begin{figure}[t]
\centering
\includegraphics[scale=0.34]{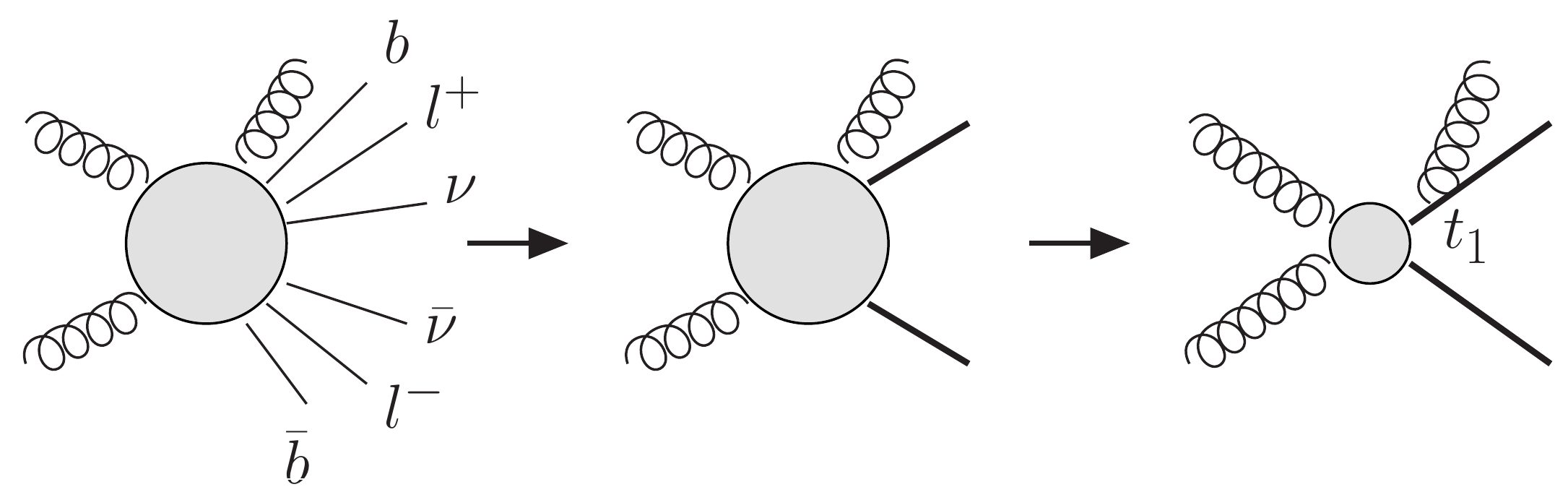}
\hspace{0.7cm}
\includegraphics[scale=0.34]{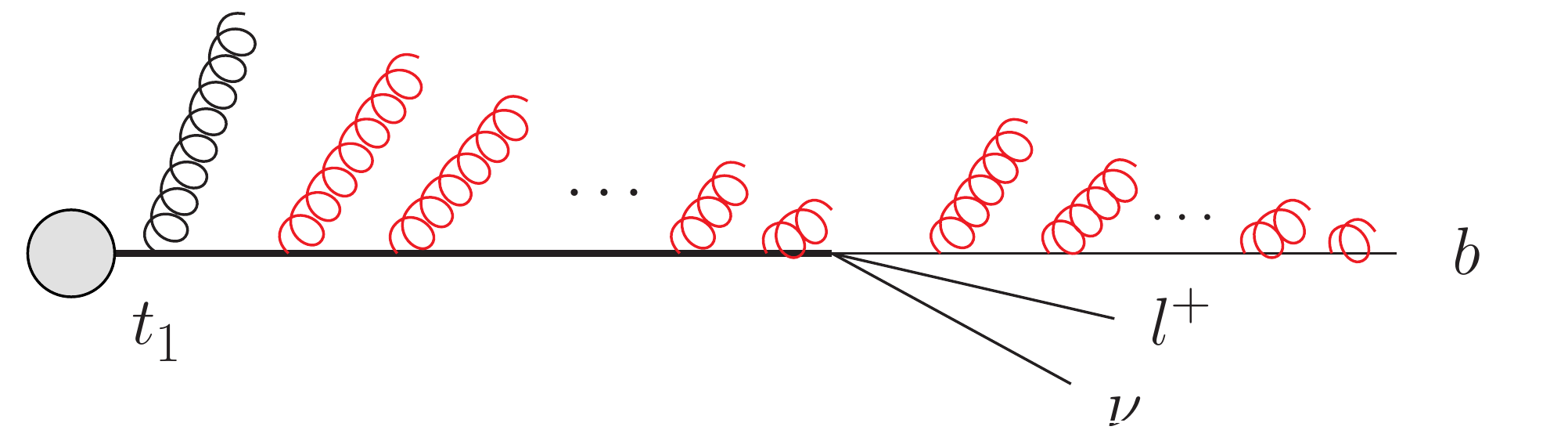}
\caption{\small {\it left}: An event sample $\{p\}_{bl^+_{}\nu\bar{b}l^-_{}\bar{\nu}+g}^{}$ reconstructs a $\{p\}_{t\bar{t}+g}^{}$ as an intermediate process, at the first allow. A shower history of the $\{p\}_{t\bar{t}+g}^{}$ is constructed, at the second allow. {\it right}: A part of the shower evolution of the $\{p\}_{bl^+_{}\nu\bar{b}l^-_{}\bar{\nu}+g}^{}$. The black lines represent the particles generated according to the tree level cross section. The red lines represent particles generated with the shower evolution.}
\label{fig:topdecay-gluon}
\end{figure}

In our study, the event generation of the $t\bar{t}$ production including a $t\bar{t}$ decay is performed as follows.
Let us consider an event sample $\{p\}_{bl^+_{}\nu\bar{b}l^-_{}\bar{\nu}+g}^{}$, which originates from a $\{p\}_{t\bar{t}+g}^{}$ and is generated with the tree level matrix elements. At first, the $\{p\}_{t\bar{t}+g}^{}$ is reconstructed as an intermediate process from the $\{p\}_{bl^+_{}\nu\bar{b}l^-_{}\bar{\nu}+g}^{}$. The top and antitop quarks are not necessarily on-shell. Next, the program for the shower history construction is executed on the $\{p\}_{t\bar{t}+g}^{}$, and hence a $\{p\}_{t\bar{t}}^{}$ and a clustering scale $t_1^{}$ are obtained as a shower history. The sequence of the procedures is illustrated in the left picture of Figure~\ref{fig:topdecay-gluon}.  
The first Sudakov form factor in the evolution equation in eq.~(\ref{improved-DGLAP-5}) will be calculated based on the $\{p\}_{t\bar{t}}^{}$ and the $t_1^{}$, i.e. $\Pi(t_X^{}, t_1^{}; \{p\}_{t\bar{t}}^{})$. The calculation of the second Sudakov form factor $\Pi( t_1^{}, \{p\}_{t\bar{t}+2}^{} < Q_{\mathrm{cut}}^2, \cdots;  \{p\}_{t\bar{t}+g}^{})$ looks a little tricky, because the shower generator must be executed on the $\{p\}_{t\bar{t}+g}^{}$ for the calculation of the second Sudakov form factor, while it must be executed on the $\{p\}_{bl^+_{}\nu\bar{b}l^-_{}\bar{\nu}+g}^{}$ for the event generation. 
In fact, when the top and antitop quarks are present as intermediate particles in a Les Houches event file of the $\{p\}_{bl^+_{}\nu\bar{b}l^-_{}\bar{\nu}+g}^{}$, the PYTHIA8 parton shower program starts the shower evolution of the $\{p\}_{t\bar{t}+g}^{}$ at first. This implementation makes the calculation of $\Pi( t_1^{}, \{p\}_{t\bar{t}+2}^{} < Q_{\mathrm{cut}}^2, \cdots;  \{p\}_{t\bar{t}+g}^{})$ possible. Once the shower evolution of the $\{p\}_{t\bar{t}+g}^{}$ is completed, the kinematic change of the $t\bar{t}$ due to the evolution is reflected into the kinematics of its decay products i.e. $bl^+_{}\nu\bar{b}l^-_{}\bar{\nu}$. Finally, the shower evolution of the decay products is performed in such a way that it does not change the invariant mass of the top quark and that of the antitop quark~\cite{Sjostrand:2004ef}. The right picture of Figure~\ref{fig:topdecay-gluon} schematically shows a part of the shower evolution of the $\{p\}_{bl^+_{}\nu\bar{b}l^-_{}\bar{\nu}+g}^{}$. The black lines represent the particles generated according to the tree level cross section. The red lines represent particles generated with the shower evolution. Note that the first radiation off the bottom quark is also corrected internally by using the differential decay width $d\Gamma(t \to bW^+g)$~\cite{Norrbin:2000uu}.

%%%%%%%%%%%%%%%%%%%%%%%%%%%%%%%%%%%%%%%%%%%%%%%%%%%%%%%%%%%%%%%%%%%%%%%%%%%%%%%%%%%%%%%%%%%%%%%%%%%%%%%%%%%%%%%%%%%%%%%%%%%%%%%%%%%%%%%%%%%%%%%%%%%%%%%%%%%%%%%%%%%%%%%%%%%%%%%%%%%%%%%%%%%%%%%%%%%%%%%%%%%%%%%%%%%%%%%%%%%%%%%%%%%%%%%%%%%%%%%%%%%%%%%%%%%%%%%%%%%%%%%%%%%%%%%%%%%%%%%%%%%%%%%%%%%%%%%%%%%%%%%%%%%%%%%%%%%%%%%%%%%%%%%%%%%%%%%%%%%%%%%%%%%%%%%%%%%%%%%%%%%%%%%%%%%%%%%%%%%%%%%%%%%%%%%%%%%%%%%%%%%%%%%%%%%%%%%%%%%%%%%%%%%%%%%%%%%%%%%%%%%%%%%%%%%%%%%%%%%%%%%%%%%%%%%%%%%%%%%%%%%%%%%%%%%%%%%%%%%%%%%%%%%%%%%%%%%%%%%%%%%%%%%%%%%%%%%%%%%%%%%%%%%%%%%%%%%%%%%%%%%%%%%%%%%%%%%%%%%%%%%%%%%%%%%%%%%%%%%%%%%%%%%%%%%%%%%%%%%%%%%%%%%%%%%%%%%%%%%%%%%%%%%%%%%%%%%%%%%%%%%%%%%%%%%%%%%%%%%%%%%%%%%%%%%%%%%%%%%%%%%%%%%%%%%%%%%%%%%%%%%%%%%%%%%%%%%%
\subsection{Event generation}\label{sec:event-generation}

In this section, we describe a procedure for the event generation of the top quark pair production in detail, by combining the knowledges given in the above sections. \\

\begin{enumerate}
\item Generate the event samples for the $t\bar{t}+0, 1, \dots, N$ partons production processes at proton proton (pp) collisions according to the tree level cross sections, i.e. $\{p\}_{t\bar{t}}^{}, \{p\}_{t\bar{t}+1}^{}, \cdots, \{p\}_{t\bar{t}+N}^{}$. When a decay of the $t\bar{t}$ is to be simulated, the event samples are generated according to the tree level cross sections including the decay of the $t\bar{t}$, i.e. $\{p\}_{bl^+_{}\nu\bar{b}l^-_{}\bar{\nu}}^{}, \{p\}_{bl^+_{}\nu\bar{b}l^-_{}\bar{\nu}+1}^{}, \cdots, \{p\}_{bl^+_{}\nu\bar{b}l^-_{}\bar{\nu}+N}^{}$.
In this paper, only the dilepton decay is studied. 
We let $N$ denotes the maximal number of partons provided by the tree level cross section. We use MadGraph5\verb|_|aMC@NLO\cite{Alwall:2014hca} version 5.2.2.1 for this purpose. The merging scale $Q_{\mathrm{cut}}^{}$ is defined by the longitudinal-boost invariant $k_{\perp}^{}$ variable \cite{Catani:1993hr}
\begin{subequations}\label{ktalgorithm}
\begin{align}
k_{\perp iB}^{}&=p_{T i}^{},\\
k_{\perp ij}^{}&=\mathrm{min}(p_{T i}^{}, p_{T j}^{})\sqrt{(y_i^{}-y_j^{})^2_{}+(\phi_i^{}-\phi_j^{})^2_{}}/R, \label{relativekt}
\end{align}
\end{subequations}
where $p_{T i}^{}$, $y_i^{}$ and $\phi_i^{}$ are the transverse momentum with respect to the beam, rapidity and azimuthal angle of outgoing particle $i$. $R$ is the radius parameter and $R=1$ is used if not otherwise specified. The merging scale cut is imposed only on the light partons, and no cut is imposed on the $t\bar{t}$ and its decay products. A fixed value $t_{\Lambda}^{}$ is used for the scales in the strong couplings and in the parton distribution functions (PDFs). The PDF set CTEQ6L1~\cite{Pumplin:2002vw} is used. 
The center of mass energy for the pp collisions is $14$ TeV, except in the case that the simulation is compared with the data at the $7$ TeV.

\item Select an event sample for the $t\bar{t}+n$ partons process, i.e. $\{p\}_{t\bar{t}+n}^{}$, or $\{p\}_{bl^+_{}\nu\bar{b}l^-_{}\bar{\nu}+n}^{}$ when the $t\bar{t}$ is decayed, with the probability proportional to its integrated tree level cross section obtained in step 1,
\begin{align}
P_n=\frac{\sigma(pp\to t\bar{t}+n)}{\sum_{i=0}^{N}\sigma(pp\to t\bar{t}+i)}.
\end{align}

\item Construct a parton shower history of the $\{p\}_{t\bar{t}+n}^{}$ by following the procedure described in Section~\ref{history}. The history consists of intermediate events $\{p\}_{t\bar{t}+(n-1)}^{}, \{p\}_{t\bar{t}+(n-2)}^{}, \cdots, \{p\}_{t\bar{t}+i}^{}, \cdots, \{p\}_{t\bar{t}+1}^{}, \{p\}_{t\bar{t}}^{}$ with the scales $t_n^{} < t_{n-1}^{} < \cdots < t_{i+1}^{} < \cdots < t_2^{}< t_1^{}$. When the $t\bar{t}$ is decayed, a $\{p\}_{t\bar{t}+n}^{}$ is reconstructed at first from the $\{p\}_{bl^+_{}\nu\bar{b}l^-_{}\bar{\nu}+n}^{}$ as described in Section~\ref{sec:algorithm-top-decays}.

\item When the CKKW-L algorithm is used, the merging scale cut is imposed on the intermediate events, too, as indicated in eq.~(\ref{improved-DGLAP-2}).
Thus, the event sample is vetoed as a whole, unless $\{p\}_{t\bar{t}+i}^{} > Q_{\mathrm{cut}}^2$ for $i=1,2,\cdots, n-2, n-1$. This is not the case in the CKKW-L+ algorithm.

\item Calculate the weight function for the $\{p\}_{t\bar{t}+n}^{}$ by using eqs.~(\ref{general-const-function}) and (\ref{total-const-function}). The scale $t_X^{}$ must be determined from the intermediate event $\{p\}_{t\bar{t}}^{}$ and we use
\begin{align}
t_X^{}=E_T^{}(t) \times E_T^{}(\bar{t}),\label{tX-define}
\end{align}
where $E_T^2=m^2_{}+p_T^2$. Note that when the $t\bar{t}$ is on-shell, $E_T^{}(t) = E_T^{}(\bar{t})$ thus $t_X^{}=E_T^{2}(t)$. This is not the case when the decay of the $t\bar{t}$ is also generated in step 1, since the reconstructed $t\bar{t}$ is not necessarily on-shell. This scale $t_X^{}$ is also used as the renormalization scales of the strong couplings $\alpha_s^2$ for the hard process, that is,
\begin{align}
\frac{\alpha_s^2(t_X^{}) }{ \alpha_s^2(t_{\Lambda}^{}) }
\end{align}
should be added in the weight function. We use $\alpha_s^{}(m_z^{})=0.13$ in the above factor. 
When the construction of the shower history was abandoned in step 3, it is not possible to calculate the weight function in the given way. In such a case, the weight function is given by, instead of eqs.~(\ref{general-const-function}) and (\ref{total-const-function}),
\begin{align}
\frac{\alpha_s^2(t_X^{}) }{ \alpha_s^2(t_{\Lambda}^{}) }
\frac{\alpha_s^{}(p_T^2(1)) \alpha_s^{}(p_T^2(2)) \cdots \alpha_s^{}(p_T^2(n))  }{ \alpha_s^{n}(t_{\Lambda}^{}) }
\frac{ f_1^{(n)}( x_1^{(n)}, t_X^{} ) }{ f_1^{(n)}( x_1^{(n)}, t_{\Lambda}^{} ) }
\frac{ f_2^{(n)}( x_2^{(n)}, t_X^{} ) }{ f_2^{(n)}( x_2^{(n)}, t_{\Lambda}^{} ) },\label{weight-abondoned}
\end{align}
where $p_T^{}(i)$ is the transverse momentum of a parton $i$ in the $\{p\}_{t\bar{t}+n}^{}$. The scale $t_X^{}$ is defined by eq.~(\ref{tX-define}) and now is determined from the $\{p\}_{t\bar{t}+n}^{}$. We use $\alpha_s^{}(m_z^{})=0.13$ for all the strong couplings in eq.~(\ref{weight-abondoned}). 
Once the weight function is obtained, the $\{p\}_{t\bar{t}+n}^{}$ is re-weighted with the function. However, the weight function is not bounded above by unity. Therefore, the upper bound of the weight function must be found at first by calculating the weight function for a large number of $\{p\}_{t\bar{t}+n}^{}$. The integrated cross section obtained in step 1 has to be multiplied by the obtained upper bound of the weight function, of course.

\item Calculate the Sudakov form factor(s) by following the procedure described in Section~\ref{sec:CKKW-L}. When the construction of the shower history was abandoned in step 3, all the Sudakov form factors are set to unity. We use the parton shower model~\cite{Sjostrand:2004ef, Corke:2010yf, Norrbin:2000uu} in PYTHIA8~\cite{Sjostrand:2007gs, Sjostrand:2006za} version 8186. The default tune of the version 8186, tune 4C~\cite{Corke:2010yf}, is basically used, while some functions are turned off. To simplify the analysis, the hadronization after the shower evolution and the multiple interaction are turned off.
The rapidity ordering in the initial state radiation is turned off as suggested in ref.~\cite{Lonnblad:2011xx}. All functions inducing azimuthal asymmetry are turned off, since azimuthal angle information of hard partons is provided by exact tree level matrix elements in our simulation. 

\item Once the event sample is accepted and thus the shower evolution is performed until the shower cutoff scale in step 6, all visible particles in the final state within a rapidity range $|y|<5.0$ including charged leptons are clustered to construct inclusive jets according to the anti-$k_T^{}$ algorithm~\cite{Cacciari:2008gp}. The $t\bar{t}$ will not be included, if it is not decayed. 
The radius parameter is $R=0.4$ if not otherwise specified. We use Fastjet~\cite{Cacciari:2011ma} version $3.1.0$ for this purpose. The rapidity and $p_T^{}$ cuts on jets will be specified in the studies of Section~\ref{result1}.

\item Repeat the above procedures from step 2 to step 7 until a large number of the accepted event samples are accumulated.
\end{enumerate}

%%%%%%%%%%%%%%%%%%%%%%%%%%%%%%%%%%%%%%%%%%%%%%%%%%%%%%%%%%%%%%%%%%%%%%%%%%%%%%%%%%%%%%%%%%%%%%%%%%%%%%%%%%%%%%%%%%%%%%%%%%%%%%%%%%%%%%%%%%%%%%%%%%%%%%%%%%%%%%%%%%%%%%%%%%%%%%%%%%%%%%%%%%%%%%%%%%%%%%%%%%%%%%%%%%%%%%%%%%%%%%%%%%%%%%%%%%%%%%%%%%%%%%%%%%%%%%%%%%%%%%%%%%%%%%%%%%%%%%%%%%%%%%%%%%%%%%%%%%%%%%%%%%%%%%%%%%%%%%%%%%%%%%%%%%%%%%%%%%%%%%%%%%%%%%%%%%%%%%%%%%%%%%%%%%%%%%%%%%%%%%%%%%%%%%%%%%%%%%%%%%%%%%%%%%%%%%%%%%%%%%%%%%%%%%%%%%%%%%%%%%%%%%%%%%%%%%%%%%%%%%%%%%%%%%%%%%%%%%%%%%%%%%%%%%%%%%%%%%%%%%%%%%%%%%%%%%%%%%%%%%%%%%%%%%%%%%%%%%%%%%%%%%%%%%%%%%%%%%%%%%%%%%%%%%%%%%%%%%%%%%%%%%%%%%%%%%%%%%%%%%%%%%%%%%%%%%%%%%%%%%%%%%%%%%%%%%%%%%%%%%%%
\subsection{Tests of our implementation}\label{check}

\begin{figure}[t]
\centering
\includegraphics[scale=0.4]{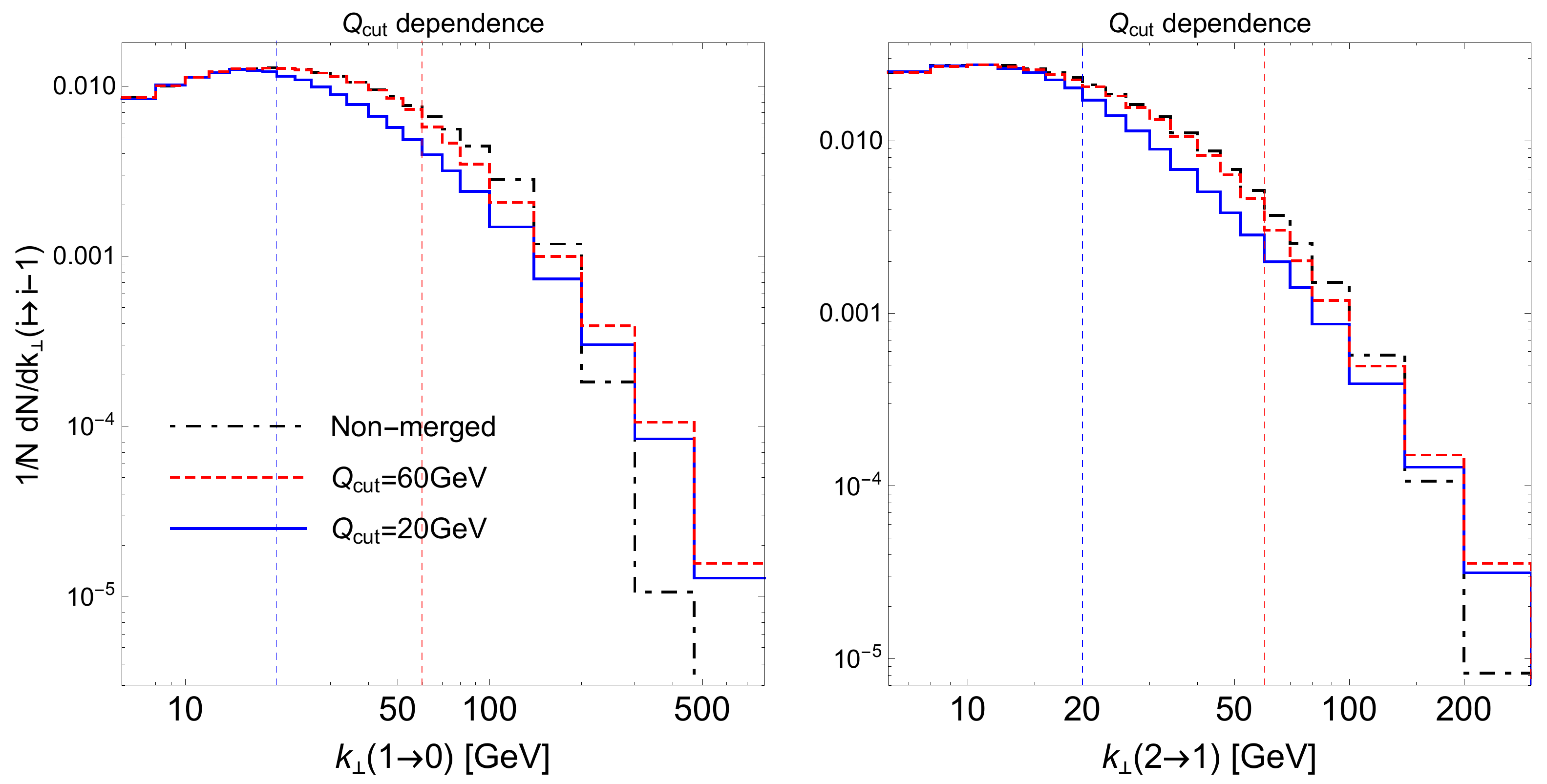}
\caption{\small The differential jet rates for $1\to0$ (left) and $2\to1$ (right) jets. The merging scale is set to $Q_{\mathrm{cut}}^{}=20$ GeV for the blue sold curve and to $Q_{\mathrm{cut}}^{}=60$ GeV for the red dashed curve. The black broken curve represents the result without merging algorithms}
\label{fig:qcut-dependence}
\end{figure}

In this section, our implementation of the CKKW-L+ merging algorithm is tested. At first, the dependence of differential jet rates on the merging scale cutoff $Q_{\mathrm{cut}}^{}$ and that on the parton shower starting scale are studied. A comparison with experimental data is also presented.\\

Differential jet rates are calculated by using the longitudinal-boost invariant $k_{\perp}^{}$ definition in eq.~(\ref{ktalgorithm}) with the radius parameter $R=1$.
In Figure~\ref{fig:qcut-dependence} the differential jet rates for $1\to0$ (left) and $2\to1$ (right) jets are plotted. 
The maximal number of partons $N$ predicted by the tree level cross section is $N=3$, i.e. the event samples are generated exactly according to eq.~(\ref{improved-DGLAP-4}).
A vertical dashed line indicates the merging scale $Q_{\mathrm{cut}}^{}$. 
The merging scale is set to $Q_{\mathrm{cut}}^{}=20$ GeV for the blue sold curve and to $Q_{\mathrm{cut}}^{}=60$ GeV for the red dashed curve. The black broken curve represents the result without merging algorithms i.e. purely the shower prediction. The three curves are set equal at the bin between 10 and 12 GeV for comparison. \\

The obtained results are distributed smoothly around the merging scales.
A clear difference is, however, observed by varying the merging scale from $20$ GeV to $60$ GeV. 
The reason for this is that the results obtained with the merging algorithm already deviate from the parton shower prediction in the soft and/or collinear region. BY comparing the blue solid curve and the black broken curve, it is clear that the result obtained with the merging algorithm starts to deviate from the shower prediction at around the merging scale ($20$ GeV). The same can be observed for the red dashed curve. This behavior is quite natural, considering the idea of merging algorithms, namely parton showers are populated below the merging scale and tree level matrix elements are populated above the merging scale.
The observation of the clear difference does not imply a fault in our implementation of the CKKW-L+ merging algorithm, but suggests that we should choose a smaller value for the merging scale for the $t\bar{t}$ pair production when the current shower model is used.\\

\begin{figure}[t]
\centering
\includegraphics[scale=0.4]{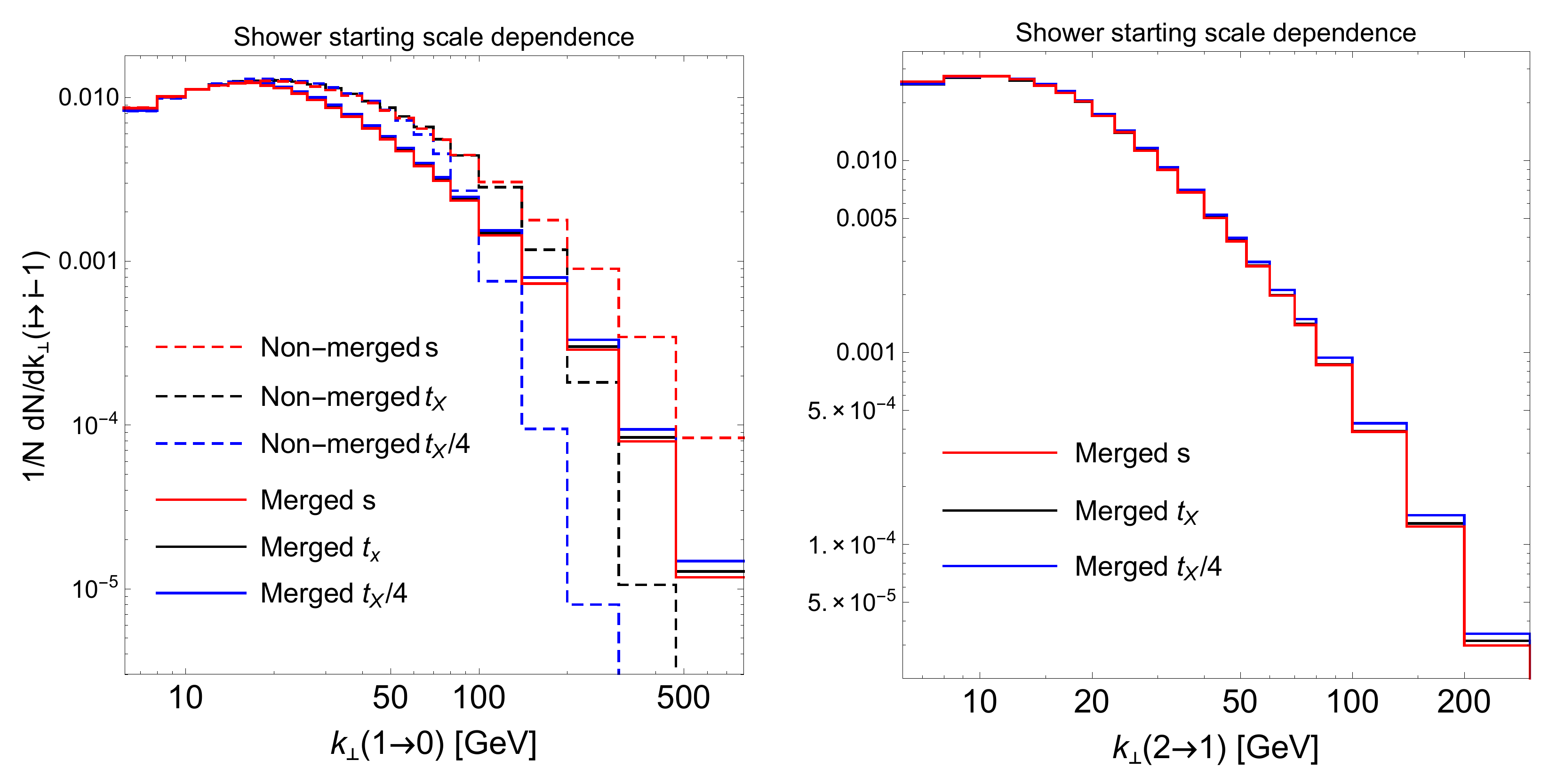}
\caption{\small The differential jet rates for $1\to0$ (left) and $2\to1$ (right) jets. The dependence on the parton shower starting scale is studied. The solid lines represent the results obtained with the merging algorithm and the dashed lines represent the results obtained without merging algorithms. 
Three different scale choices are considered for the shower starting scale, namely $s=(14)^2_{}$ $\mathrm{TeV}^2_{}$, $t_X^{}$ and $t_X^{}/4$.}
\label{fig:ttjj-starting-dependence}
\end{figure}

As the second test, the dependence on the parton shower starting scale is studied. The parton shower starting scale has been explained at the end of Section~\ref{sec:CKKW-L}. Although the most natural choice for the parton shower starting scale is the scale $t_X^{}$, we have argued that predictions such as a jet $p_T^{}$ distribution should be insensitive to the parton shower starting scale. We have also mentioned that the inclusive cross section can be sensitive to it. We confirm these statements in the following. We consider three different scale choices for the parton shower starting scale, namely $s=(14)^2_{}$ $\mathrm{TeV}^2_{}$, $t_X^{}$ and $t_X^{}/4$. We use $N=3$ and $Q_{\mathrm{cut}}^{}=20$ GeV. 
The results are shown as the differential jet rates for $1\to0$ (left) and $2\to1$ (right) jets in Figure~\ref{fig:ttjj-starting-dependence}. The solid lines represent the results obtained with the merging algorithm and the dashed lines represent the results obtained without merging algorithms. 
It is clearly shown that the dependence on the parton shower starting scale is reduced significantly by the merging algorithm. \\

\begin{table}[t]
\centering
{\footnotesize
\begin{tabular}{|c|c|c|c|c|c|c|}
\hline
 & $\sigma_{\mathrm{inc}}^{}(t\bar{t})$ (pb) & $\sigma_{\mathrm{exc}}^{}( t\bar{t}+0 )/\sigma_{\mathrm{inc}}^{}(t\bar{t})$ & $\sigma_{\mathrm{exc}}^{}( t\bar{t}+1 )/\sigma_{\mathrm{inc}}^{}(t\bar{t})$ & $\sigma_{\mathrm{exc}}^{}( t\bar{t}+2 )/\sigma_{\mathrm{inc}}^{}(t\bar{t})$ & $\sigma_{\mathrm{exc}}^{}( t\bar{t}+3 )/\sigma_{\mathrm{inc}}^{}(t\bar{t})$ \\
\hline
$s$  & 346  & 0.28 & 0.33 & 0.22  & 0.17  \\
\hline
$t_X^{}$ & 422 & 0.28 & 0.33 & 0.22 & 0.17  \\
\hline
$t_X^{}/4$ & 556 & 0.27 & 0.33 & 0.22 & 0.18  \\
\hline 
\end{tabular}
}
\caption{\small The inclusive cross section and the ratio of the exclusive cross section to the inclusive cross section, with the different choices for the shower starting scale. The inclusive cross section without merging algorithms is 562 pb.}
\label{table:inclusive-cross-section}
\end{table}

The inclusive cross section and the ratio of the exclusive cross section to the inclusive cross section are introduced in eqs.~(\ref{inc-cross-sec-1}) and (\ref{inc-cross-sec-2}). They are shown in Table~\ref{table:inclusive-cross-section} with the different scale choices for the parton shower starting scale. Note that the inclusive cross section without merging algorithms is 562 pb.
While the inclusive cross section is sensitive to the shower starting scale, the ratios are little affected. This is exactly what we have argued at the end of Section~\ref{sec:CKKW-L}. The small increase of $\sigma_{\mathrm{exc}}^{}( t\bar{t}+3 )/\sigma_{\mathrm{inc}}^{}$ for the $t_X^{}/4$ may explain the small enhancement at the high $k_{\perp}^{}$ region in Figure~\ref{fig:ttjj-starting-dependence}. \\

\begin{figure}[t]
\centering
\includegraphics[scale=0.41]{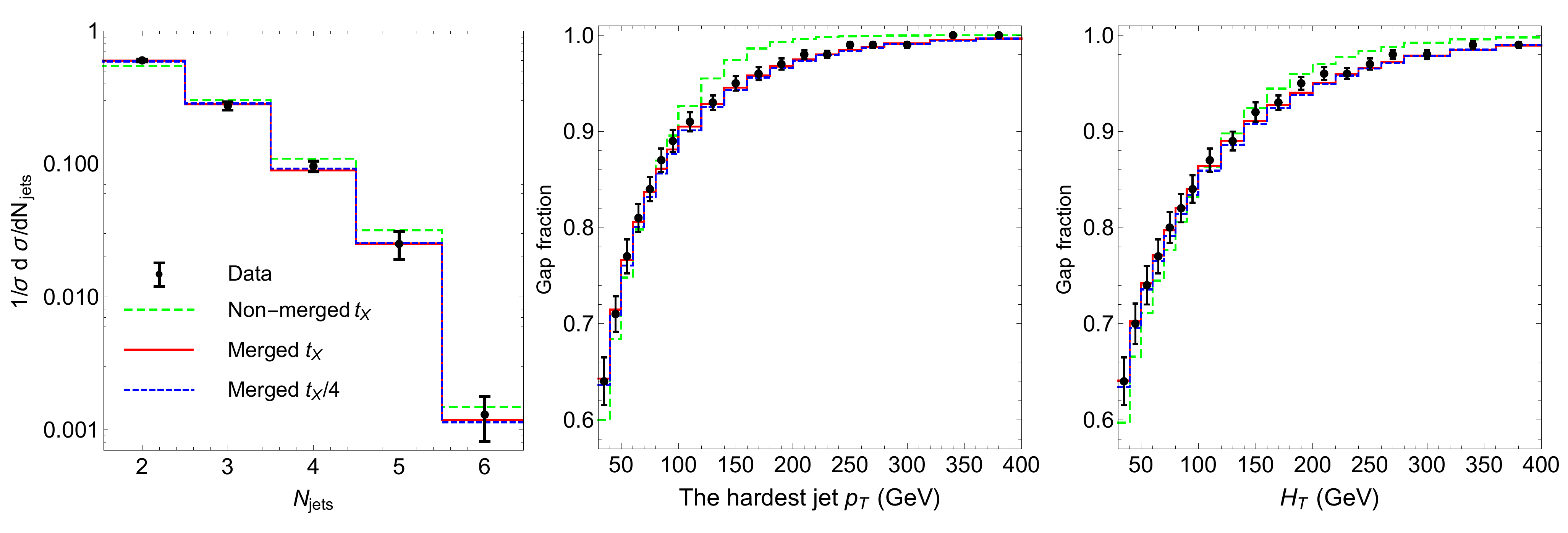}
\caption{\small The differential cross section as a function of jet multiplicity (left panel). The gap fraction as a function of the $p_T^{}$ of the highest $p_T^{}$ additional jet (middle panel) and that as a function of the scalar sum of the $p_T^{}$ of the additional jets (right panel). The green dashed curve shows the result without merging algorithms. The red solid and the blue dotted curves show the results obtained with the merging algorithm. The shower starting scale is set to the $t_X^{}$ for the red solid and to the $t_X^{}/4$ for the blue dotted.
The merging parameters are set to $N=2$ and $Q_{\mathrm{cut}}^{}=20$ GeV. 
Compared to the data from the CMS experiment~\cite{Chatrchyan:2014gma}.}
\label{figure:exp}
\end{figure}

The discussion and the results up to now have assumed that the $t\bar{t}$ is stable. Three jet observables are produced from the event samples including the leptonic decays of the top and antitop quarks and plotted in Figure~\ref{figure:exp}.
We use $N=2$ and $Q_{\mathrm{cut}}^{}=20$ GeV. The center of mass energy for the pp collisions is $7$ TeV. The differential cross section as a function of jet multiplicity is shown in the left panel. The last bin includes the contribution of $N_{\mathrm{jets}}^{} \ge 6$. 
The gap fraction is defined as
\begin{align}
f(x)=\frac{N(x)}{N_{\mathrm{total}}^{}},
\end{align}
where $N(x)$ is the number of events that give $x$ less than the given value of $x$. The gap fraction as a function of the $p_T^{}$ of the highest $p_T^{}$ additional jet is shown in the middle panel and that as a function of the scalar sum of the $p_T^{}$ of the additional jets is shown in the right panel. The additional jets are defined as all jets not including the two highest $p_T^{}$ b jets. The green dashed curve shows the result without merging algorithms. The red solid and the blue dotted curves show the results obtained with the merging algorithm. The parton shower starting scale is set to the $t_X^{}$ for the red solid curve and to the $t_X^{}/4$ for the blue dotted curve.
Our predictions are compared to the data from the CMS experiment~\cite{Chatrchyan:2014gma}.
It is shown that the merging algorithm gives the better description of the data in all the three jet observables. It can also be confirmed that the predictions are stable under the variation of the shower starting scale between the $t_X^{}$ and the $t_X^{}/4$.\\

Smaller inclusive cross sections obtained with merging algorithms imply less efficient event generation. We therefore choose the scale $t_X^{}/4$ for the parton shower starting scale for generating the event samples to be analyzed in the following sections.

%%%%%%%%%%%%%%%%%%%%%%%%%%%%%%%%%%%%%%%%%%%%%%%%%%%%%%%%%%%%%%%%%%%%%%%%%%%%%%%%%%%%%%%%%%%%%%%%%%%%%%%%%%%%%%%%%%%%%%%%%%%%%%%%%%%%%%%%%%%%%%%%%%%%%%%%%%%%%%%%%%%%%%%%%%%%%%%%%%%%%%%%%%%%%%%%%%%%%%%%%%%%%%%%%%%%%%%%%%%%%%%%%%%%%%%%%%%%%%%%%%%%%%%%%%%%%%%%%%%%%%%%%%%%%%%%%%%%%%%%%%%%%%%%%%%%%%%%%%%%%%%%%%%%%%%%%%%%%%%%%%%%%%%%%%%%%%%%%%%%%%%%%%%%%%%%%%%%%%%%%%%%%%%%%%%%%%%%%%%%%%%%%%%%%%%%%%%%%%%%%%%%%%%%%%%%%%%%%%%%%%%%%%%%%%%%%%%%%%%%%%%%%%%%%%%%%%%%%%%%%%%%%%%%%%%%%%%%%%%%%%%%%%%%%%%%%%%%%%%%%%%%%%%%%%%%%%%%%%%%%%%%%%%%%%%%%%%%%%%%%%%%%%%%%%%%%%%%%%%%%%%%%%%%%%%%%%%%%%%%%%%%%%%%%%%%%%%%%%%%%%%%%%%%%%%%%%%%%%%%%%%%%%%%%%%%%%%%%%%%%%%%
\section{Azimuthal angle correlation}\label{result1}

In this section, the azimuthal angle difference between the two hardest jets in the top quark pair production process is studied, by using the event samples generated with the merging algorithm described in Section~\ref{algorithm}.
In Section~\ref{setup}, the detailed definition of the azimuthal angle difference is specified. 
In Section~\ref{mergingscale}, we discuss the requirements on the merging parameters in order to obtain an accurate prediction of the azimuthal angle difference.
A numerical comparison of our merging algorithm with the CKKW-L algorithm is also presented. 
In Section~\ref{compare}, the distribution of the azimuthal angle difference is shown. The result obtained by merging the matrix elements for the $t\bar{t}$ plus up to 2 partons is compared to the one obtained by merging the matrix elements for the $t\bar{t}$ plus up to 3 partons. The roles of the $t\bar{t}+3$ partons matrix elements are studied in detail. 
It is assumed that the top and antitop quarks are stable in Section~\ref{compare}.
In Section~\ref{sec:results-with-topdecay}, the distribution of the azimuthal angle difference is produced from the event samples including the dilepton decay of the $t\bar{t}$ and the effect of the decay is studied.

%%%%%%%%%%%%%%%%%%%%%%%%%%%%%%%%%%%%%%%%%%%%%%%%%%%%%%%%%%%%%%%%%%%%%%%%%%%%%%%%%%%%%%%%%%%%%%%%%%%%%%%%%%%%%%%%%%%%%%%%%%%%%%%%%%%%%%%%%%%%%%%%%%%%%%%%%%%%%%%%%%%%%%%%%%%%%%%%%%%%%%%%%%%%%%%%%%%%%%%%%%%%%%%%%%%%%%%%%%%%%%%%%%%%%%%%%%%%%%%%%%%%%%%%%%%%%%%%%%%%%%%%%%%%%%%%%%%%%%%%%%%%%%%%%%%%%%%%%%%%%%%%%%%%%%%%%%%%%%%%%%%%%%%%%%%%%%%%%%%%%%%%%%%%%%%%%%%%%%%%%%%%%%%%%%%%%%%%%%%%%%%%%%%%%%%%%%%%%%%%%%%%%%%%%%%%%%%%%%%%%%%%%%%%%%%%%%%%%%%%%%%%%%%%%%%%%%%%%%%%%%%%%%%%%%%%%%%%%%%%%%%%%%%%%%%%%%%%%%%%%%%%%%%%%%%%%%%%%%%%%%%%%%%%%%%%%%%%%%%%%%%%%%%%%%%%%%%%%%%%%%%%%%%%%%%%%%%%%%%%%%%%%%%%%%%%%%%%%%%%%%%%%%%%%%%%%%%%%%%%%%%%%%%%%%%%%%%%%%%%%%%%
\subsection{Definition}\label{setup}

An event sample with two or more jets is picked up and the following requirements which are often called vector boson fusion (VBF) cuts are applied to the two hardest jets,
\begin{align}
y_1^{} \times y_2^{} < 0,\ \ \ \ |y_1^{}-y_2^{}|>4. \label{vbf1}
\end{align}
The transverse momentum $p_T^{}$ with respect to the beam of an object describes the hardness of the object. Therefore a jet which has the highest $p_T^{}$ is called the hardest jet and another jet which has the second highest $p_T^{}$ is called the second hardest jet, and these jets are assigned to the two hardest jets. One of the two jets which has a positive rapidity is chosen for an azimuthal angle $\phi_1^{}$ and the other jet which has a negative rapidity is chosen for an azimuthal angle $\phi_2^{}$. The azimuthal angle difference between the two hardest jets is defined by
\begin{align}
\Delta \phi=\phi_1^{}-\phi_2^{}.\label{delta-phi-def}
\end{align} 
Therefore a jet for the $\phi_1^{}$ is not necessarily the hardest jet in our definition. 
To enhance the correlation in $\Delta \phi$, a cut is applied on the invariant mass of the top quark pair~\cite{Hagiwara:2013jp}, 
\begin{align}
m_{t\bar{t}}^{}<500\ \mathrm{GeV}. \label{vbf2}
\end{align}
No other cuts are applied to the top and anti-top quarks. \\

All the event samples analyzed in the following sections satisfy the above cuts in eqs.~(\ref{vbf1}) and (\ref{vbf2}), even when it is not stated explicitly. This is also the case, when we analyze exclusive parton level event samples in Section~\ref{compare}. There, the cuts in eq.~(\ref{vbf1}) are applied on partons, not jets.

%%%%%%%%%%%%%%%%%%%%%%%%%%%%%%%%%%%%%%%%%%%%%%%%%%%%%%%%%%%%%%%%%%%%%%%%%%%%%%%%%%%%%%%%%%%%%%%%%%%%%%%%%%%%%%%%%%%%%%%%%%%%%%%%%%%%%%%%%%%%%%%%%%%%%%%%%%%%%%%%%%%%%%%%%%%%%%%%%%%%%%%%%%%%%%%%%%%%%%%%%%%%%%%%%%%%%%%%%%%%%%%%%%%%%%%%%%%%%%%%%%%%%%%%%%%%%%%%%%%%%%%%%%%%%%%%%%%%%%%%%%%%%%%%%%%%%%%%%%%%%%%%%%%%%%%%%%%%%%%%%%%%%%%%%%%%%%%%%%%%%%%%%%%%%%%%%%%%%%%%%%%%%%%%%%%%%%%%%%%%%%%%%%%%%%%%%%%%%%%%%%%%%%%%%%%%%%%%%%%%%%%%%%%%%%%%%%%%%%%%%%%%%%%%%%%%%%%%%%%%%%%%%%%%%%%%%%%%%%%%%%%%%%%%%%%%%%%%%%%%%%%%%%%%%%%%%%%%%%%%%%%%%%%%%%%%%%%%%%%%%%%%%%%%%%%%%%%%%%%%%%%%%%%%%%%%%%%%%%%%%%%%%%%%%%%%%%%%%%%%%%%%%%%%%%%%%%%%%%%%%%%%%%%%%%%%%%%%%%%%%%%%%%%%%%%%%%%%%%%%%%%%%%%%%%%%%%%%%%%%%%%%%%%%%%%%%%%%%%%%%%%%%%%%%%%%%%%%%%%%%%%%%%%%%%%%%%%%
\subsection{Contamination}\label{mergingscale}

In this section, we discuss the requirements on the merging parameters in order to obtain an accurate prediction on the azimuthal angle difference. 
We explore a relation between the merging scale $Q_{\mathrm{cut}}^{}$ and a lower cut on the $p_{T}^{}$ of jets by which the contamination is negligible. 
A numerical comparison of our merging algorithm with the CKKW-L algorithm is also presented.\\

In order for each of the two hardest jets to have the correct azimuthal angle information, each of them must have its origin in a parton generated with matrix elements.  
If one or both of them originate from partons generated with parton showers, angular correlations between them are not correct. One of the requirements on the merging parameters is therefore 
\begin{align}
N\geq 2.
\end{align}
\\

\begin{table}[t]
\centering
\begin{tabular}{|c|c|c|c|}
\hline
$p_{T \mathrm{cut}}^{}$ (GeV) & $20$ & $25$ & $30$ \\
\hline
CKKW-L & 12.13  & 7.00 & 4.50 \\
\hline
CKKW-L+ & 11.65 & 6.70 & 4.29 \\
\hline
\end{tabular}
\caption{\small The contamination (\%) for the two different merging algorithms and for the three different $p_{T \mathrm{cut}}^{}$ values, defined in eq.~(\ref{contamination-rate}). The rapidity cut on jets is $|y|<4.5$.
The merging parameters are set to $N=3$ and $Q_{\mathrm{cut}}^{}=20$ GeV. 
}
\label{tab:contamination}
\end{table}

Another requirement is that the merging scale $Q_{\mathrm{cut}}^{}$ is chosen smaller than the scale of a jet definition. The anti-$k_T^{}$ algorithm~\cite{Cacciari:2008gp} includes two parameters, namely the radius parameter $R$ and a lower cutoff on the transverse momentum of jets $p_{T \mathrm{cut}}^{}$. The radius parameter will not be restricted so much, since the two hardest jets which are well separated with each other are of interest to us, see eq.~(\ref{vbf1}). We choose $R=1$ in the merging scale definition in eq.~(\ref{ktalgorithm}), while $R=0.4$ is used in the jet definition. The $p_{T \mathrm{cut}}^{}$ has to satisfy~\cite{Buckley:2014fqa}
\begin{align}
Q_{\mathrm{cut}}^{} \leq p_{T \mathrm{cut}}^{}.
\end{align}
As we have already mentioned several times, the contamination is the contribution from the $t\bar{t}+0,1$ parton matrix elements to the event samples with two or more jets. The contamination can be written by using the inclusive cross section and the exclusive cross sections as
\begin{align}
\frac{ \sigma_{\mathrm{exc}}^{}( t\bar{t}+0 ) + \sigma_{\mathrm{exc}}^{}( t\bar{t}+1 ) }{ \sigma_{\mathrm{inc}}^{}( t\bar{t} )  }.\label{contamination-rate}
\end{align}
This notation is introduced in eq.~(\ref{inc-cross-sec-1}). 
The $\Delta \phi$ prediction will not be reliable, unless the contamination is small. The event samples are generated with the two different merging algorithms, namely the CKKW-L and the CKKW-L+. Then, the contamination is calculated for the three different $p_{T \mathrm{cut}}^{}$ values, namely $20$, $25$ and $30$ GeV. The rapidity cut on jets is set to $|y|<4.5$. 
We use $N=3$ and $Q_{\mathrm{cut}}^{}=20$ GeV, thus the above two requirements are satisfied. The result is shown in units of percentage in Table~\ref{tab:contamination}. The contamination decreases with a rise in the $p_{T \mathrm{cut}}^{}$, as expected. However, the contamination is not so suppressed in the CKKW-L+ algorithm, compared to the CKKW-L algorithm. \\

Our introduction of the CKKW-L+ algorithm has been motivated by our numerical finding that the contamination is not negligible in the CKKW-L algorithm when the $Q_{\mathrm{cut}}^{}$ is set equal to or slightly smaller than the $p_{T \mathrm{cut}}^{}$. However, it is found that a large suppression of the contamination cannot be achieved in the CKKW-L+ algorithm. \\

In the following sections, we carry on the further analyses on the event samples generated with the CKKW-L+ algorithm.
We choose $Q_{\perp \mathrm{cut}}^{}=20$ GeV and $p_{T \mathrm{cut}}^{}=30$ GeV, in order to suppress the contamination reasonably while avoiding the event generation which is too inefficient. Note that about $4\%$ of the correlation in $\Delta \phi$ is already lost due to the contamination in the results presented in the following sections.

%%%%%%%%%%%%%%%%%%%%%%%%%%%%%%%%%%%%%%%%%%%%%%%%%%%%%%%%%%%%%%%%%%%%%%%%%%%%%%%%%%%%%%%%%%%%%%%%%%%%%%%%%%%%%%%%%%%%%%%%%%%%%%%%%%%%%%%%%%%%%%%%%%%%%%%%%%%%%%%%%%%%%%%%%%%%%%%%%%%%%%%%%%%%%%%%%%%%%%%%%%%%%%%%%%%%%%%%%%%%%%%%%%%%%%%%%%%%%%%%%%%%%%%%%%%%%%%%%%%%%%%%%%%%%%%%%%%%%%%%%%%%%%%%%%%%%%%%%%%%%%%%%%%%%%%%%%%%%%%%%%%%%%%%%%%%%%%%%%%%%%%%%%%%%%%%%%%%%%%%%%%%%%%%%%%%%%%%%%%%%%%%%%%%%%%%%%%%%%%%%%%%%%%%%%%%%%%%%%%%%%%%%%%%%%%%%%%%%%%%%%%%%%%%%%%%%%%%%%%%%%%%%%%%%%%%%%%%%%%%%%%%%%%%%%%%%%%%%%%%%%%%%%%%%%%%%%%%%%%%%%%%%%%%%%%%%%%%%%%%%%%%%%%%%%%%
\subsection{The effects of the $t\bar{t}+$3 partons matrix elements}\label{compare}

The requirement on the maximal number of partons $N$ provided by the tree level matrix elements (MEs) i.e. $N\geq 2$ is briefly discussed in Section~\ref{mergingscale}. In general the more accurate description of multi-jet processes is expected by larger values for $N$. 
Thus, the prediction on the $\Delta \phi$ distribution is also expected to be more accurate for $N=3$ than for $N=2$. 
In this section, we produce the $\Delta \phi$ distribution by using the event samples generated with the merging algorithm for $N=2$ and by using those for $N=3$. Then, the two results are compared. The effects of the $t\bar{t}+3$ partons matrix elements on $\Delta \phi$ are studied in detail.\\

\begin{figure}[t]
\centering
\includegraphics[scale=0.55]{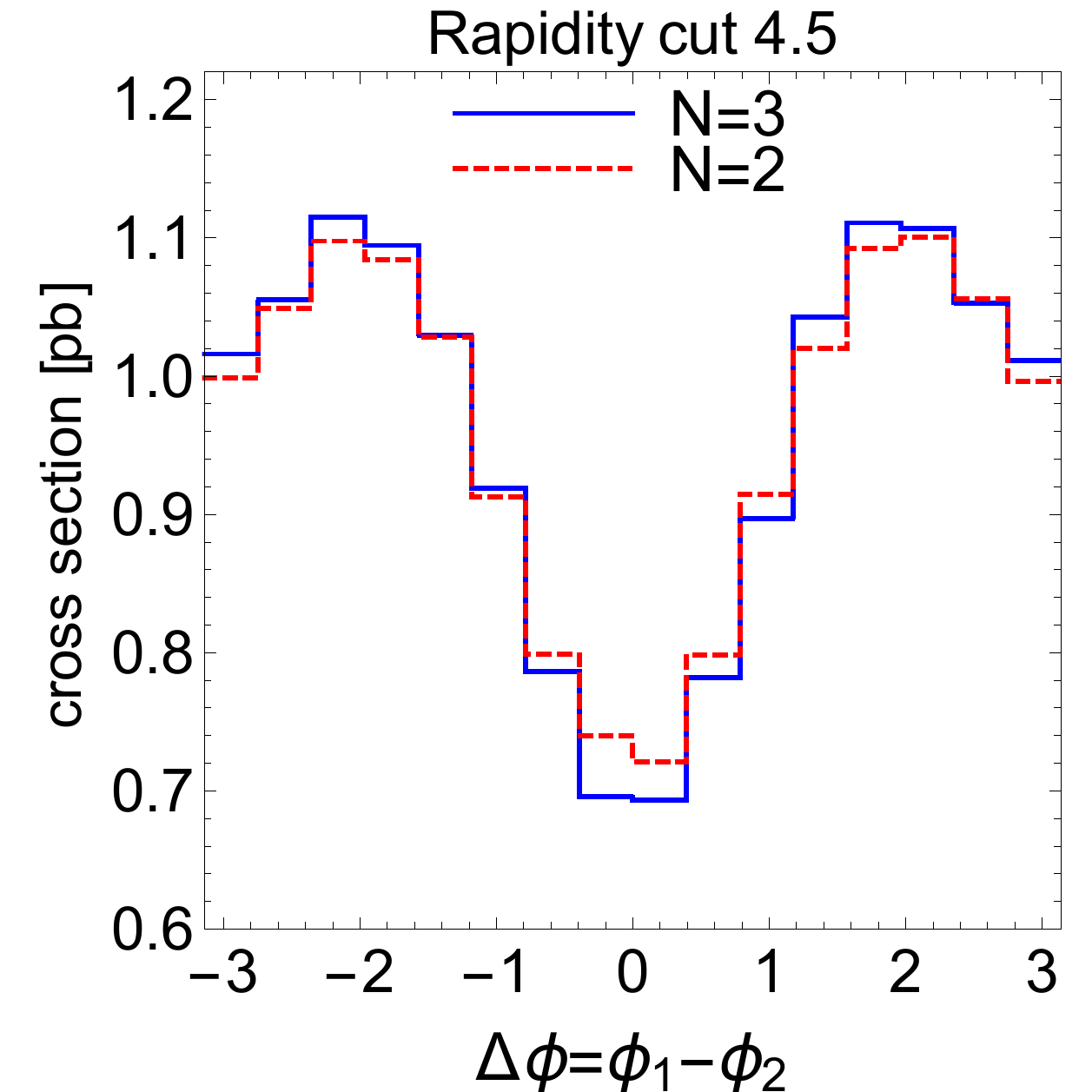}
\includegraphics[scale=0.55]{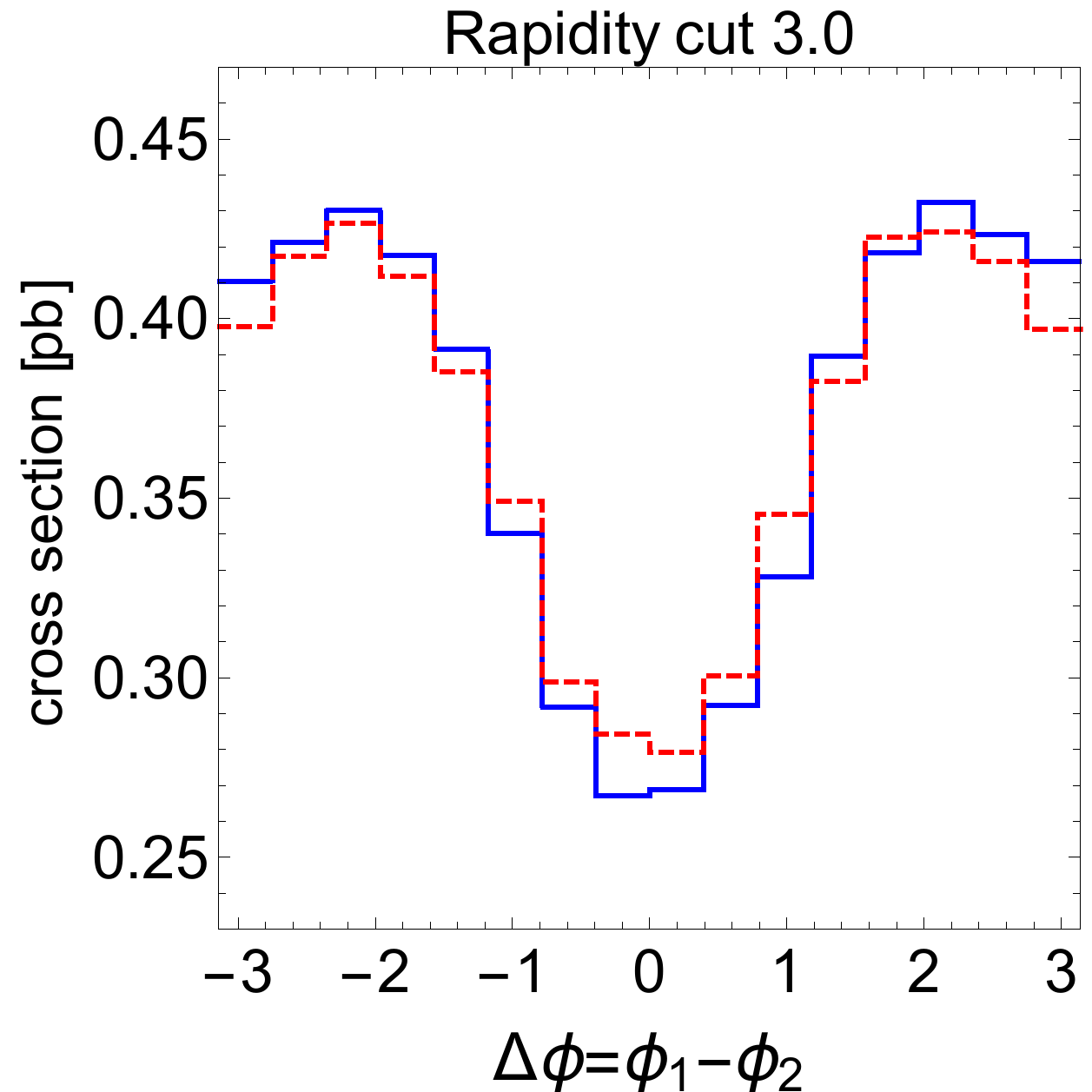}
\caption{\small The $\Delta \phi$ distribution produced from the event samples generated with the merging algorithm for $N=2$ (red dotted curve) and from those for $N=3$ (blue solid curve). The rapidity cut on jets is $|y|<4.5$ in the left panel and $|y|<3.0$ in the right panel.}
\label{fig:delta-phi-result}
\end{figure}

We show the $\Delta \phi$ distribution produced from the event samples generated with the merging algorithm for $N=2$ (red dotted curve) and from those for $N=3$ (blue solid curve) in Figure~\ref{fig:delta-phi-result}. The rapidity cut on jets is $|y|<4.5$ in the left panel and $|y|<3.0$ in the right panel. The total inclusive cross section for $m_t^{}=172.5$ GeV is estimated to be $960$pb from ref.~\cite{Czakon:2013goa}. From this value, the cross section at each bin is calculated. The results show strong correlations in $\Delta \phi$, as predicted in the analysis based on the $t\bar{t}+2$ partons tree level matrix elements~\cite{Hagiwara:2013jp}. 
This observation indicates that the correlation found in the $t\bar{t}+2$ partons tree level matrix elements~\cite{Hagiwara:2013jp} is still present after including the dominant higher order corrections and thus can be observed in the experiments. 
We can find a clear difference in the $\Delta \phi$ distribution between the result for $N=2$ and the one for $N=3$. The difference looks slightly larger in the right panel, where the rapidity range is more restricted. The origins of the difference are studied in the following.\\

\begin{figure}[t]
\centering
\includegraphics[scale=0.45]{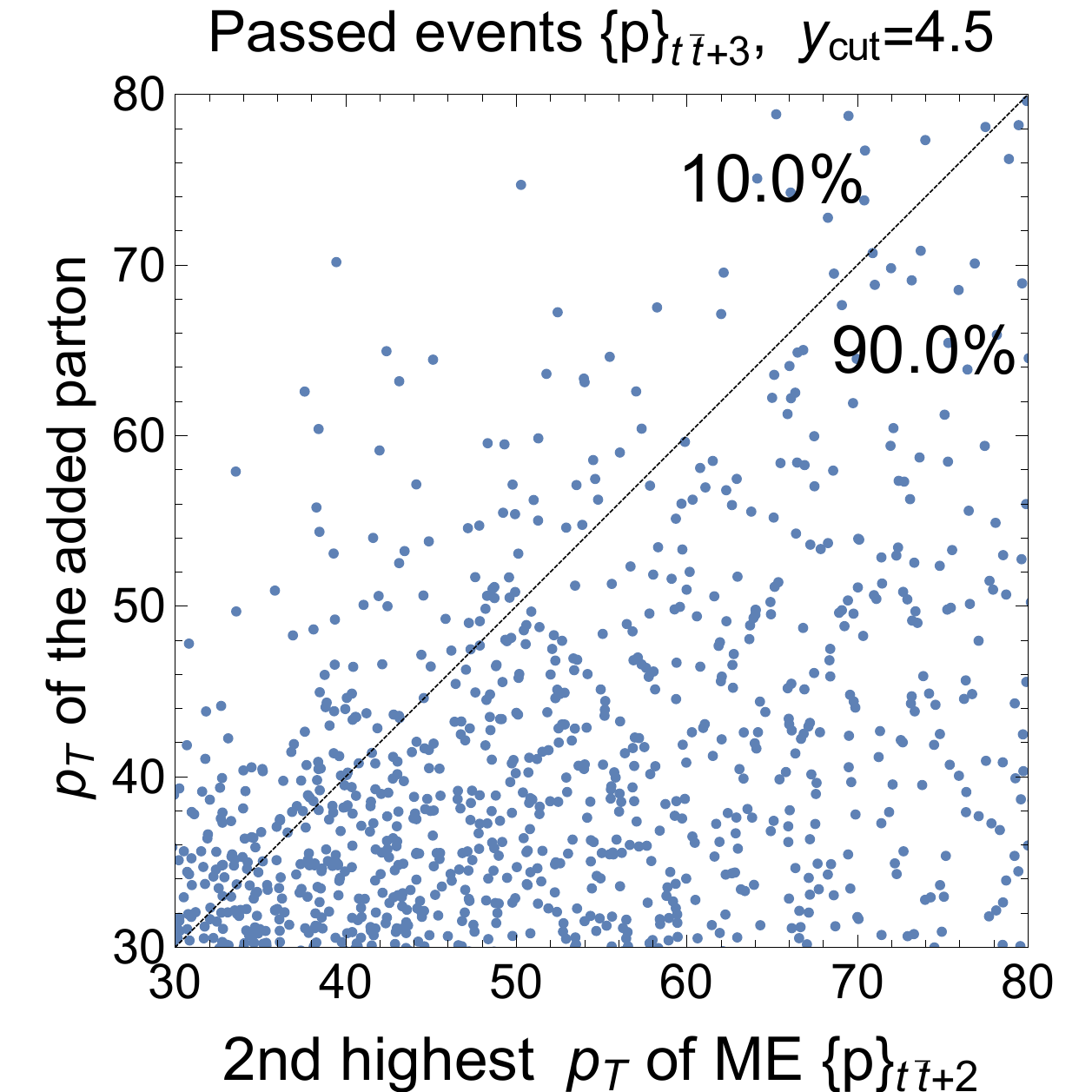}
\hspace{0.7cm}
\includegraphics[scale=0.45]{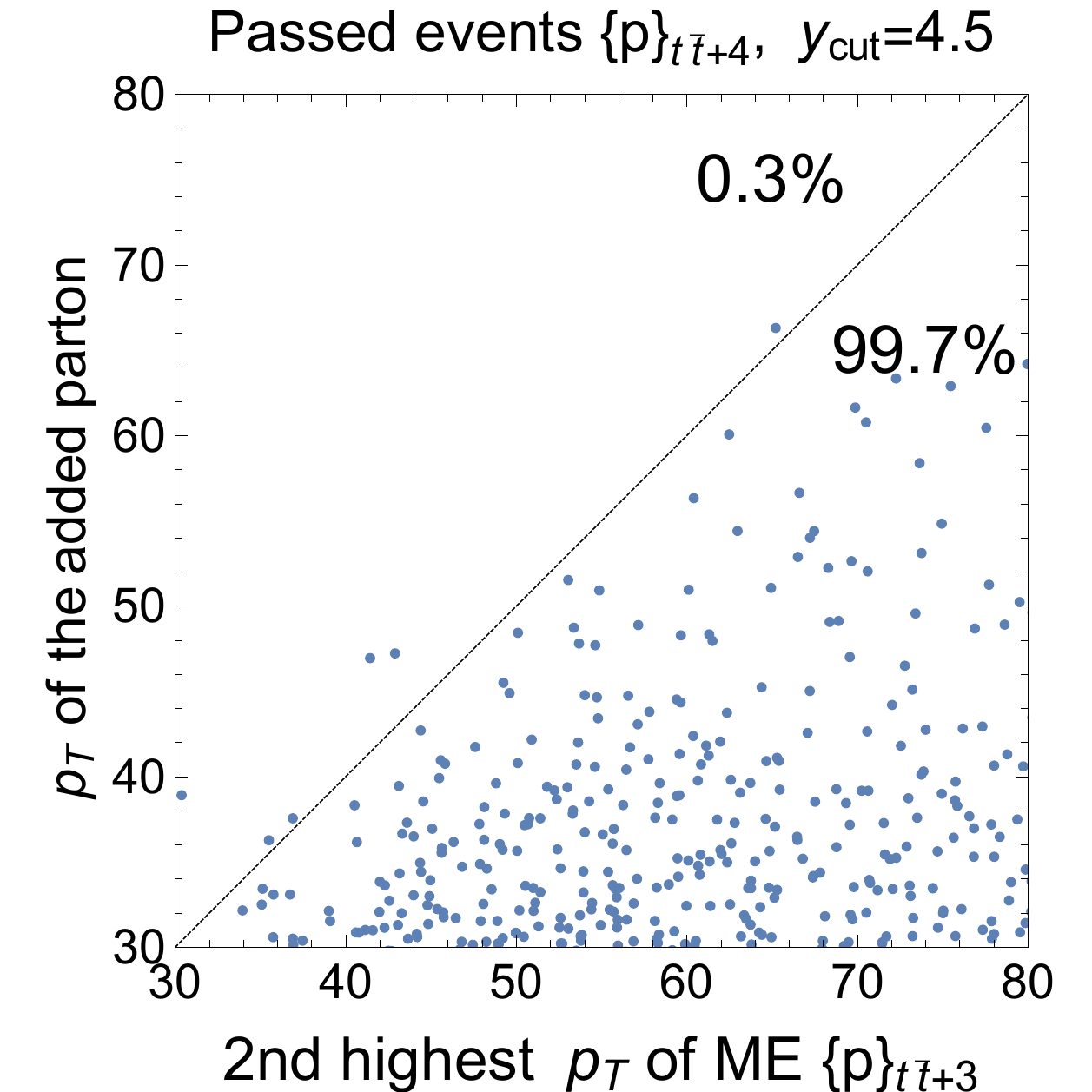}
\caption{\small {\it left}: Event samples $\{p\}_{t\bar{t}+3}^{}$ obtained by one initial state shower evolution of the ME samples $\{p\}_{t\bar{t}+2}^{}$ are plotted. The vertical axis represents $p_T^{}$ of the added parton. The horizontal axis represents the $p_T^{}$ of the second hardest parton of the $\{p\}_{t\bar{t}+2}^{}$ after the evolution. 
{\it right}: Event samples $\{p\}_{t\bar{t}+4}^{}$ obtained by one initial state shower evolution of the ME samples $\{p\}_{t\bar{t}+3}^{}$ are plotted. The vertical axis represents $p_T^{}$ of the added parton. The horizontal axis represents the $p_T^{}$ of the second hardest parton of the $\{p\}_{t\bar{t}+3}^{}$ after the evolution.}
\label{fig:scatter-plot-pT}
\end{figure}

First of all, let us remind us of one point in the merging algorithm. 
When the parton shower (PS) generator is executed on a matrix element (ME) event sample for the highest parton multiplicity $N$ i.e. $\{p\}_{t\bar{t}+N}^{}$, the shower is constrained to be softer than the $N$ partons of the $\{p\}_{t\bar{t}+N}^{}$ in terms of the shower evolution variable. More precisely, following a parton shower history of the $\{p\}_{t\bar{t}+N}^{}$ consisting of $\{p\}_{t\bar{t}+(N-1)}^{}, \{p\}_{t\bar{t}+(N-2)}^{}, \cdots, \{p\}_{t\bar{t}+i}^{}, \cdots, \{p\}_{t\bar{t}+1}^{}, \{p\}_{t\bar{t}}^{}$ with the corresponding scales $t_N^{} < t_{N-1}^{} < \cdots < t_{i+1}^{} < \cdots < t_2^{}< t_1^{}$, the evolution scale of the shower is restricted to be below the $t_N^{}$. \\

Now let us consider the merging algorithm for $N=2$ and suppose that the shower evolution of a ME event sample $\{p\}_{t\bar{t}+2}^{}$ is performed and thus an event sample $\{p\}_{t\bar{t}+3}^{}$ is generated. When the shower evolution generates an initial state radiation, one parton will be added in the $\{p\}_{t\bar{t}+2}^{}$ and accordingly the kinematics of the two partons of the $\{p\}_{t\bar{t}+2}^{}$ will be changed. 
In the left panel of Figure~\ref{fig:scatter-plot-pT}, the $p_T^{}$ of the added parton by the initial state radiation is assigned on the vertical axis and
the $p_T^{}$ of the second hardest parton of the $\{p\}_{t\bar{t}+2}^{}$ after the initial state radiation is assigned on the horizontal axis. 
It must be noted that the kinematics of the $\{p\}_{t\bar{t}+2}^{}$ after the initial state radiation is not identical to that of the original ME event sample $\{p\}_{t\bar{t}+2}^{}$ anymore, since it has been changed by the radiation.
The rapidity range for the two partons which construct $\Delta \phi$ is $|y|<4.5$. It is not required, however, that all of the three partons in the $\{p\}_{t\bar{t}+3}^{}$ are within the range $|y|<4.5$. 
From the left panel, we can find that the added parton does not have a lower $p_T^{}$ than the two partons of the $\{p\}_{t\bar{t}+2}^{}$ in the considerable fraction ($10.0\%$), which is shown by the dots in the upper left of the panel. This observation implies a non-negligible loss of the correlation between the two hardest jets, because the added parton which has the highest or second highest $p_T^{}$ may give rise to one of the two hardest jets at the end of the shower evolution.\\

The above problem can be solved by merging the matrix elements for the $t\bar{t}+$3 partons. Let us consider the merging algorithm for $N=3$ and suppose that the shower evolution of a ME event sample $\{p\}_{t\bar{t}+3}^{}$ is performed and thus an event sample $\{p\}_{t\bar{t}+4}^{}$ is generated. When the shower evolution generates an initial state radiation, one parton will be added in the $\{p\}_{t\bar{t}+3}^{}$ and accordingly the kinematics of the three partons of the original $\{p\}_{t\bar{t}+3}^{}$ will be changed. In the right panel of Figure~\ref{fig:scatter-plot-pT}, the $p_T^{}$ of the added parton by the initial state radiation is assigned on the vertical axis and the $p_T^{}$ of the second hardest parton of the $\{p\}_{t\bar{t}+3}^{}$ after the initial state radiation is assigned on the horizontal axis. The rapidity range is the same as above. The panel shows that the probability that the added parton has the highest or second highest $p_T^{}$ is quite suppressed (0.3\%). \\

Our discussion so far has been based on the exclusive parton level event samples, not on the inclusive jet level event samples. However, we believe that our numerical findings reasonably explain the difference observed in Figure~\ref{fig:delta-phi-result}. In the event samples generated with the merging algorithm for $N=2$, a non-negligible loss ($5\sim10 \%$) of the correlation between the two hardest jets can be unavoidable, because a jet originating from hard parton showers can have a higher $p_T^{}$ than one of the two jets originating from the two hard partons of the $t\bar{t}+$2 partons matrix elements. In the event samples generated with the merging algorithm for $N=3$, the loss of the correlation due to the above reason can be avoided as follows. Jets originating from hard parton showers cannot have higher $p_T^{}$ than two of the three jets originating from the three hard partons of the $t\bar{t}+$3 partons matrix elements.\\

\begin{figure}[t]
\centering
\includegraphics[scale=0.45]{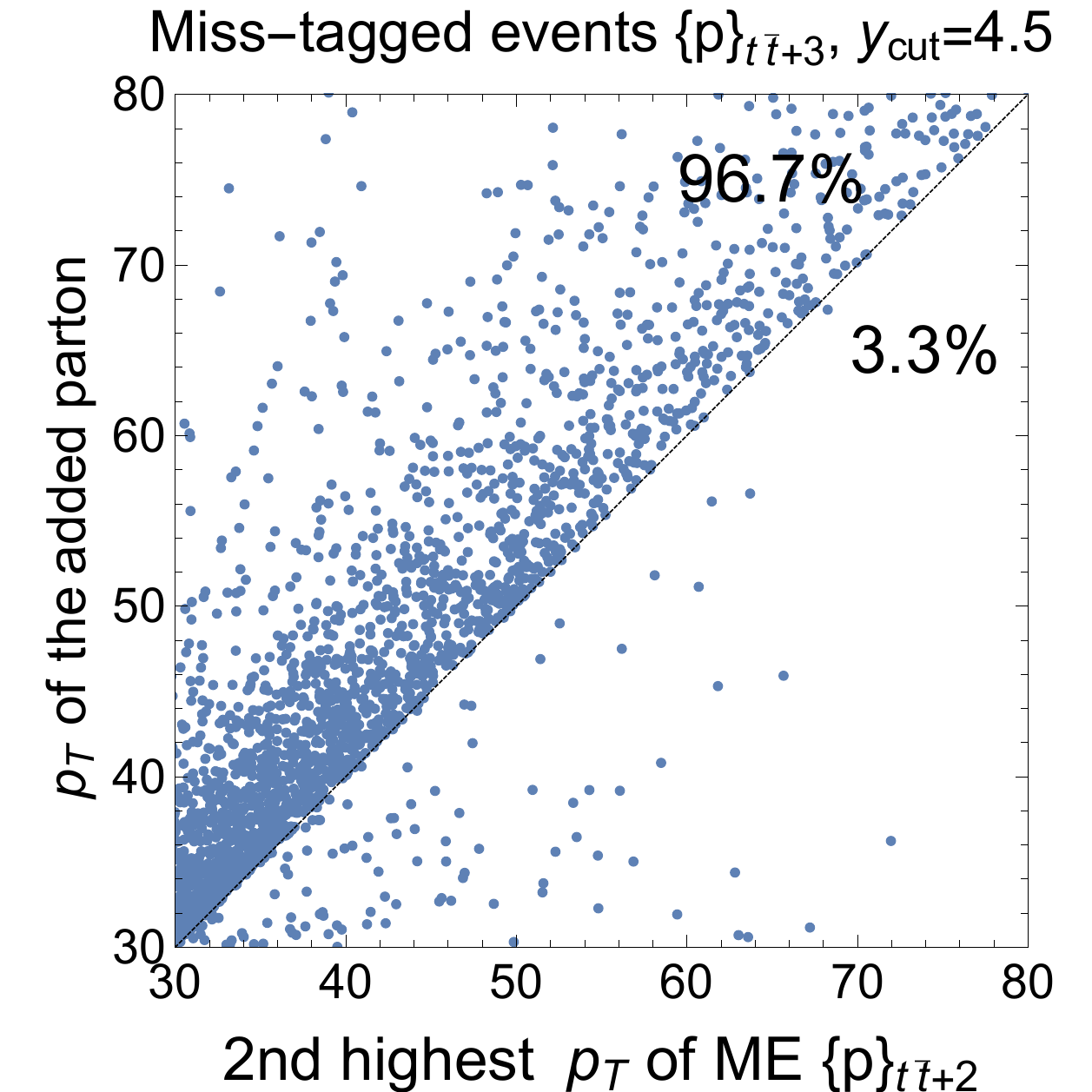}
\hspace{0.7cm}
\includegraphics[scale=0.45]{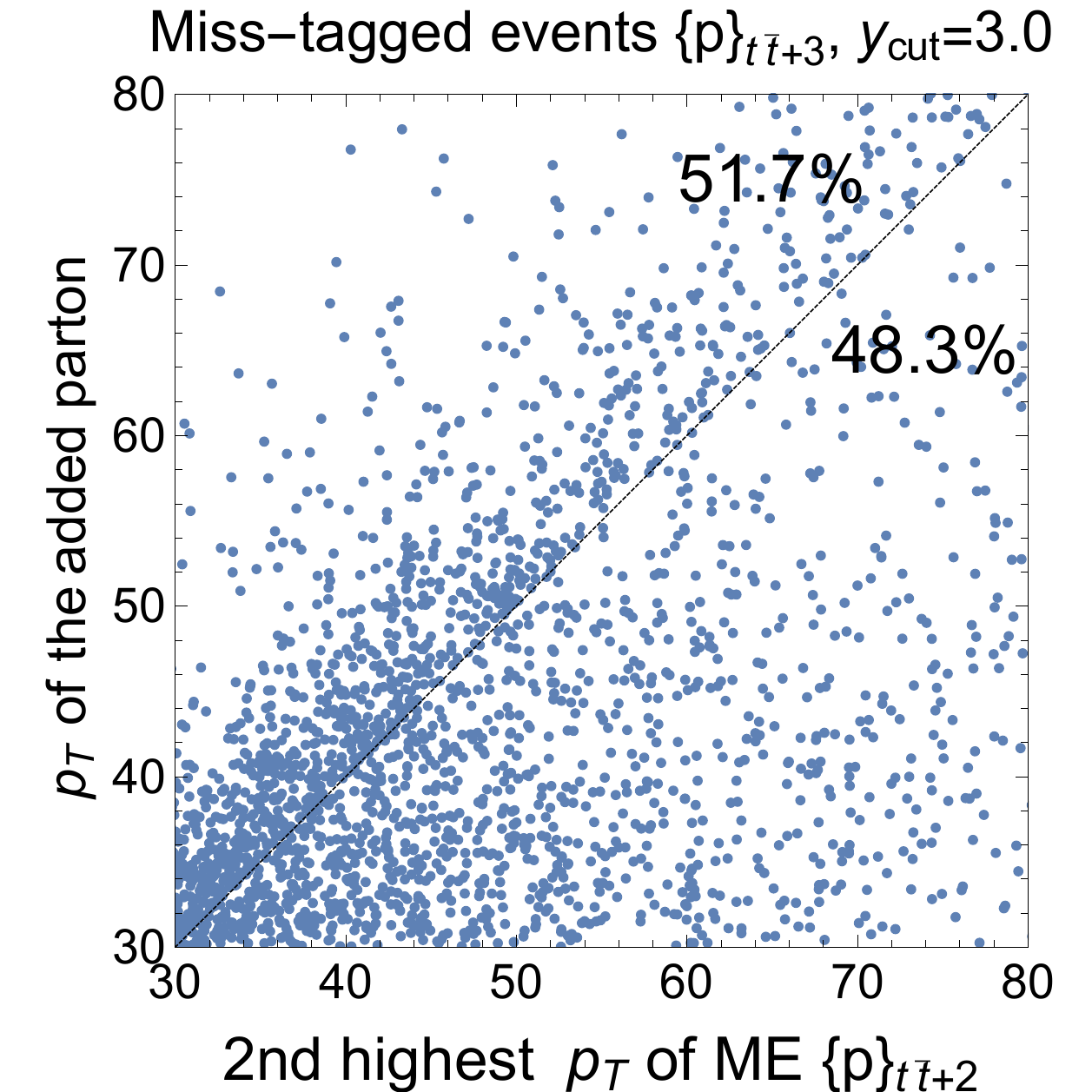}
\caption{\small As in the merging algorithm for $N=2$, event samples $\{p\}_{t\bar{t}+3}^{}$ are generated at first by the initial state shower evolution of the ME event samples $\{p\}_{t\bar{t}+2}^{}$. And then only the miss-tagged event samples are plotted. 
The axes are identical to those in the left panel of Figure~\ref{fig:scatter-plot-pT}. The rapidity range for the two partons which construct $\Delta \phi$ is $|y|<4.5$ in the left panel and $|y|<3.0$ in the right panel.}
\label{fig:scatter-plot-pT-misstag}
\end{figure}

When the rapidity cut on jets is more stringent, a further loss of the correlation may arise, because jets originating from the hard partons of matrix elements may be removed by the rapidity cut. 
In order to examine the possibility of this, let us define a miss-tagged event sample as an event sample in which a parton added by parton showers or a jet originating from parton showers is picked up for constructing the azimuthal angle difference $\Delta \phi$. In miss-tagged event samples, the $\Delta \phi$ distribution will not be produced correctly, of course. \\

As in the merging algorithm for $N=2$, event samples $\{p\}_{t\bar{t}+3}^{}$ are generated by the initial state shower evolution of the ME event samples $\{p\}_{t\bar{t}+2}^{}$. Then, only the miss-tagged event samples are plotted in the left panel of Figure~\ref{fig:scatter-plot-pT-misstag}. The rapidity range for partons is $|y|<4.5$. It must be noted that the rapidity range $|y|<4.5$ is for the two partons which construct $\Delta \phi$ and thus it is not required that all of the three partons in the $\{p\}_{t\bar{t}+3}^{}$ are within the range $|y|<4.5$.
The vertical and horizontal axes are identical to those in the left panel of Figure~\ref{fig:scatter-plot-pT}.
The miss-tagged event samples in the upper left of the panel correspond to the samples in which the added parton has a higher $p_T^{}$ than one of the two partons of the $\{p\}_{t\bar{t}+2}^{}$ and thus the added parton is picked up for $\Delta \phi$. 
The miss-tagged event samples in the lower right of the panel correspond to the samples in which one of the two partons of the $\{p\}_{t\bar{t}+2}^{}$ is removed by the rapidity cut and thus the added parton is picked up for $\Delta \phi$, despite that the added parton has a lower $p_T^{}$. The panel shows that the former possibility is dominant (96.7\%). The miss-tagged fraction in the samples $\{p\}_{t\bar{t}+3}^{}$ is $10.3\%$. \\

The miss-tagged event samples $\{p\}_{t\bar{t}+3}^{}$ with a more restricted rapidity range $|y|<3.0$ are plotted in the right panel of Figure~\ref{fig:scatter-plot-pT-misstag}. 
The panel shows that one of the two partons of the $\{p\}_{t\bar{t}+2}^{}$ is removed by the rapidity cut in the larger fraction (48.3\%) of the samples, as expected. 
As a result, the miss-tagged fraction in the samples $\{p\}_{t\bar{t}+3}^{}$ is increased to $16.6\%$. \\

In event samples $\{p\}_{t\bar{t}+4}^{}$ generated by the initial state shower evolution of the ME event samples $\{p\}_{t\bar{t}+3}^{}$ as in the merging algorithm for $N=3$, the miss-tagged fraction is $0.4\%$ for a rapidity cut $|y|<4.5$ and $2.0\%$ for a rapidity cut $|y|<3.0$. 
The miss-tagged fraction in the $\{p\}_{t\bar{t}+3}^{}$ is more increased, compared to the $\{p\}_{t\bar{t}+4}^{}$, as the rapidity cut is set more stringent. \\

Although the above discussion is again based on the exclusive parton level event samples, we believe that our findings reasonably explain the slightly larger difference observed in the right panel of Figure~\ref{fig:delta-phi-result}.
In the event samples generated with the merging algorithm for $N=2$, a further loss of the correlation between the hardest two jets arises when the rapidity cut on jets is set stringent, because one of the two jets originating from the two hard partons of the $t\bar{t}+$2 partons matrix elements can be removed by the rapidity cut. In the event samples generated with the merging algorithm for $N=3$, the loss of the correlation due to the above reason can be reduced as follows. Even though one of the three jets originating from the three hard partons of the $t\bar{t}+$3 partons matrix elements is removed by the rapidity cut, there are still two jets which correctly predict the correlation. 
When the rapidity range for jets is $|y|<4.5$, the loss of the correlation due to the above reason can be negligible. However, when the rapidity range is more restricted such as $|y|<3.0$, the loss of the correlation cannot be negligible anymore in the event samples generated with the merging algorithm for $N=2$, while it can be much reduced in the event samples generated with the merging algorithm for $N=3$. This observation can explain the larger difference in the right panel of Figure~\ref{fig:delta-phi-result}. \\

To summarize, a non-negligible loss of the correlation between the two hardest jets will be unavoidable in the event samples generated with the merging algorithm for $N=2$. The loss of the correlation can be reduced significantly in the event samples generated with the merging algorithm for $N=3$.
The role of the $t\bar{t}+$3 partons matrix elements can be more important as the rapidity range for jets is set more restricted.

%%%%%%%%%%%%%%%%%%%%%%%%%%%%%%%%%%%%%%%%%%%%%%%%%%%%%%%%%%%%%%%%%%%%%%%%%%%%%%%%%%%%%%%%%%%%%%%%%%%%%%%%%%%%%%%%%%%%%%%%%%%%%%%%%%%%%%%%%%%%%%%%%%%%%%%%%%%%%%%%%%%%%%%%%%%%%%%%%%%%%%%%%%%%%%%%%%%%%%%%%%%%%%%%%%%%%%%%%%%%%%%%%%%%%%%%%%%%%%%%%%%%%%%%%%%%%%%%%%%%%%%%%%%%%%%%%%%%%%%%%%%%%%%%%%%%%%%%%%%%%%%%%%%%%%%%%%%%%%%%%%%%%%%%%%%%%%%%%%%%%%%%%%%%%%%%%%%%%%%%%%%%%%%%%%%%%%%%%%%%%%%%%%%%%%%%%%%%%%%%%%%%%%%%%%%%%%%%%%%%%%%%%%%%%%%%%%%%%%%%%%%%%%%%%%%%%%%%%%%%%%%%%%%%%%%%%%%%%%%%%%%%%%%%%%%%%%%%%%%%%%%%%%%%%%%%%%%%%%%%%%%%%%%%%%%%%%%%%%%%%%%%%%%%%%%%
\subsection{The effect of a $t\bar{t}$ decay}\label{sec:results-with-topdecay}

\begin{figure}[t]
\centering
\includegraphics[scale=0.55]{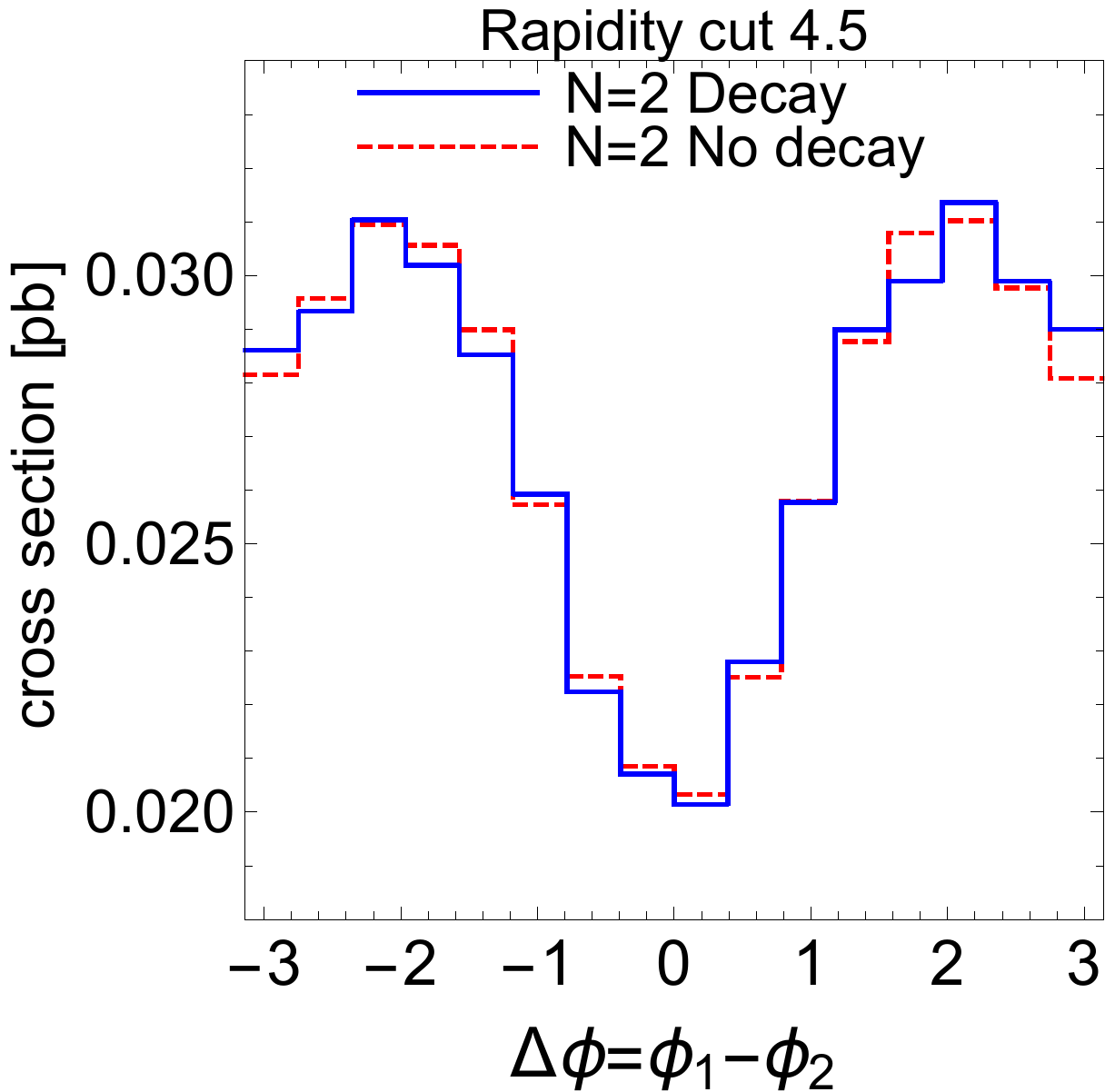}
\caption{\small 
The $\Delta \phi$ distribution produced from the event samples generated by the merging algorithm for $N=2$, including the dilepton $t\bar{t}$ decay (blue solid) and not including $t\bar{t}$ decays (red dotted).
}
\label{fig:delta-phi-result-decay}
\end{figure}

In the studies of the previous section, it is assumed that the top and antitop quarks are stable. This assumption can be justified, since the purposes of the previous section are to confirm that the correlation in $\Delta \phi$ found in the $t\bar{t}+2$ partons matrix elements~\cite{Hagiwara:2013jp} is still visible after including the dominant QCD higher order corrections, and to study the effects of the $t\bar{t}+3$ partons matrix elements on $\Delta \phi$. In this section, the effect of a $t\bar{t}$ decay on $\Delta \phi$ is studied. \\

We produce the $\Delta \phi$ distribution from the event samples including the dilepton $t\bar{t}$ decay generated by the merging algorithm for $N=2$. 
Following ref.~\cite{Chatrchyan:2014gma}, the following kinematic constraints are imposed. Two oppositely charged leptons (muon or electron) are required to have $p_T^{}>20$ GeV within the rapidity range $|y|<2.4$. Jets are rejected if the selected leptons are within a cone of $\Delta R=0.4$ with respect to the jet. 
A jet is identified as a b jet if it contains at least one b quark. 
At least two b jets which fulfill $p_T^{}>30$ GeV and $|y|<2.4$ are required. The additional jets are defined as all jets not including the two highest $p_T^{}$ b jets. The two hardest jets for $\Delta \phi$ are picked up from the additional jets. We set the rapidity range for the additional jets as $|y|<4.5$. In our result, the $t\bar{t}$ invariant mass is calculated from the kinematic information of the top and antitop quarks obtained just before their decays. \\

The result is shown by the blue solid curve in Figure~\ref{fig:delta-phi-result-decay}. The red dotted curve is the result without $t\bar{t}$ decays and is given for comparison. Note that the cross section for the result without $t\bar{t}$ decays is set equal to the one for the result with the $t\bar{t}$ decay.
The figure shows that the effect of the $t\bar{t}$ decay on $\Delta \phi$ is small and the correlation is still visible. \\

%%%%%%%%%%%%%%%%%%%%%%%%%%%%%%%%%%%%%%%%%%%%%%%%%%%%%%%%%%%%%%%%%%%%%%%%%%%%%%%%%%%%%%%%%%%%%%%%%%%%%%%%%%%%%%%%%%%%%%%%%%%%%%%%%%%%%%%%%%%%%%%%%%%%%%%%%%%%%%%%%%%%%%%%%%%%%%%%%%%%%%%%%%%%%%%%%%%%%%%%%%%%%%%%%%%%%%%%%%%%%%%%%%%%%%%%%%%%%%%%%%%%%%%%%%%%%%%%%%%%%%%%%%%%%%%%%%%%%%%%%%%%%%%%%%%%%%%%%%%%%%%%%%%%%%%%%%%%%%%%%%%%%%%%%%%%%%%%%%%%%%%%%%%%%%%%%%%%%%%%%%%%%%%%%%%%%%%%%%%%%%%%%%%%%%%%%%%%%%%%%%%%%%%%%%%%%%%%%%%%%%%%%%%%%%%%%%%%%%%%%%%%%%%%%%%%%%%%%%%%%%%%%%%%%%%%%%%%%%%%%%%%%%%%%%%%%%%%%%%%%%%%%%%%%%%%%%%%%%%%%%%%%%%%%%%%%%%%%%%%%%%%%%%%%%%%%%%%%%%%%%%%%%%%%%%%%%%%%%%%%%%%%%%%%%%%%%%%%%%%%%%%%%%%%%%%%%%%%%%%%%%%%%%%%%%%%%%%%%%%%%%%%%%%%%%%%%%%%%%%%%%%%%%%%%%%%%%%%%%%%%%%%%%%%%%%%%%%%%%%%%%%%%%%%%%%%%%%%%%%%%%%%%%%%%%%%%%%
\section{Conclusion}\label{conclusion}

In this work, the azimuthal angle difference between the two hardest jets (i.e. the two highest $p_T^{}$ jets), $\Delta \phi=\phi_1^{}-\phi_2^{}$ in the top quark pair production process at the 14 TeV LHC has been studied. The event samples are generated by merging the tree level matrix elements for the $t\bar{t}$ plus up to 2 or 3 partons with the parton shower model in PYTHIA8. \\

As tree level merging algorithms, we have implemented the CKKW-L algorithm and a new algorithm. Our new algorithm differs from the CKKW-L algorithm in the strategy for phase space separation and it is designed so that the contribution from the $t\bar{t}+0, 1$ parton matrix elements to the event samples with two or more jets, which we call the contamination, can be more suppressed above the merging scale. Although it has been confirmed that the contamination is more suppressed in our algorithm by numerically comparing the two algorithms, the difference is found not drastic. We find that the contamination is not negligible when the merging scale is set equal to or slightly smaller than the scale of the anti-$k_T^{}$ jet definition. \\

The $\Delta \phi$ distribution is produced from the generated event samples. 
The distribution shows a strong correlation in $\Delta \phi$, as predicted in the previous analysis~\cite{Hagiwara:2013jp} based on the $t\bar{t}+2$ partons tree level matrix elements. This observation confirms that the correlation found in the $t\bar{t}+2$ partons tree level matrix elements~\cite{Hagiwara:2013jp} is still visible after including the dominant QCD higher order corrections and thus can be observed in the experiments. \\

We find a clear difference in the $\Delta \phi$ distribution between the event samples generated by merging the tree level matrix elements for the $t\bar{t}$ plus up to 2 partons and those generated by merging the tree level matrix elements for the $t\bar{t}$ plus up to 3 partons. Furthermore, the difference is found slightly larger, when the rapidity range for jets is more restricted. 
We have studied the origins of the difference, or in other words the effects of the $t\bar{t}+3$ partons matrix elements on $\Delta \phi$. 
When the matrix elements for the $t\bar{t}$ plus up to 2 partons are merged, a non-negligible fraction ($5-10\%$) of the correlation between the two hardest jets can be lost because a jet originating from hard parton showers can have a higher $p_T^{}$ than one of the two jets originating from the two hard partons of the $t\bar{t}+2$ partons matrix elements. 
When the rapidity range for jets is more restricted, a further loss of the correlation arises because one of the two hard jets can be removed by the rapidity cut. 
When the matrix elements for the $t\bar{t}$ plus up to 3 partons are merged, the loss of the correlation due to the above two reasons can be avoided as follows. At first, jets originating from hard parton showers cannot have higher $p_T^{}$ than two of the three jets originating from the three hard partons of the $t\bar{t}+3$ partons matrix elements. Second, even though one of the three hard jets is removed by the rapidity cut, there are still two jets which correctly predict the correlation.
Therefore, it can be concluded that, the $t\bar{t}+3$ partons matrix elements play important roles in predicting $\Delta \phi$ accurately, since they effectively reduce the loss of the correlation. They can be more important as the rapidity cut on jets is set more stringent. \\

We present a method for merging the matrix element event samples which include a $t\bar{t}$ decay as a part of the hard process with the parton shower.  
The effect of the $t\bar{t}$ dilepton decay on $\Delta \phi$ is studied. We have shown that the effect is small when the two hardest jets for $\Delta \phi$ are picked up from all jets not including the two hardest b jets.\\

We note that our findings should be applicable equally to other heavy particle production processes by gluon fusion. We hope that our findings help experimentalists perform the proposal of ref.~\cite{Hagiwara:2013jp} and achieve precise measurement of the CP property of the Higgs boson by using the azimuthal angle correlation between the two hardest jets at the LHC.

%%%%%%%%%%%%%%%%%%%%%%%%%%%%%%%%%%%%%%%%%%%%%%%%%%%%%%%%

\section*{Acknowledgments}
JN is grateful to Rikkert Frederix for many helpful discussions and his encouragement, to Junichi Kanzaki for giving valuable comments on the plots and to Yasuhito Sakaki for fruitful discussions on simulating decays of a top quark. JN would also like to acknowledge the warm hospitality of the Theory Division at CERN, where the first step of the present work was carried out. 
The authors are grateful to KEK Computing Research Center for the management of KEKCC on which most of the simulation in this study was executed. This work is supported in part by JSPS KAKENHI Grant No.~25.4461 of JN.

%\appendix
% \renewcommand{\theequation}{A-\arabic{equation}}
  % redefine the command that creates the equation no.
%  \setcounter{equation}{0}  % reset counter 

%\section*{Appendix}

%\section{a}\label{app1}

%%%%%%%%%%%%%%%%%%%%%%%%%%%%%%%%%%%%%%%%%%%%%%%%%%%%%%%%%%%%%%%%%%%%%%%%%%%%%%%%%%%%%%%%%%%%%%%%%%%%%%

\end{document}